%% file: 2016-05-05-arXiv.tex
\documentclass[10pt,draftcls,onecolumn,romanappendices]{IEEEtran}
\input{Preamble}

\title{On the Aloha throughput-fairness tradeoff}
\author{
Nan Xie,~\IEEEmembership{Member,~IEEE,}
and
Steven Weber,~\IEEEmembership{Senior~Member, IEEE}
\thanks{The authors are with the Department of Electrical and Computer Engineering, Drexel University, Philadelphia, PA 19104, USA. \hfill \newline E-mail: N.~Xie {\sf nx23@drexel.edu}, S.~Weber {\sf sweber@coe.drexel.edu} (contact author).}  
}

\begin{document}

\maketitle
\begin{abstract}
A well-known inner bound of the stability region of the slotted Aloha protocol on the collision channel with $n$ users assumes worst-case service rates (all user queues non-empty).  Using this inner bound as a feasible set of achievable rates, a characterization of the throughput--fairness tradeoff over this set is obtained, where throughput is defined as the sum of the individual user rates, and two definitions of fairness are considered: the Jain-Chiu-Hawe function and the sum-user $\alpha$-fair (isoelastic) utility function.  This characterization is obtained using both an equality constraint and an inequality constraint on the throughput, and properties of the optimal controls, the optimal rates, and the fairness as a function of the target throughput are established.  A key fact used in all theorems is the observation that all contention probability vectors that extremize the fairness functions take at most two non-zero values. 
\end{abstract}


\input{Sec1-Introduction}

\input{Sec2-Model}

\input{Sec3-PropOptCont}
\input{Sec4-Jain}
\input{Sec5-AlphaFair}
\input{Sec6-Conclusion}

\bibliographystyle{../IEEEtran}
\bibliography{../AlohaThroughputFairnessTradeoff-2015-06-27}

\appendices
\input{AppA-PropertiesProofs}
\input{AppB-JainProofs}
\input{AppC-AlphaFairProofs}

\end{document}

%% file: Preamble.tex
\usepackage{mathdefs}
\usepackage{amsmath}
\usepackage{amssymb}
\usepackage[pdftex]{graphicx} 
\usepackage{float}
\usepackage{algpseudocode}
\usepackage{algorithm}

\newtheorem{remark}{Remark}
\newtheorem{corollary}{Corollary}
\newtheorem{definition}{Definition}

\newtheorem{lemma}{Lemma}
\newtheorem{proposition}{Proposition}
\newtheorem{theorem}{Theorem}

\newcommand{\corref}[1]{Cor.~\ref{cor:#1}}
\newcommand{\defref}[1]{Def.~\ref{def:#1}}

\newcommand{\lemref}[1]{Lem.~\ref{lem:#1}}
\newcommand{\prpref}[1]{Prop.~\ref{prp:#1}}
\newcommand{\thmref}[1]{Thm.~\ref{thm:#1}}
\newcommand{\eqnref}[1]{(\ref{eq:#1})}
\newcommand{\secref}[1]{\S\ref{sec:#1}}
\newcommand{\appref}[1]{Appendix~\ref{app:#1}}
\newcommand{\figref}[1]{Fig.~\ref{fig:#1}}

\newcommand{\algoref}[1]{Alg.~\ref{alg:#1}}

%% file: Sec1-Introduction.tex
%%%%%%%%%%%%%%%%%%%%%%%%%%%%%%%%%%%%%%%%%%%%%%
% Section 1. Introduction
%%%%%%%%%%%%%%%%%%%%%%%%%%%%%%%%%%%%%%%%%%%%%%

\section{Introduction}
\label{sec:Introduction}
We investigate the throughput--fairness tradeoff for the slotted Aloha medium access control (MAC) protocol \cite{Abr1977,Rob1975} serving $n$ users contending on a shared collision channel.  Throughput--fairness tradeoffs naturally arise in settings of shared access to a constrained resource, where maximum use of the resource is at odds with fair access to the resource, on account of the inefficiency incurred in resource contention.  In the setting of Aloha, this incurred inefficiency takes the form of wasted slots in which either no user contends (idle) or multiple users contend (collision).  Trivially, maximum throughput of one successful packet per time slot is achieved by the unfair allocation granting one user access and shutting out all other users, while the maximally fair allocation granting each user equal access achieves a throughput that decays to zero in the number of users.  Our focus is on characterizing the tradeoff connecting these two extreme points.

Although modern MAC protocols in use today are far more complex and more sophisticated than Aloha, many of them nonetheless retain at their core the notion of random access, which is the defining characteristic of Aloha.  It is therefore natural, in our opinion, to first analyze the throughput--fairness tradeoff in random access in the canonical setting of slotted Aloha before seeking to characterize such tradeoffs under more complicated protocols.  

One difficulty precluding this goal from being achieved is that the stability region for slotted Aloha on the collision channel remains unknown, in spite of 40+ years of effort.  Because of this, we employ a well-known inner bound on the stability region, obtained by assuming each of the user's queues is nonempty, thereby yielding a worst-case effective service rate seen by each user.  This inner bound is known to be tight for all special cases for which the stability region of slotted Aloha is known.  Even with this simplifying assumption, however, the throughput--fairness problem is still nontrivial on account of the fact that the inner bound cannot be described explicitly.  Rather, the inner bound is given as the image of the function mapping contention probability vectors (controls) to (worst-case) packet transmission rates, over the set of all possible controls.  

\subsection{Related work}

The throughput--fairness tradeoff literature is quite large and diverse, stemming from its relevance to a wide variety of disciplines, including queueing theory, communication networks, optimization, and economics.  As such, we restrict our discussion to only the most pertinent prior work.  Specifically, we summarize prior work on each of the two fairness metrics used in this paper, namely, the Jain-Chiu-Hawe function and the $\alpha$-fair utility function.  

The Jain-Chiu-Hawe fairness measure \cite{JaiChi1984}, hereafter simply Jain's fairness, measures the fairness of an $n$-vector $\xbf = (x_1,\ldots,x_n)$, representing in our context the vector of user rates, as the normalized distance from $\xbf$ to the ``all-rates-equal'' ray passing from the origin through the point $\mathbf{1}$.  This metric has been widely adopted, e.g., \cite{RadLeB2004,GuoShe2013}. 

The $\alpha$-fair parameterized family of utility functions was introduced to the networking community in \cite{MoWal2000}, but is nearly identical to the classic isoelastic utility function in economics \cite{Atk1970}.  The $\alpha$-fair family of utility functions has found profitable use in characterizing throughput--fairness tradeoffs and resource allocation policies in wired and wireless networks, and in that sense may be viewed as part of the larger body of work termed network utility maximization (NUM), e.g., \cite{KelMau1998,LeeChi2006,LeeChi2007,ChiLow2007}.  The basic concept in NUM is to associate with each user a utility (often assumed to be concave increasing) that depends upon the resources allocated to the user, and seek a feasible resource allocation that maximizes the sum-user utility.  In essence, the concavity of the utility function captures the law of diminishing returns for each user, and thus optimizing sum utility over all feasible allocations yields a solution that is ``fair'' in the sense that all users enjoy a common marginal utility.  Returning to $\alpha$-fair utility functions, the parameter $\alpha \geq 0$ controls the ``concavity'' of the utility function, where $\alpha = 0$ corresponds to a linear utility function (no diminishing returns), $\alpha = 1$ is a logarithmic utility function (so-called proportional fair utility), and as $\alpha \to \infty$ the utility-optimal resource allocation is the so-called max-min fair allocation.  Given this, it is natural to think that increasing $\alpha$ would trade sum-user throughput for fairness, although recent work \cite{TanWan2006,BerFar2010,BerFar2011,SedGoh2013} has identified counter-examples. 

Recent work has addressed throughput--fairness tradeoffs using both these fairness measures in the context of {\em downlink} scheduling \cite{SedGoh2013,GuoShe2013}.  In contrast, our focus is on {\em uplink}, and this fundamental difference limits the applicability of many of the results in \cite{SedGoh2013,GuoShe2013} to our setting.  An axiomatic approach to fairness is given in \cite{LanKao2010}, with an insightful discussion contrasting Jain's fairness and $\alpha$-fairness.  

\subsection{Outline and contributions}

The primary contribution of this paper is a characterization of the throughput--fairness (T-F) tradeoff for $n$ users employing slotted Aloha on a collision channel.  This is done through six theorems:
\begin{itemize}
\itemsep=-2pt
\item Theorem 1 (2) gives the T-F tradeoff under Jain's fairness with a throughput equality (inequality) constraint and Theorem 3 gives properties of the optimal controls, optimal rates, and the T-F tradeoff itself.
\item Theorem 4 (5) gives the T-F tradeoff under $\alpha$-fairness with a throughput equality (inequality) constraint, and Theorem 6 gives properties of the optimal controls, optimal rates, and the T-F tradeoff itself.
\end{itemize}

This rest of the paper is organized as follows. The model and problem statement are introduced in \secref{ModelAndProblemStatement}, while \secref{PropOptCont} contains results common to both fairness measures. Building upon \secref{PropOptCont}, the next two sections (\secref{TFtradeoffUsingJain'sFairness}, \secref{TFtradeoffUsingAlphaFairUtilityFunctionsWithThroughputConstraint}) address the Aloha throughput-fairness tradeoff under Jain's and $\alpha$-fairness respectively. Finally \secref{Conclusion} offers a brief conclusion.  Three appendices follow the references, holding long proofs from \secref{PropOptCont}, \secref{TFtradeoffUsingJain'sFairness}, and \secref{TFtradeoffUsingAlphaFairUtilityFunctionsWithThroughputConstraint} respectively. Table \ref{tab:summaryofresults} lists all the results in the paper, and Table \ref{tab:notationgeneral} provides general notation.

\begin{table}
\centering
\caption{Summary of results}
\begin{tabular}{cll}
{\bf \S $\textbf{\#}$}/Result & {\bf Title}/Description \\ \hline \hline
{\bf \secref{ModelAndProblemStatement}}  &  {\bf Model and problem statement} \\ 
\lemref{AllRatesEqualRay} &  ``All-rates'' equal ray's geometric and algebraic properties \\ \hline
{\bf \secref{PropOptCont}} &  {\bf Properties of optimal controls} \\ 
\prpref{SchurConcavityCJFairnessAndAlphaFairUtiityFunction} &  Schur-concavity of fairness measures in rate space \\ 
\prpref{MajorizationProperties} &  Majorization properties under throughput constraint \\ 
\corref{MajorizationCorollary} &  Sufficiency to optimize over $\partial \Lambda$ (or $\partial \Smc$ in control space) \\
\prpref{tputeqconst} &  Properties of controls in $\overline{\partial \Smc_{2}}$ under throughput constraint \\ 
 \prpref{AtMostTwoDistinctNonzeroComponentValues} &  Sufficiency to optimize over the restricted set in \defref{restrictedControls} \\ \hline
{\bf \secref{TFtradeoffUsingJain'sFairness}} &  {\bf Jain-Chiu-Hawe fairness tradeoff} \\ 
\prpref{TFtradeoffNequals2} &  T-F tradeoff under Jain's fairness when $n = 2$ \\ 
\prpref{mono1} &  Monotonicity properties of the Jain's objective over $\partial \Smc_{2}$ \\ 
\thmref{mainresultChiuJain} &  T-F tradeoff under Jain's fairness for general $n \geq 2$ \\ 
\thmref{2} &  No change under throughput inequality constraint \\
\algoref{tfplot} &  Incremental plotting of T-F tradeoff for a sequence of $n$'s  \\
\thmref{ChiuJainTFtradeoffProperties} &  Properties of the Jain T-F tradeoff \\ \hline
{\bf \secref{TFtradeoffUsingAlphaFairUtilityFunctionsWithThroughputConstraint}} &  {\bf $\alpha$-fair network utility maximization} ($\alpha \geq 1$) \\ 
\prpref{AlphaFair-TFtradeoffNequals2} &  T-F tradeoff under $\alpha$-fairness when $n = 2$ \\ 
\prpref{mono2} &  Monotonicity property of the $\alpha$-fair objective over $\partial \Smc_{2}$ \\ 
\thmref{mainresultAlphaFairWhenAlphaGreaterThanOrEqualToOne} &  T-F tradeoff under $\alpha$-fairness for general $n \geq 2$\\ 
\thmref{5} &  Change under throughput inequality constraint \\
\thmref{AlphaFairTFtradeoffProperties} &  Properties of the $\alpha$-fair T-F tradeoff \\ \hline \hline
\end{tabular}
\label{tab:summaryofresults}
\end{table}

%% file: Sec2-Model.tex
%%%%%%%%%%%%%%%%%%%%%%%%%%%%%%%%%%%%%%%%%%%%%%
% Section II. Model and problem statement
%%%%%%%%%%%%%%%%%%%%%%%%%%%%%%%%%%%%%%%%%%%%%%

\section{Model and problem statement}
\label{sec:ModelAndProblemStatement}

This section is divided into the following subsections: an introduction of some general notation in \secref{generalnotation}, a discussion of the Aloha protocol and the collision channel in \secref{AlohaProtocol}, definition of the Aloha stability region $\Lambda_A$ and its inner bound $\Lambda$ in \secref{Lambda}, and the definitions of throughput and fairness in \secref{fairnessmeasures}.

%%%%%%%%%%%%%%%%%%%%%%%%%%%%%%%%%%%%%%%%%%%%%%
\subsection{General notation} 
\label{sec:generalnotation}
All vectors are lowercase and bold and are by default of length $n$. Inequalities between two vectors are understood to hold component-wise. We write $[n]$ to denote $\{1,\ldots,n\}$ for $n \in \Nbb$. The unit vector with a one in position $i$ is denoted $\ebf_i$, for $i \in [n]$. The all-one vector is denoted by $\mathbf{1}$, the uniform distribution $\frac{1}{n} \mathbf{1}$ is denoted $\ubf$, and the all-zero vector is denoted by $\mathbf{0}$.  Euclidean distance is denoted $d (\xbf, \ybf)$.  Cardinality of a set $\Vmc$ is denoted $\vert \Vmc \vert$. We sometimes write $\bar{z}$ to denote $1 - z$. Table \ref{tab:notationgeneral} lists frequently used notation; additional notation will be explained at first use.

\begin{table}
\centering
\caption{General notation}
\begin{tabular}{ll}
Symbol & Meaning \\ \hline
$n$ & number of users; default vector length \\
$[n]$ & positive integers up to $n$ \\
$\xbf$ & vector of user arrival rates\\
$\pbf$ & vector of user contention probabilities\\
$\xbf(\pbf)$ & worst case service rates under control $\pbf$ \eqnref{xofp} \\
$\ubf = \frac{1}{n}\mathbf{1}$ & uniform contention probability vector \\
$\mbf$ & rate vector for $\pbf = \ubf$ (\secref{fairnessmeasures}) \\
$\ebf_{i}$ & unit vector with $1$ in position $i \in [n]$ \\
$d(\xbf,\ybf)$ & Euclidean distance between $\xbf$ and $\ybf$ \\ \hline
$\Lambda$ & Aloha stability region inner bound \eqnref{Lambda} \\
$\partial \Lambda$ & the boundary of the set $\Lambda$ \eqnref{LambdaBoundary} \\
$\Smc$ & closed standard unit simplex (\secref{Lambda}) \\ 
$\partial \Smc$ & probability vectors \eqnref{dS}; efficient controls, c.f., \eqnref{LambdaBoundary} \\ \hline
$T(\xbf)$ & sum-user throughput of $\xbf$ \eqnref{ThroughputOfxbf}\\
$F(\xbf)$ & fairness measure of $\xbf$: $F_{J}$ \eqnref{Chiu-JainFairnessDefinition} or $F_{\alpha}$ \eqnref{falphadef} \\
$\{\theta_{t}\}_{t=1}^{n}$ & critical throughputs \eqnref{criticalthroughputs} \\
\hline
$\Vmc(\pbf)$ & the set of non-zero values in $\pbf$ (\defref{restrictedControls}) \\
$\pbf(p_{s}, k, n')$\!\!\!\!\!&restricted control vectors in \defref{restrictedControls} \\
$\partial\Smc_1$ & efficient controls with $\vert \Vmc(\pbf) \vert = 1$ (\defref{restrictedControls}) \\
$\partial\Smc_2$ & efficient controls with $\vert \Vmc(\pbf) \vert = 2$ (\defref{restrictedControls}) \\
$\partial \Smc_{1,2}$ & efficient controls with $\vert \Vmc(\pbf) \vert \in \{1, 2\}$ (\defref{restrictedControls}) \\
\hline
$\alpha$ & parameter in $\alpha$-fair utility functions \eqnref{AlphaFairUtilityFunction} \\
$\theta$ & target throughput \\
$F^{*}(\theta)$ & optimized fairness given target throughput $\theta$ \\ \hline
\end{tabular}
\label{tab:notationgeneral}
\end{table} 

%%%%%%%%%%%%%%%%%%%%%%%%%%%%%%%%%%%%%%%%%%%%%%
\subsection{The Aloha protocol and the collision channel} 
\label{sec:AlohaProtocol}

Recall a MAC protocol specifies a mechanism to coordinate competing users' access to the shared channel; we consider the finite-user slotted Aloha MAC protocol operating on a collision channel. The protocol parameters are $(n, \xbf, \pbf)$, where $i)$ $n \in \Nbb$ is the number of users, $ii)$ $\xbf \in \Rbb_+^n$ is an $n$-vector denoting the independent arrival rates of users' data packets, which we henceforth call the {\em rate} vector, and $iii)$ $\pbf \in [0,1]^n$ is an $n$-vector indicating the user contention (or channel access) probabilities, which we henceforth call the {\em control} vector.  Each user has an associated packet queue that can hold an infinite number of packets, stored in order of arrival. Each packet will be removed from the queue if and only if it has just been successfully transmitted. The channels are error-free. Time is slotted and synchronized. At the beginning of each time slot, every user with a non-empty queue, say user $i \in [n]$, contends for channel access to the common base station by transmitting its head-of-line packet with a fixed probability $p_i$, independent of anything else. The collision channel assumption means the state of the channel in each time slot may be classified as $i)$ {\em idle} (no one attempts to transmit, either because of having an empty queue or electing not to transmit), $ii)$ {\em collision} (more than one user transmits, and all attempted transmissions fail), or $iii)$ {\em success} (precisely one user transmits, and this attempted transmission succeeds). This ternary feedback is error-free and instantaneous at the end of each time slot. 

\subsection{The stability region $\Lambda_A$ and its inner bound $\Lambda$}
\label{sec:Lambda}

An important yet still open problem is the queueing-theoretic \textit{stability region} (also called the {\em network layer capacity region} \cite[pp.\ 28]{GeoNee2006}) of this model, denoted $\Lambda_A$ ($A$ for Aloha), which contains all arrival rate vectors $\xbf$ that can be stabilized by the protocol, i.e., for each $\xbf \in \Lambda_A$ there exists a control vector $\pbf$ that stabilizes each of the $n$ queues.  The stability region is open even for the case of independent arrival process and $n > 2$ users.  A summary of the history of this problem is provided in \cite{LuoEph2006}, with compelling recent work including \cite{BorMcD2008,KomMaz2013} among others.  

As $\Lambda_A$ is unknown, we employ a suitable {\em inner bound} on $\Lambda_A$ as a {\em proxy} for the stability region of slotted Aloha.  This inner bound, denoted $\Lambda$ below, has been proved to coincide with the exact stability region for all special cases for which the stability region is known (\cite{TsyMik1979,Ana1991}), and has been conjectured (\cite[\S V]{RaoEph1988}, \cite[\S V Thm.\ 2]{LuoEph2006}) to in fact {\em be} the stability region, $\Lambda_A$.  The set $\Lambda$ is defined as:
\begin{equation} 
\label{eq:Lambda}
\Lambda \equiv \left\{ \xbf \in \Rbb_{+}^{n} : \exists \pbf \in [0,1]^n : ~ x_i \leq p_i \prod_{j \neq i} (1-p_j), ~ \forall i \in \left\{1, \ldots, n\right\} \right\}.
\end{equation}
The expression $p_i \prod_{j \neq i} (1-p_j)$ is the worst-case service rate for user $i$'s queue, namely the service rate assuming all other users have non-empty queues and thus all users are eligible for channel contention.  In particular, user $i$'s transmission is successful in such a time slot if user $i$ elects to contend (with probability $p_i$) and each other user $j \neq i$ does not contend (each with independent probability $1-p_j$).  Clearly, $\Lambda$ is an inner bound, since an arrival rate that is stabilizable under the worst-case service rate is certainly stabilizable under a better service rate.  It may be shown \cite[\S II, Prop.\ 2]{XieWeb2014sub} that an equivalent definition of $\Lambda$ is to change all the inequalities to equality, i.e., $\xbf \in \Lambda$ if and only if there exists a $\pbf \in [0,1]^n$ for which $\xbf = \xbf(\pbf)$, where 
\begin{equation}
\label{eq:xofp}
x_i(\pbf) \equiv p_i \prod_{j \neq i} (1- p_j), ~ i \in [n].
\end{equation}
We refer to such a $\pbf$ as a {\em (critical compatible) control} for $\xbf$.\footnote{More generally, we define a {\em compatible control} for $\xbf$ as a control vector $\pbf$ for which $\xbf \leq \xbf(\pbf)$.  In this paper we only employ critical compatible controls, and as such we often refer to $\pbf$ satisfying $\xbf = \xbf(\pbf)$ simply as a {\em control} for $\xbf$.}  Based on the above definition of $\Lambda$, testing whether or not a candidate $\xbf$ is or is not in $\Lambda$ is equivalent to the solvability of $\xbf = \xbf(\pbf)$ over $\pbf \in [0,1]^n$.  The definition of $\Lambda$ is therefore {\em implicit}, in the sense that testing membership $\xbf \in \Lambda$ requires establishing the existence (or not) of a suitable control $\pbf$.  When addressing throughput--fairness tradeoffs we will be optimizing an objective function over $\Lambda$, which thus becomes the {\em feasible set} for the optimization.  The implicit characterization of $\Lambda$ is what makes the corresponding throughput--fairness tradeoff optimization problem non-trivial.  The natural solution, which we employ, is to make $\pbf \in [0,1]^n$ the optimization variable, thereby requiring the corresponding nonlinear compositions on both the throughput and fairness functions, i.e., $T(\xbf(\pbf))$ and $F(\xbf(\pbf))$, defined below.  To emphasize this distinction, we refer to $\xbf$ as a rate vector in {\em rate space}, and $\pbf$ as a control vector in {\em control space}.

The boundary of $\Lambda$ in $\Rbb_{+}^n$ is denoted $\partial\Lambda$ and is characterized \cite{Pos1985} as
\begin{equation} 
\label{eq:LambdaBoundary}
\partial \Lambda = \left\{ \xbf \in \Rbb_{+}^{n} : \exists \pbf \in \partial \Smc : ~ x_i = p_i \prod_{j \neq i} (1-p_j), ~ \forall i \in \left\{1, \ldots, n\right\} \right\}, 
\end{equation}
where $\Smc \equiv \{\zbf \in \Rbb_{+}^{n}: \sum_{i=1}^{n} z_{i} \leq 1\}$ denotes the ``standard'' unit simplex, and its ``face'', denoted 
\begin{equation}
\label{eq:dS}
\partial \Smc \equiv \{\zbf \in \Rbb_{+}^{n}: \sum_{i=1}^{n} z_{i} = 1\},
\end{equation}
is the set of \textit{probability vectors} on $[n]$.  Thus, Pareto efficient throughputs, i.e., $\xbf \in \partial\Lambda$, are achieved by and only by controls that are probability vectors, i.e., $\pbf \in \partial\Smc$.  For this reason, we call $\partial\Smc$ the set of {\em efficient controls}.

It may be helpful to visualize $\Lambda$ and its boundary $\partial \Lambda$ using \figref{JainTwoUsers} (\secref{JainPrelim}) for the $n = 2$ case, where they are shown as the light blue shaded area and the brown curve respectively. In addition, the following lemma (the proof of which is straightforward and is omitted), used in some proofs, is relevant to $\Lambda$ in that it implies: $a)$ geometrically, the ray from the origin through $\mathbf{1}$ (the ``all-rates equal'' ray) resides inside $\Lambda$ until it hits the boundary $\partial \Lambda$ at $\xbf = \frac{\theta_{n}}{n} \mathbf{1}$ (see \eqnref{criticalthroughputs} and the discussion below), shown in \figref{JainTwoUsers} as the black dot, and $b)$ there only exist(s) two (one) control(s) $\pbf$ for any rate vector $\xbf$ on this ray segment that lies inside (on the boundary of) $\Lambda$, in the sense of \eqnref{xofp}.

\begin{lemma}
\label{lem:AllRatesEqualRay}
Let an integer $n \geq 2$ be given. The function $p (1-p)^{n-1}$ for $p \in [0,1]$ is increasing when $p \in [0, 1/n]$ and decreasing when $p \in [1/n,1]$, with the maximum $\frac{1}{n} \left(1 - \frac{1}{n} \right)^{n-1}$ attained at $p = 1/n$.
\end{lemma}

%%%%%%%%%%%%%%%%%%%%%%%%%%%%%%%%%%%%%%%%%%%%%%
\subsection{Throughput and two fairness measures}
\label{sec:fairnessmeasures}

The sum-user throughput of any rate vector $\xbf \in \Lambda$ is defined as:
\begin{equation}
\label{eq:ThroughputOfxbf}
T(\xbf) \equiv \sum_{i=1}^{n} x_i.
\end{equation}
Note $T(\xbf) \in [0,1]$ since, by the definition of the collision channel, there is at most one successful transmission on the channel in each time slot.  We define the vector $\boldsymbol\theta = (\theta_1,\ldots,\theta_n)$ with $\theta_1 = 1$ and 
\begin{equation}
\label{eq:criticalthroughputs}
\theta_t \equiv \left( 1-1/t \right)^{t-1}, ~ t \in \{2,\ldots,n\}
\end{equation}
as the vector of {\em critical throughputs}.  Observe $1 = \theta_1 > \cdots > \theta_n > 1/\erm$.  Define the rate vector $\mbf \equiv \frac{\theta_n}{n}\mathbf{1} = \xbf(\ubf)$ associated with $\theta_n$, i.e., $\mbf$ is the rate vector for the uniform control $\ubf$, with corresponding throughput $T(\mbf) = \theta_n$.  Geometrically, $\mbf$ is the unique intersection of the ray from the origin through $\mathbf{1}$ (the ``all-rates equal'' ray) with $\partial\Lambda$.

The fairness of $\xbf$ is denoted $F(\xbf)$; we will employ the following two fairness definitions in this paper.  The first, {\em Jain-Chiu-Hawe fairness} \cite{JaiChi1984}, henceforth referred to simply as Jain's fairness and denoted $F_J(\xbf)$, is a now classic means of quantifying the fairness of a resource allocation $\xbf$: 
\begin{equation} 
\label{eq:Chiu-JainFairnessDefinition}
F_J(\xbf) = \frac{T(\xbf)^2}{n \|\xbf\|^2}. 
\end{equation}
The Jain's fairness function has the following properties: $i)$ scale invariance, i.e., $F_J(\beta \xbf) = F_J(\xbf)$ for any $\beta \in \Rbb_{++}$; and $ii)$ boundedness, i.e., $F_J \in [1/n,1]$, with $F_J(\beta \ebf_i) = 1/n$ for any $i \in [n]$ and $F_J(\beta \mathbf{1}) = 1$ for any $\beta \in \Rbb_{++}$.  

The second fairness measure, the \textit{$\alpha$-fair} sum-user utility function, defined as 
\begin{equation}
\label{eq:falphadef}
F_{\alpha}(\xbf) \equiv \sum_{i=1}^n U_{\alpha}(x_i),
\end{equation}
for $\alpha \geq 0$, is the sum-user utility of the allocation $\xbf$, where the (common) per-user utility functions are defined, for $\alpha \in \Rbb$, as:\footnote{Note that $\lim_{\alpha \to 1} U_{\alpha}(x) = \pm 1/0$, i.e., is undefined, and not equal to $U_1(x) = \log x$.  One way to rectify this discrepancy is to modify the definition to include a constant shift, e.g., $\tilde{U}_{\alpha}(x) \equiv \frac{1}{1-\alpha}\left(x^{1-\alpha} - 1\right)$, which is known as the \textit{isoelastic utility function} in economics. As is conventional in the networking literature, we omit this constant as it has no effect on the extremizers.} 
\begin{equation}
\label{eq:AlphaFairUtilityFunction}
U_{\alpha}(x) = \left\{ \begin{array}{ll}
\log (x), & \alpha = 1 \\
\frac{1}{1-\alpha} x^{1-\alpha}, & \alpha \neq 1
\end{array} \right..
\end{equation}
Maximization of sum-user utility over a set of feasible allocations, for any  concave increasing utility function $U_{\alpha}(x)$, often {\em implicitly} enforces a throughput--fairness tradeoff.  For example, the cases $\alpha = 0,1,\infty$ have corresponding optimal solutions that maximize throughput, proportional fairness (log-utility), and max-min fairness, respectively.  It is for this reason that we refer to $F_{\alpha}(\xbf)$ as a fairness function.

Observe that under the throughput equality constraint $T(\xbf) =\theta$, the objective $F_J(\xbf)$ is inversely proportional to $F_{-1}(\xbf)$, i.e., $F_{\alpha}(\xbf)$ in \eqnref{falphadef} with $\alpha=-1$, and as such maximizing $F_J(\xbf)$ under $T(\xbf) =\theta$ is equivalent, in the sense of having the same extremizers, to minimizing $F_{-1}(\xbf)$. Even though $F_{\alpha}$ only possesses the desirable properties of a utility function for $\alpha \geq 0$, this equivalence allows us to study extremizers of $F_J$ and $F_{\alpha}$ ($\alpha \geq 0$) under a unified framework, as in \prpref{AtMostTwoDistinctNonzeroComponentValues} in \secref{PropOptCont}.

The general throughput-fairness tradeoff for slotted Aloha, using the proxy stability region $\Lambda$ as the feasible set of arrival rate vectors, is the Pareto frontier of the parametric plot $(T(\xbf), F(\xbf))$ over $\xbf \in \Lambda$.  An equivalent alternate formulation of the throughput--fairness tradeoff is to seek to maximize $F(\xbf)$ over $\xbf \in \Lambda$ such that $T(\xbf) = \theta$, for $\theta \in (0,1)$ a target throughput constraint.  We omit $\theta = 0$ and $\theta = 1$ as target throughputs as both correspond to trivial edge cases. In fact, we will address two types of throughput constraints in this paper: $i)$ a throughput equality constraint $T(\xbf) = \theta$, and $ii)$ a throughput inequality constraint $T(\xbf) \geq \theta$.  The equality constraint is used, as mentioned above, to characterize the throughput--fairness tradeoff, while the inequality constraint admits a natural operational interpretation: allocate ``resources'' as fairly as possible subject to the sum throughput exceeding a minimum requirement.  As we will show, there are parameter regimes wherein these two problems are the same, and regimes where they are different.  

Finally, observe that $\Lambda$, $F(\xbf)$, and $T(\xbf)$ are each permutation invariant, and as such any extremizer $\xbf^*$ that maximizes fairness under a throughput constraint is permutation invariant, meaning any permutation of $\xbf^*$ is likewise an extremizer.  

{\bf Further notes about notation}. Auxiliary functions (typically named as $f_{1}$, $f_{2}$, etc.) used in proofs are understood to be \textit{internal} meaning a different function with the same name might be used in a different proof. The following inequality about the natural logarithm function is frequently used in the paper:
\begin{equation}
\label{eq:InequalityOnLog}
\log (1 + z) \leq z, ~ \text{ for all } z > -1,
\end{equation}
which is strict unless $z =0$. Finally, we use $F^{*}(\theta)$ to represent the maximum fairness for a given target throughput $\theta$, which is not to be confused with $F(\xbf)$ defined in \eqnref{Chiu-JainFairnessDefinition} and \eqnref{falphadef}.

%% file: Sec3-PropOptCont.tex
%%%%%%%%%%%%%%%%%%%%%%%%%%%%%%%%%%%%%%%%%%%%%%
% Properties of optimal controls
%%%%%%%%%%%%%%%%%%%%%%%%%%%%%%%%%%%%%%%%%%%%%%
\section{Properties of optimal controls}
\label{sec:PropOptCont}

We use the framework of majorization in \secref{MajorizationAttempt} to establish that it suffices to restrict the control space from $[0,1]^n$ to the set of {\em efficient controls}, namely $\partial\Smc$ \eqnref{dS}, and then use Karush-Kuhn-Tucker (KKT) conditions in \secref{opttputconst} to establish structural properties of those controls that extremize $F_{\alpha}(\xbf)$ for $\alpha \in (-\infty, -1] \cup [1, \infty)$ under a throughput constraint.

\subsection{A majorization approach} 
\label{sec:MajorizationAttempt}
We address the Aloha T-F tradeoff problem through the lens of \textit{majorization} \cite{MarOlk2011}, the origins of which are rooted in questions of fairness.
Majorization defines a partial order on the set of vectors with the same length and sum of components. More precisely, $\abf$ is majorized by $\bbf$, denoted $\abf \prec \bbf$, if $\sum_{i=1}^k a_{[k]} \leq \sum_{i=1}^k b_{[k]}$ for all $k \in [n]$, where $a_{[k]}$ is the $k^{\rm th}$ component of $\abf$ sorted in nonincreasing order.  For example, the ``quasi--uniform'' probability vectors (in $\partial\Smc$) below are majorized as \cite[pp.\ 9]{MarOlk2011}: 
\begin{equation}
\left(\frac{1}{n}, \ldots, \frac{1}{n} \right) \prec \left(\frac{1}{n-1}, \ldots, \frac{1}{n-1}, 0\right) \prec \left(\frac{1}{2}, \frac{1}{2}, 0, \ldots, 0\right) \prec \cdots \prec  \left(1, 0, \ldots, 0 \right).
\end{equation}
As the above example suggests, in many contexts the statement $\xbf \prec \ybf$ may be interpreted as $\xbf$ is {\em more fair} than $\ybf$, in the sense that the components of vector $\xbf$ are more nearly equal than those of $\ybf$.  It is therefore natural to try to study our T-F tradeoff within the framework of majorization.  The class of Schur (concave or convex) functions are symmetric functions that preserve majorization, i.e., $F$ is Schur concave (convex) if $F(\xbf) \geq F(\ybf)$ ($F(\xbf) \leq F(\ybf)$) for all $(\xbf,\ybf)$ such that $\xbf \prec \ybf$.  The following result, taken from \cite{LanKao2010} (c.f. Thm.\  A.\ 4 in Ch.\ 3 of \cite{MarOlk2011}), indicates the relevance of Schur concavity to our problem (note Schur concavity is preserved under summation, c.f.\ \eqnref{falphadef}).
\begin{proposition} 
\label{prp:SchurConcavityCJFairnessAndAlphaFairUtiityFunction}
The Jain's fairness function \eqref{eq:Chiu-JainFairnessDefinition} and $\alpha$-fair utility function \eqref{eq:AlphaFairUtilityFunction} for $\alpha \geq 0$ are Schur concave in $\xbf$.
\end{proposition}
\begin{remark}
\label{rem:controlrestrictFromOutset}
An immediate consequence of this result is that it allows us to restrict the set of feasible controls from $[0,1]^n$ to $[0,1)^n$.  First, observe that if there are multiple users contending with probability one, then the corresponding rate vector is $\xbf = \mathbf{0}$, and as such $T(\xbf) = 0$, meaning such points cannot achieve any target throughput $\theta \in (0,1)$.  Second, if there is a unique user, say $i$, with $p_i = 1$ (i.e., $p_j \in [0,1)$ for all $j \neq i$), then $\xbf = \pi_i \ebf_i$, where $\pi_i = \prod_{j \neq i} (1-p_j)$.  But, such an $\xbf$ majorizes every other feasible point in rate space, and thus will not maximize either of our fairness objectives.  
\end{remark}
The following result establishes two key facts.  First, it suffices to consider only efficient controls, $\pbf \in \partial\Smc$, for maximizing fairness under a throughput (equality) constraint.  Second, there is no majorization relationship among any two efficient controls that both satisfy the throughput constraint. Thus, majorization does not by itself solve the T-F tradeoff optimization problem.

\begin{proposition} 
\label{prp:MajorizationProperties}
Fix the number of users $n$ and the target throughput $\theta \in (\theta_n,1)$.  Define the hyperplane $\Hmc_{\theta} = \{ \xbf \in \Rbb^n_+ : T(\xbf) = \theta \}$ of rate vectors with throughput $\theta$.  Define $\Lambda_{\theta} = \Lambda \cap \Hmc_{\theta}$, $\partial\Lambda_{\theta} = \partial \Lambda \cap \Hmc_{\theta}$, and $\Lambda_{\theta}^{\rm int} = \Lambda_{\theta} \setminus \partial\Lambda_{\theta}$ as the set of stable, stable efficient, and stable inefficient rate vectors with throughput $\theta$, respectively. Then
\begin{enumerate}
\item for any $\xbf \in \Lambda_{\theta}^{\rm int}$, there exists some $\xbf' \in \partial \Lambda_{\theta}$ such that $\xbf' \prec \xbf$;
\item for any distinct $\xbf, \xbf'$ both in $\partial \Lambda_{\theta}$, it holds that $\xbf \not \prec \xbf'$ and $\xbf' \not \prec \xbf$.
\end{enumerate}
\end{proposition}
The proof is found in \appref{MajorizationAttemptPf}. One consequence is the following.
\begin{corollary}
\label{cor:MajorizationCorollary}
When maximizing either Jain's fairness \eqnref{Chiu-JainFairnessDefinition} or the $\alpha$-fair objective \eqnref{falphadef} over $\Lambda$ subject to a throughput equality constraint $T(\xbf) = \theta$ for $\theta \in [\theta_{n}, 1)$, it suffices to restrict the feasible set the set of points on the boundary of $\Lambda$ that satisfy the throughput constraint, i.e., to $\partial \Lambda_{\theta}$ (defined in \prpref{MajorizationProperties}). This then implies an optimal control, $\pbf^*$, defined in \secref{JainMain}, is in $\partial \Smc$.
\end{corollary}
This corollary follows almost immediately from \prpref{SchurConcavityCJFairnessAndAlphaFairUtiityFunction} and \prpref{MajorizationProperties} (item $1)$) taking into account the fact that $\pbf \in \partial \Smc$ iff $\xbf(\pbf) \in \partial \Lambda$ \cite{Pos1985}. An independent proof is given in \appref{MajorizationAttemptPf} for the case of Jain's fairness, highlighting the geometric intuition behind the result.

%%%%%%%%%%%%%%%%%%%%%%%%%%%%%%%%%%%%%%%%%%%%%%
\subsection{Optimal controls under a throughput constraint}
\label{sec:opttputconst}

In this subsection we present two results that apply to both the Jain's fairness analysis in \secref{TFtradeoffUsingJain'sFairness} and the $\alpha$-fair analysis in \secref{TFtradeoffUsingAlphaFairUtilityFunctionsWithThroughputConstraint}.  First, we define some useful restrictions of the feasible set of controls in \defref{restrictedControls}; this restriction is an essential component in most of our subsequent proofs.  Second, in \prpref{tputeqconst} we present some properties associated with the throughput constraint $T(\xbf(\pbf)) = \theta$ over this restricted set.  Finally, \prpref{AtMostTwoDistinctNonzeroComponentValues} establishes that the optimal controls for both fairness objectives will lie in the restricted set in \defref{restrictedControls}.  

\begin{definition}
\label{def:restrictedControls}
Let $\pbf \in [0,1)^n$ be a control, and define the following:
\begin{enumerate}
\itemsep=-2pt
\item $\Vmc(\pbf) = \bigcup_{i \in [n]} \{p_i\} \setminus \{0\}$.
Thus $\Vmc(\pbf)$ ($|\Vmc(\pbf)|$) denotes the set (number) of distinct nonzero values\footnote{$|\Vmc(\pbf)|$ is the number of distinct nonzero values, not the number of indices taking nonzero values.} in $\pbf$.
\item $\partial\Smc_1 = \{ \pbf \in \partial\Smc : |\Vmc(\pbf)| = 1\}$ denotes the set of efficient controls with exactly {\em one} distinct nonzero value.  Note $\partial\Smc_1$ consists of all vectors $\pbf$ (and their permutations) of the form $p_i = 1/n'$ for $i \in [n']$ and $p_i = 0$ for $i \in \{n'+1,\ldots,n\}$, for $n' \in [n]$. 
\item $\partial\Smc_2 = \{ \pbf \in \partial\Smc : |\Vmc(\pbf)|=2\}$ denotes the set of efficient controls with exactly {\em two} distinct nonzero values.  These two values are denoted $p_s, p_l$ (for ``small'' and ``large'', respectively) with $0 < p_s < p_l < 1$.  Moreover, any such $\pbf$ has a total of $n'$ nonzero components, of which $k$ take value $p_s$ and $n'-k$ take value $p_l$, for some $k \in [n'-1]$ and some $n' \in \{2,\ldots,n\}$, and $p_s \in (0,1/n')$.  Since $\pbf \in \partial\Smc$, it follows that $k p_s + (n'-k) p_l = 1$, or equivalently, 
\begin{equation}
\label{eq:plofpskn}
p_l = p_l(p_s,k,n') \equiv \frac{1 - k p_{s}}{n' - k}.
\end{equation}
We call $(p_s,k,n')$ the three free parameters which together characterize a $\pbf \in \partial\Smc_2$, and write $\pbf(p_s,k,n')$ to denote a $\pbf$ with those parameters.  The rates associated with controls $p_s,p_l$ are denoted $x_s,x_l$, respectively, with
\begin{eqnarray}
x_s &=& x_s(p_s,k,n') \equiv p_s (1-p_s)^{k-1}(1-p_l)^{n'-k} \nonumber \\
x_l &=& x_l(p_s,k,n') \equiv p_l (1-p_s)^k (1-p_l)^{n'-k-1} \label{eq:xsxl}
\end{eqnarray}
and it is easily shown that $x_s < x_l$.  

\item $\partial \Smc_{1,2} = \{ \pbf \in \partial\Smc : |\Vmc(\pbf)| \leq 2\} $ denotes the set of efficient controls with \textit{at most two} distinct nonzero values. Because $\pbf \in \partial \Smc$ it follows that $ |\Vmc(\pbf)| \neq 0$, and thus $\partial \Smc_{1,2}= \partial \Smc_1 \cup \partial \Smc_2$. Observe $\partial \Smc_{1}$ may be viewed as the limiting case of $\partial \Smc_{2}$ as $p_s \uparrow 1/n'$.  Therefore $\partial \Smc_{1,2}$ may equivalently be defined as the closure of $\partial \Smc_{2}$ and thus $\pbf \in \partial \Smc_{1,2}$ may also be parameterized by $(p_{s}, k, n')$ with the modification that $p_{s} \in (0, 1/n']$. In fact, we will use $\partial \Smc_{1, 2}$ and $\overline{\partial \Smc_{2}}$ interchangeably with the former highlighting $\vert \Vmc(\pbf) \vert \in \{1, 2\}$ and the latter emphasizing $p_{s}$ can take the boundary value $1/n'$.
\end{enumerate}
\end{definition}

Following the $\pbf(p_{s}, k, n')$ parameterization in \defref{restrictedControls}, we further define the following shorthands to be used:
\begin{eqnarray} 
\label{eq:ratioShortHands}
& & r_{x} = r_{x} (p_{s}, k, n') \equiv \frac{x_{l}}{x_{s}} = \frac{p_{l} (1- p_{s})}{p_{s} (1- p_{l})} \nonumber \\
& & r_{\bar{p}} = r_{\bar{p}} (p_{s}, k, n') \equiv \frac{1 - p_{s}}{1 - p_{l}}.
\end{eqnarray}

The following proposition gives properties of the solution of the throughput equality constraint $T(\xbf(\pbf)) = \theta$ over $\pbf \in \overline{\partial\Smc_2}$.  Leveraging the $(p_s,k,n')$ parameterization in \defref{restrictedControls}, we define (for fixed $n' \in \{2,\ldots,n\}$):
\begin{eqnarray}
\label{eq:signedtputgap}
T(p_s,k,n') & \equiv & T(\xbf(\pbf(p_s,k,n'))) \\
\label{eq:tputgaprange}
\Rmc(k,n') & \equiv & \{T(p_s,k,n') : p_s \in (0,1/n']\}
\end{eqnarray}
for $p_s \in (0,1/n']$ and $k \in [n'-1]$.  Note $\Rmc(k,n')$ is the set of achievable throughputs over $\pbf \in \overline{\partial\Smc_2}$ with fixed $(k,n')$, i.e., the image of $T(p_s,k,n')$ over $p_s \in (0,1/n']$.  This image is a subinterval of $[0,1]$ on account of the continuity of $T(p_s,k,n')$ in $p_s$.

\begin{figure}[!h]
\centering
\includegraphics[width=\textwidth ]{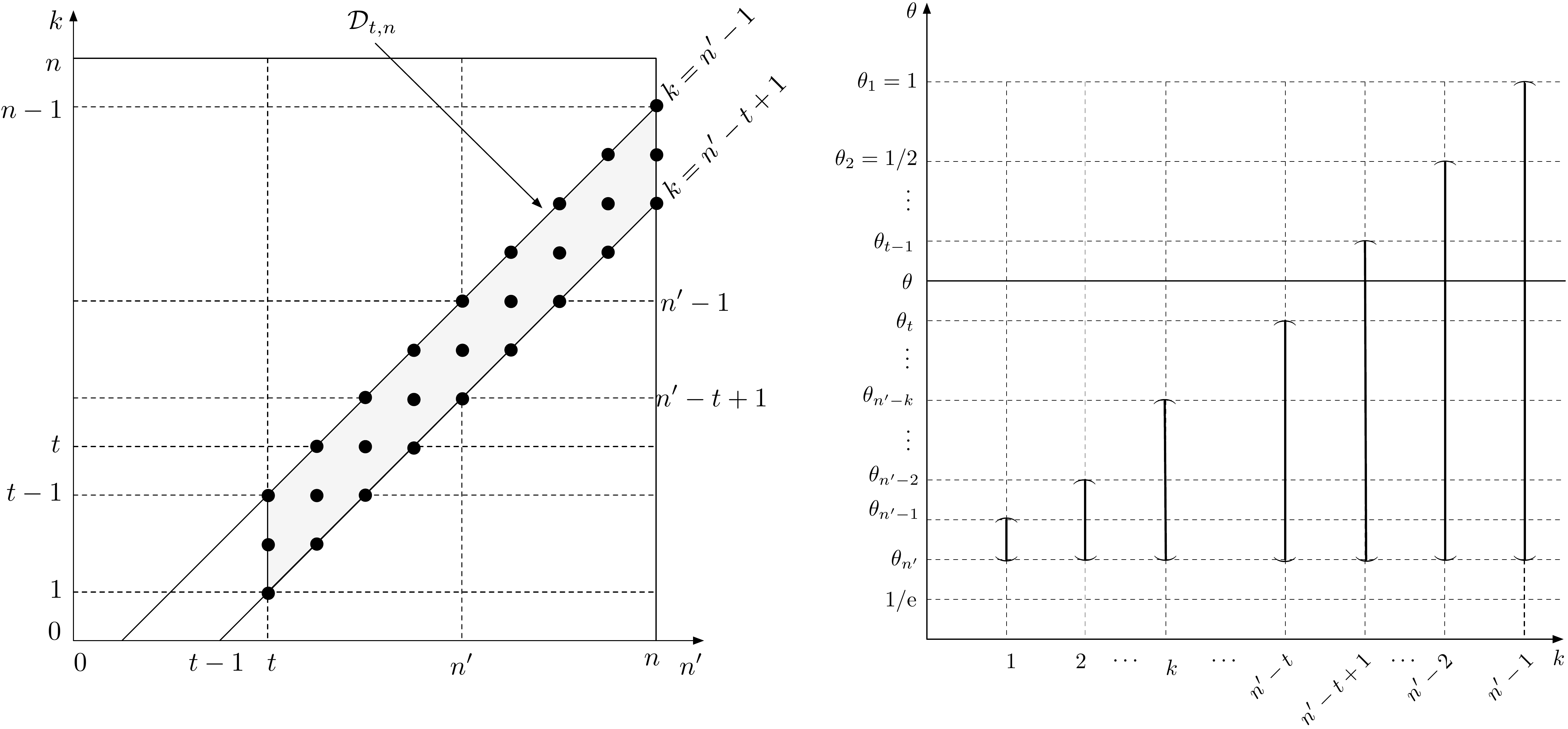}
\caption{{\bf Left:} Illustration of the region $\Dmc_{t,n}$ \eqnref{DtnpRegion} (to scale, the figure shows the case $t=4$ and $n=12$, with the value $n'=8$ selected on the $n'$ axis).
{\bf Right:} Illustration that $k \in \{n'-t+1,\ldots,n'-1\}$ is necessary and sufficient (when $n' \geq t$) for $\theta \in [\theta_t,\theta_{t-1})$ to intersect $\Rmc(k,n') = [\theta_{n'},\theta_{n'-k})$ (shown as solid vertical intervals) in \eqnref{tputgaprange}.} 
\label{fig:DtnpRegion}
\end{figure}

\begin{proposition}
\label{prp:tputeqconst}
Assume $\pbf \in \overline{\partial \Smc_{2}}$ is parameterized using $(p_s,k,n')$ as in \defref{restrictedControls}.
\begin{enumerate}
\itemsep=-2pt
\item Fix $k$, $n'$. The throughput $T(p_s,k,n')$ is monotone decreasing in $p_s \in (0,1/n']$, and as such at most one $p_s \in (0,1/n']$ will solve $T(p_s,k,n') = \theta$.  This unique $p_s$, when it exists, is denoted by $p_s(k,n',\theta)$, and is the solution to
\begin{equation}
\label{eq:psfunctheta}
T(p_s(k,n',\theta),k,n') = \theta,
\end{equation}
which can be expressed as an order-$n'$ polynomial (in $p_{s}$) equation.
\item Now only fix $n'$. The range of achievable throughputs for a given $k$ is $\Rmc(k,n') = [\theta_{n'},\theta_{n'-k})$, which is an increasing (in $k$) nested sequence of intervals: $\Rmc(1,n') \subseteq \cdots \subseteq \Rmc(n'-1,n')$.
\item For $\theta \in [\theta_t,\theta_{t-1})$, for some $t \in \{2,\ldots,n\}$, the set of $(k,n')$ pairs for which there exists $p_s \in (0,1/n']$ such that $T(p_s,k,n') = \theta$ is 
\begin{equation}
\label{eq:DtnpRegion}
\Dmc_{t,n} \equiv \bigcup_{n' \in \{t,\ldots,n\}} \{ (k,n') \in \Nbb^2 : k \in \{n'-t+1,\ldots,n'-1\}\},
\end{equation}
and is illustrated in \figref{DtnpRegion} (left).
\end{enumerate}
\end{proposition}

The proof is in \appref{opttputconstPf}.  The following proposition shows that optimal controls for both the Jain's fairness and $\alpha$-fair objectives will lie in the restricted set of \defref{restrictedControls}.

\begin{proposition} 
\label{prp:AtMostTwoDistinctNonzeroComponentValues}
Consider the following two extremization (maximization or minimization) problems, each parameterized by $\alpha \in (-\infty, -1] \cup [1, \infty)$ and $\theta \in (0,1)$:
\begin{equation}
\label{eq:extremization}
\underset{\pbf \in [0,1)^n}{\text{extremize~}} F_{\alpha}(\xbf(\pbf)) : T(\xbf(\pbf)) \geq \theta, ~~~~
\underset{\pbf \in [0,1)^n}{\text{extremize~}} F_{\alpha}(\xbf(\pbf)) : T(\xbf(\pbf)) = \theta.
\end{equation}
$i)$ For {\em both} the inequality and equality constrained problems above, a necessary condition for $\pbf$ to extremize \eqnref{extremization} is $\vert \Vmc(\pbf) \vert \leq 2$.
$ii)$ For the inequality constrained problem: if an optimizer $\pbf^*$ of \eqnref{extremization} (left) has the property that $\vert \Vmc(\pbf^*) \vert = 2$, then the throughput constraint holds with equality, i.e., $T(\xbf(\pbf^*)) = \theta$.
\end{proposition}

The proof is in \appref{opttputconstPf}.

%% file: Sec4-Jain.tex
%%%%%%%%%%%%%%%%%%%%%%%%%%%%%%%%%%%%%%%%%%%%%%
% Section III. Jain-Chiu-Hawe fairness tradeoff
%%%%%%%%%%%%%%%%%%%%%%%%%%%%%%%%%%%%%%%%%%%%%%

\section{Jain-Chiu-Hawe fairness tradeoff} 
\label{sec:TFtradeoffUsingJain'sFairness}

Recall from \secref{fairnessmeasures} that maximizing $F_J(\xbf)$ \eqnref{Chiu-JainFairnessDefinition} under a throughput equality constraint $T(\xbf) = \theta$ is equivalent, in the sense of having the same extremizers, to minimizing $F_{-1}(\xbf)$ \eqnref{falphadef}, i.e., $\alpha=-1$, under the same constraint.  As mentioned in \secref{Lambda}, any $\xbf \in \Lambda$ may be expressed as $\xbf(\pbf)$ \eqnref{xofp} for some $\pbf \in [0,1]^n$.  Thus, an equivalent formulation of the Jain throughput--fairness optimization problem for $n$ users with target throughput $\theta \in (0,1)$ is:
\begin{equation}
\label{eq:TFtradeoffProblemFormulation}
\boxed{
\min_{\pbf \in [0,1)^n} F_{-1}(\xbf(\pbf)) = \frac{1}{2} \sum_{i=1}^{n} x_{i}(\pbf)^{2} ~\mbox{s.t.}~ T(\xbf(\pbf)) = \theta.}
\end{equation}
This section is comprised of three subsections. We give: $i)$ preliminary results in \secref{JainPrelim}, $ii)$ the main results in \secref{JainMain}, and $iii)$ some additional properties of the Jain throughput-fairness tradeoff in \secref{JainProp}.

%%%%%%%%%%%%%%%%%%%%%%%%%%%%%%%%%%%%%%%%%%%%%%
\subsection{Preliminary results}
\label{sec:JainPrelim}

\begin{figure}[!h]
\centering
\includegraphics[width=0.49 \textwidth ]{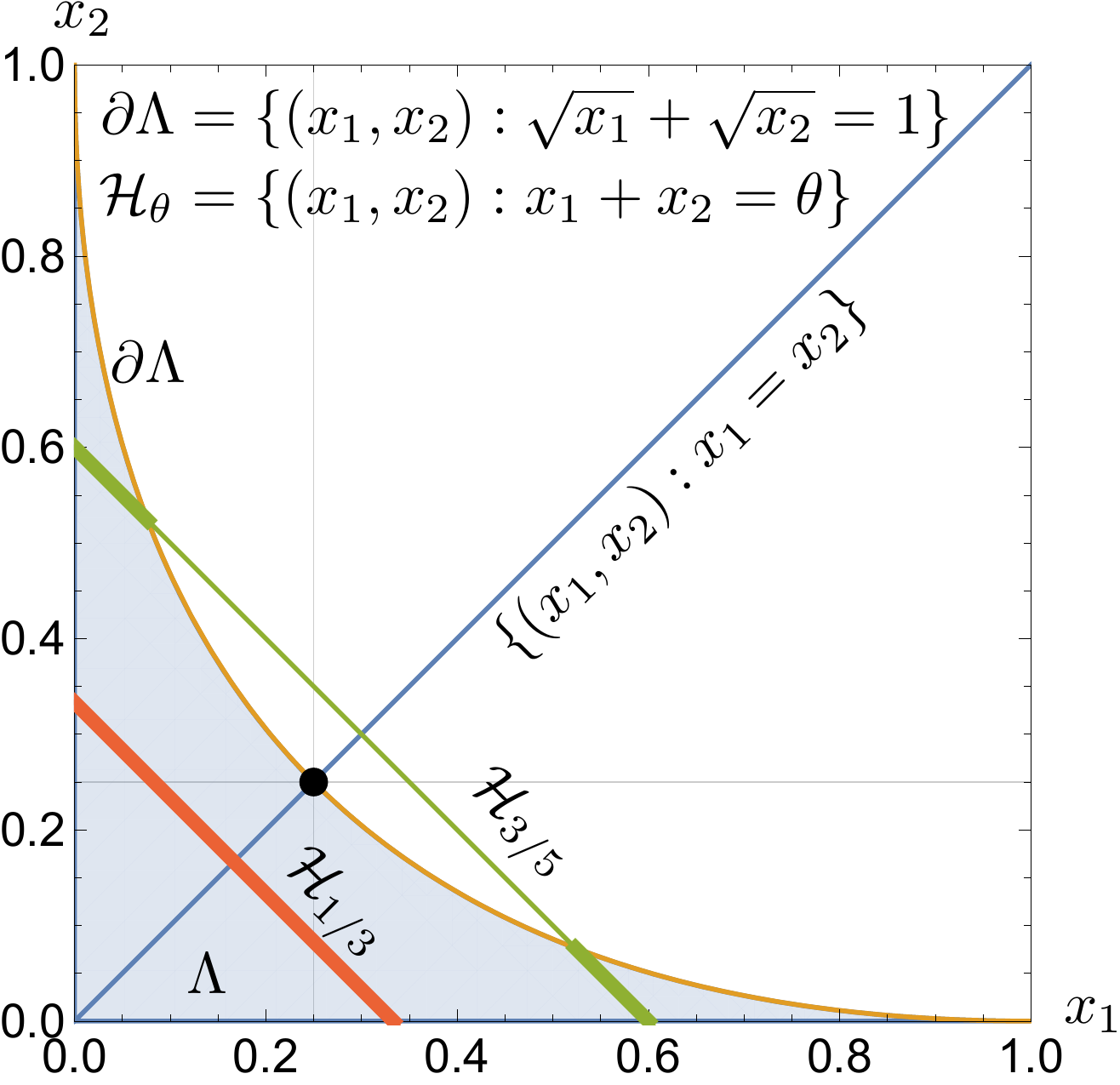}
\caption{Illustration of the proof of \prpref{TFtradeoffNequals2}, the Jain throughput--fairness tradeoff for $n=2$ users.  Shown are the set $\Lambda$, its boundary $\partial\Lambda$, two throughput constraint hyperplanes $\Hmc_{\theta}$ for $\theta \in \{1/3,3/5\}$, and the maximum fairness line $\{(x_1,x_2) : x_1 = x_2\}$.  The constrained feasible set $\Lambda \cap \Hmc_{\theta}$ (bold line segments) intersects the maximum fairness line (on which $F_J(\xbf) = 1$) for $\theta \leq 1/2$, but not for $\theta > 1/2$.} 
\label{fig:JainTwoUsers}
\end{figure}

We start with the special case $n = 2$.
\begin{proposition} 
\label{prp:TFtradeoffNequals2}
The throughput--fairness tradeoff under Jain's fairness metric, for $n=2$ users, is 
\begin{equation} 
\label{eq:TFtradeoffNequals2}
F_J^*(\theta) = \left\{ \begin{array}{ll}
1, \; & \theta \in  (0, \frac{1}{2}] \\
\frac{\theta^2}{\theta^2 + 2\theta-1}, \; & \theta \in  (1/2, 1)
\end{array} \right. .
\end{equation}
\end{proposition}
\begin{IEEEproof}
For the $n=2$ case we may use a direct approach (instead of solving \eqnref{TFtradeoffProblemFormulation}), since the set $\Lambda$ may be written explicitly (i.e., parameter-free) as $\Lambda = \{ \xbf \in \Rbb_{+}^{2}: \sqrt{x_1} + \sqrt{x_2} \leq 1\}$ \cite{TsyMik1979}, illustrated in \figref{JainTwoUsers}.\footnote{ As an aside, the stability inner bound $\Lambda$ is known to be exact, i.e., $\Lambda_{A} = \Lambda$, for the case $n=2$ \cite{TsyMik1979}.}  As evident from the figure, the constrained feasible set is the intersection of the throughput constraint line (for general $n$, a hyperplane) $\Hmc_{\theta} = \{\xbf : x_1 + x_2 = \theta\}$ with $\Lambda$.  Define the maximum fairness line $\{\xbf : x_1 = x_2\}$ (for general $n$, the ray emanating from the origin $\mathbf{0}$ passing through $\mathbf{1}$), on which $F_J(\xbf) = 1$.  In the case of $\theta \in (0,1/2]$, we see $\Lambda \cap \Hmc_{\theta}$ intersects  this ray, i.e., $F_J(\xbf) = 1$ is feasible.  In the case of $\theta \in (1/2,1)$, $F_J(\xbf) = 1$ is not feasible, but the fairness is easily shown to be monotone increasing on $\Hmc_{\theta}$ as $\xbf$ moves towards $x_1 = x_2$ (c.f., \figref{ProofThm1} in the proof of \corref{MajorizationCorollary}  in \secref{MajorizationAttempt} for general $n$), and as such, the optimal fairness is achieved at the two points for which $\Hmc_{\theta}$ intersects $\partial\Lambda = \{ \xbf \in \Rbb_{+}^{2}: \sqrt{x_1} + \sqrt{x_2} = 1\}$.  These two equations together yield the solutions $(x_{1}^{*}, x_{2}^{*}) = \left( \frac{\theta \pm \sqrt{2\theta-1}}{2}, \frac{\theta \mp \sqrt{2\theta-1}}{2}\right)$, from which the maximum fairness may be computed to be the second expression in \eqnref{TFtradeoffNequals2}. 
\end{IEEEproof}

The basic idea in establishing the Jain throughput-fairness tradeoff (\thmref{mainresultChiuJain}) is to first apply \corref{MajorizationCorollary} in \secref{MajorizationAttempt} to restrict the feasible set from $\pbf \in [0,1)^n$ to $\partial\Smc$, then apply \prpref{AtMostTwoDistinctNonzeroComponentValues} in \secref{opttputconst} to further restrict it to $\partial\Smc_{1,2}$, and finally \thmref{mainresultChiuJain} is proved by employing \prpref{tputeqconst} in \secref{opttputconst} and \prpref{mono1} below, the proof of which is found in \appref{JainPrelimPf}. 

Leveraging the $(p_s,k,n')$ parameterization of $\pbf$ in \defref{restrictedControls}, recall the definition of $T(p_s,k,n')$ in \eqnref{signedtputgap} in \secref{PropOptCont} and observe the Jain objective $F_{-1}(\xbf(\pbf))$ in \eqnref{TFtradeoffProblemFormulation} may be written as

\begin{equation}
\label{eq:JainObjectiveShorthand}
F_{-1}(p_s,k,n') \equiv F_{-1}(\xbf(\pbf(p_s,k,n'))).
\end{equation}
\prpref{mono1} establishes two key monotonicity properties of the objective \eqnref{JainObjectiveShorthand} under the throughput equality constraint over the restricted set $\pbf \in \partial\Smc_2$.

\begin{proposition} 
\label{prp:mono1}
Under the constraints $\pbf \in \partial\Smc_2$ (with $\pbf = \pbf(p_s,k,n')$) and $T(p_s,k,n') = \theta$, the objective $F_{-1}(p_s,k,n')$ \eqnref{JainObjectiveShorthand} obeys the following two monotonicity properties for all $(k,n') \in \Dmc_{t,n}$ defined in \eqnref{DtnpRegion}:
\begin{enumerate}
\itemsep=-2pt
\item $F_{-1}(p_s,k,n') < F_{-1}(p_s,k+1,n')$
\item $F_{-1}(p_s,k,n') < F_{-1}(p_s,k+1,n'+1)$.
\end{enumerate}
\end{proposition} 
In \figref{DtnpRegion} (left), the two monotonicity results show $F_{-1}$ is decreasing in $k$ along any vertical line (fixed $n'$), and along any diagonal line with unit slope  (fixed $n_l = n'-k$).

%%%%%%%%%%%%%%%%%%%%%%%%%%%%%%%%%%%%%%%%%%%%%%
\subsection{Main results}
\label{sec:JainMain}
For general $(n, \theta)$, where $n > 2$ and $\theta \in (0,1)$, we are not able to obtain an {\em explicit} expression for the throughput--fairness tradeoff, primarily because there is no known explicit characterization of $\Lambda$ for $n > 2$.  If $\xbf^*$ is an optimal rate vector, i.e., a minimizer of \eqnref{TFtradeoffProblemFormulation}, then we refer to any $\pbf^*$ satisfying $\xbf(\pbf^*) = \xbf^*$ as a corresponding {\em optimal control}.  The main theorem of this subsection is an {\em implicit} characterization of this tradeoff, meaning we characterize $\pbf^*$ for each $\theta$ (as the solution of a polynomial equation), from which we can compute $F_{-1}(\xbf(\pbf^*))$.  We reiterate the permutation invariance of both $\xbf^*$ and $\pbf^*$.  

\begin{figure}[!h]
\centering
\includegraphics[width=0.35 \textwidth ]{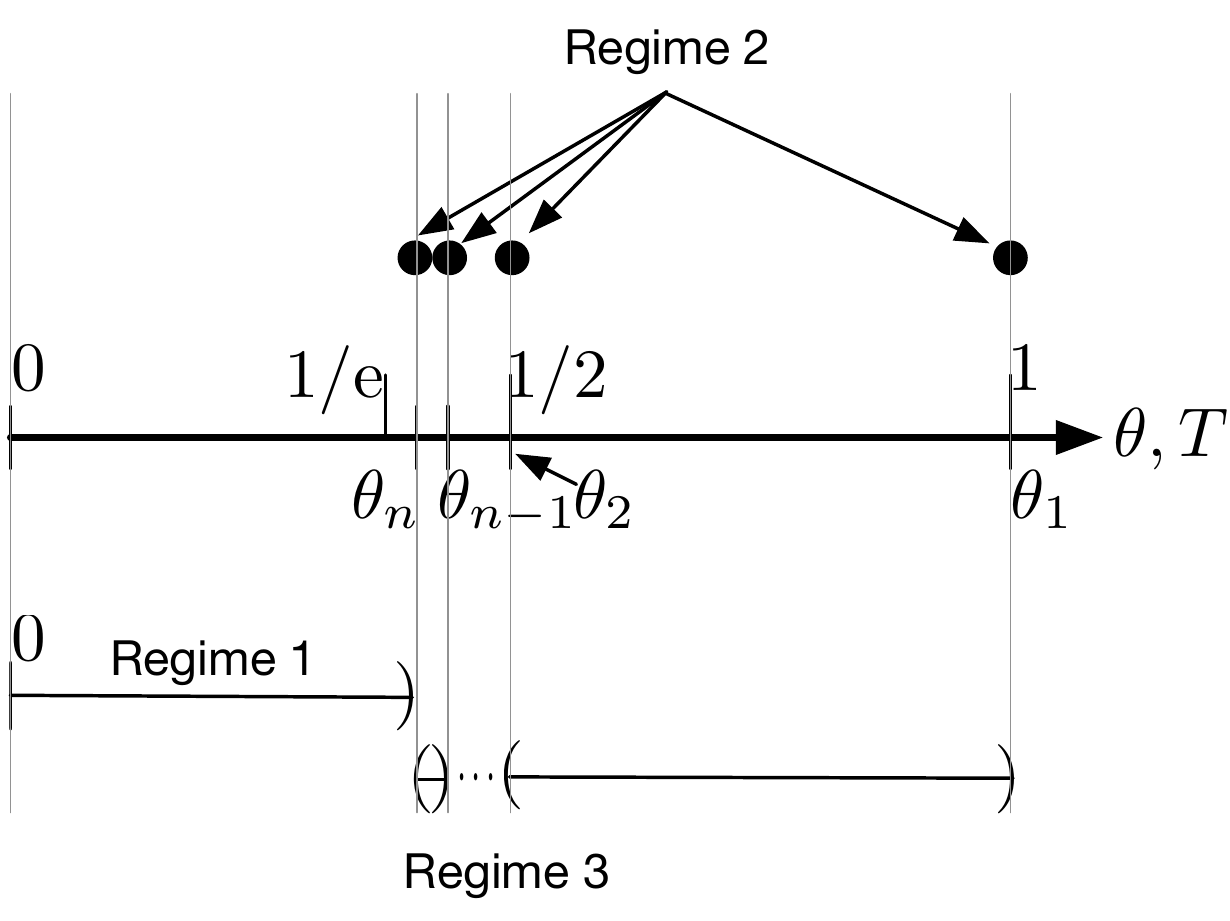}
\caption{Illustration of the three regimes, parameterized by $\theta$, in \thmref{mainresultChiuJain}: regime $1$ is $\theta \in (0,\theta_n)$, regime $2$ is $\boldsymbol\theta = (\theta_1,\ldots,\theta_n)$, and regime $3$ is $\bigcup_{t=2}^n (\theta_t,\theta_{t-1})$.} 
\label{fig:Thm1Regimes}
\end{figure}

\begin{theorem}[Throughput--fairness tradeoff under Jain's fairness] 
\label{thm:mainresultChiuJain}
The throughput--fairness tradeoff for $n \geq 2$ users under Jain's fairness metric, with a throughput equality constraint $T(\xbf) = \theta$, for $\theta \in (0,1)$, includes three regimes, illustrated in \figref{Thm1Regimes}, parameterized by $\theta$: 
\begin{enumerate}
\item if $\theta < \theta_{n}$, then the maximum fairness is $F_J^* = 1$, achieved when every user receives equal rate: $x_{i}(\pbf^{*}) = \theta / n$.
\item if $\theta = \theta_{t}$ for some $t \in [n]$, then $\pbf^{*} = (1/t) \sum_{i=1}^{t} \ebf_{i}$, with the corresponding maximum fairness $F_J^{*} = t/n$.  The function 
\begin{equation}
\label{eq:tfjinterpolation}
\tilde{T}(F) = \left(1-\frac{1}{n F}\right)^{n F -1}
\end{equation}
is a monotone, differentiable, and convex interpolation between the points $\{(\theta_t,\frac{t}{n})\}_{t \in [n]}$.
\item if $\theta \in (\theta_{t}, \theta_{t-1})$ for some $t \in \{2, \ldots, n \}$, then $\pbf^{*} = p_{s}^{*} \ebf_{1} + p_{l}^{*}\sum_{i=2}^{t} \ebf_{i}$ where $p_{l}^{*} = p_{l}(p_{s}^{*}, k^{*}, n'^{*})$ according to \eqref{eq:plofpskn} with $k^{*} = 1$, $n'^{*} = t$, and $p_{s}^{*}$ the unique real root on $(0, 1/t)$ of the following (order-$t$) polynomial (in $p_{s}$) equation:
\begin{equation} 
\label{eq:ps-star-inMainProp}
p_{s} \left(1 - p_{l}^{*} \right)^{t -1} + \left(1 - p_{s}\right)^{2} \left(1 - p_{l}^{*} \right)^{t - 2} = \theta.
\end{equation}
\end{enumerate}
\end{theorem}
The proof is found in \appref{JainMainPf}. The T-F tradeoff plots for $n = \{1, \ldots, 4\}$ users are illustrated in \figref{CJ1} (right) where regime $1)$ is omitted.
\begin{remark}
\label{rem:JainMergingRegimes2And3}
It can be verified that in the statement of \thmref{mainresultChiuJain}, regime $2)$ can be merged into $3)$ by allowing \eqnref{ps-star-inMainProp} to be solved for $p_{s}^{*}$ on $(0, 1/t]$. They are stated separately for conceptual clarity and better consistency with the proof of \thmref{2}. In addition, regime $2)$ is where we have a closed-form expression for both the extremizer and the optimized objective.  
\end{remark}

As motivated in \secref{fairnessmeasures}, the throughput inequality constraint is natural from the operational perspective of wishing to maximize fairness subject to a minimum throughput requirement.  As may be intuitive, this modification to the constraint (feasible set) has no effect on the solution, as shown in the following theorem.

\begin{theorem} 
\label{thm:2}
%cor:mainresultChiuJainInequalityThroughputConstraint
The solution in \thmref{mainresultChiuJain} of the Jain's throughput--fairness tradeoff \eqnref{TFtradeoffProblemFormulation} is unaffected by changing the throughput equality constraint to an inequality constraint $T(\xbf(\pbf)) \geq \theta$.
\end{theorem}

The proof is found in \appref{JainMainPf}.

%%%%%%%%%%%%%%%%%%%%%%%%%%%%%%%%%%%%%%%%%%%%%%
\subsection{Properties of the Jain T-F tradeoff}
\label{sec:JainProp}

As can be seen from \thmref{mainresultChiuJain}, the extremizer $\pbf^{*} = \pbf(p_{s}^{*}, k^{*}, n'^{*})$, with $p_s^*$ solving $T(p_s,k^*,n'^*) = \theta$ in \eqnref{ps-star-inMainProp}, has the property that $n'^*$, the total number of {\em active} users (i.e., users with nonzero contention probabilities), equals $t$, where $\theta \in [\theta_t,\theta_{t-1})$, for $t \in \{2, \ldots, n \}$.  In fact, because \eqnref{ps-star-inMainProp} does not depend on $n$, the total number of users in the system, one can easily verify that, if $\theta \geq \theta_{n-l}$ for some integer $l \in \{1, \ldots, n-2 \}$, then the extremizer $\pbf^{*}$ is as if the total number of users in the system were $n-l$, except that $l$ zeros need to be padded in order to make $\pbf^{*}$ an $n$-dimensional vector.  It follows that the maximum Jain's fairness satisfies
\begin{equation}
\label{eq:RecusionJain}
F_{J}^{*}(\theta; n) = \left(1 - \frac{l}{n}\right) F_{J}^{*}(\theta; n-l), ~ \theta \geq \theta_{n-l},
\end{equation}
where our notation highlights $F_{J}^{*}$ is a function of $\theta$ and is parameterized by $n$.

\begin{figure}[!ht]
\centering
\includegraphics[width=0.49 \textwidth ]{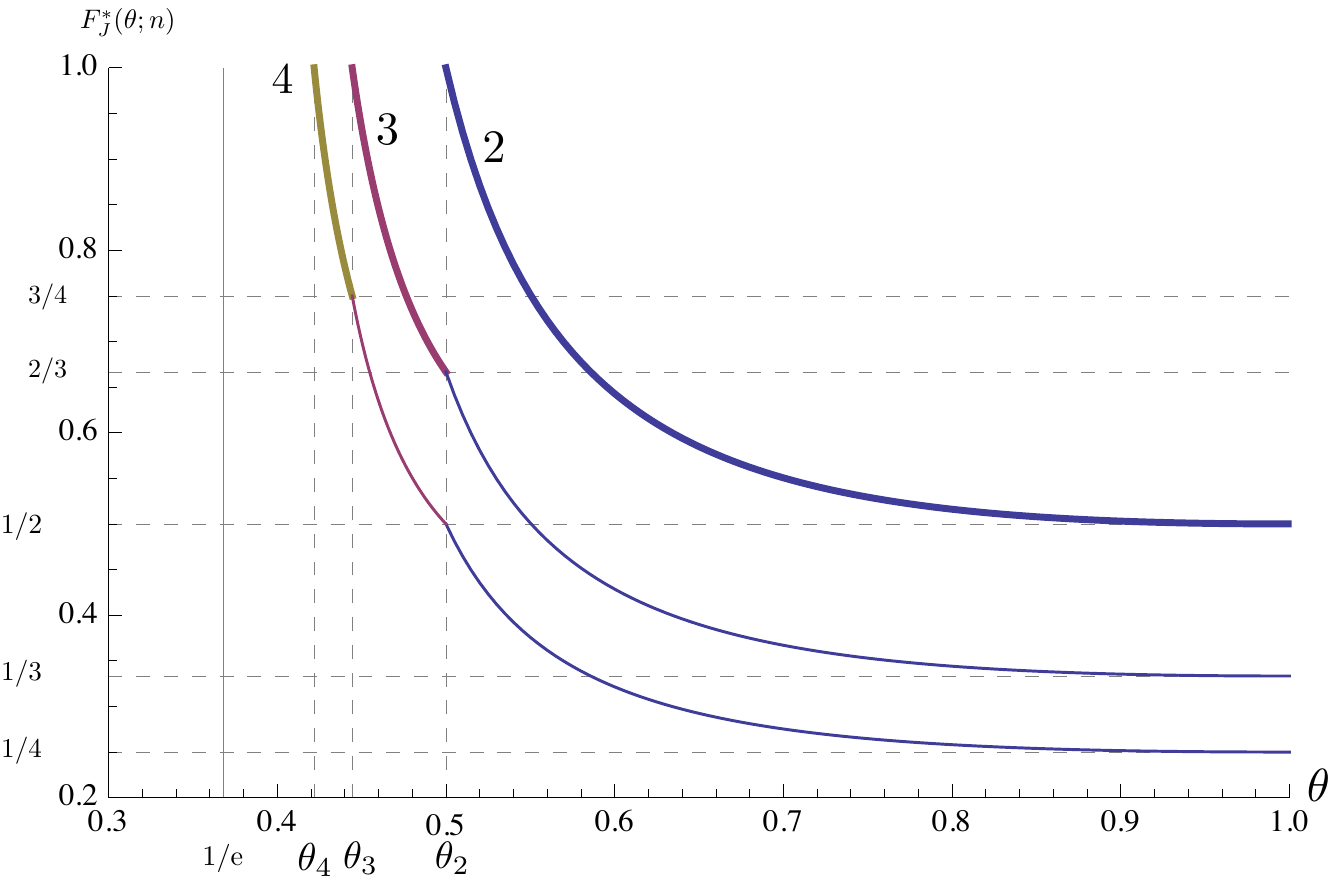}
\includegraphics[width=0.49 \textwidth ]{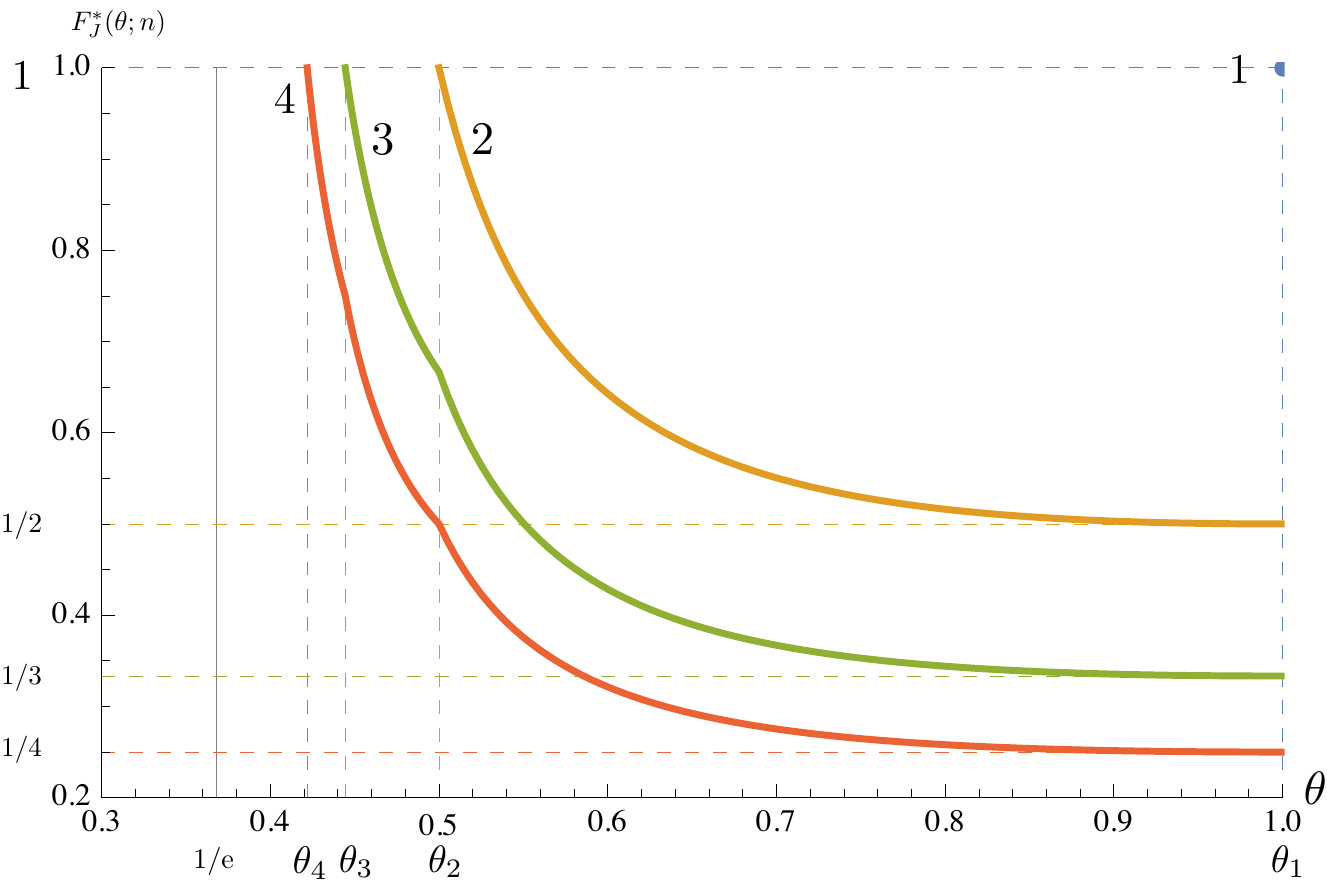}
\caption{{\bf Left:} Illustration of using \algoref{tfplot}, leveraging the Jain fairness recursion \eqnref{RecusionJain}, to incrementally plot the Jain T-F tradeoff for $n_{\rm max} = 4$ users. Vertical gridlines indicate the $\theta_{t}$'s for $t \in \{2, 3, 4\}$. Horizontal gridlines indicate the maximum fairness at the $\theta_{t}$'s for each $t \in [n]$ and each $n \in n_{\max}$. The T-F tradeoff for the active throughput intervals (thick curves) need to be computed first, after which the rest parts (thin curves) can be obtained by scaling. {\bf Right:} \thmref{mainresultChiuJain} (regimes $2)$ and $3)$): T-F tradeoff under Jain's fairness for $n = 1$ (blue), $2$ (orange), $3$ (green), and $4$ (red).} 
\label{fig:CJ1}
\end{figure}

One use of the recursive relationship \eqnref{RecusionJain} is that it enables incremental plotting of the T-F tradeoff for a sequence of values of $n \in \{2,\ldots,n_{\rm max}\}$. From \thmref{mainresultChiuJain} if $\theta \in [\theta_{n}, \theta_{n-1})$ then $n'^{*} = n$, meaning, at the optimum, every user in the system is active.  We therefore call the interval $[\theta_{n}, \theta_{n-1})$, for each $n \in \Nbb$, the {\em active throughput interval}, meaning all $n$ users are actively contending under the optimal control for any target throughput $\theta$ in this interval.  This observation is the root idea in the Jain T-F plotting algorithm (\algoref{tfplot}), which returns a plot of the Jain T-F tradeoff over $\theta \in (0,1)$ for all $n \in \{2,\ldots,n_{\rm max}\}$.  Naturally, the interval $[\theta_n,\theta_{n-1})$ must be discretized for each $n$.  \figref{CJ1} (left) illustrates \algoref{tfplot} for $n_{\rm max} = 4$ users.  First, the plot of $F_J^*(\theta;2)$ over $\theta \in [\theta_2,\theta_1)$ (i.e., the active interval for $n=2$, thick blue) is scaled using \eqnref{RecusionJain} to obtain $F_J^*(\theta;3)$ and $F_J^*(\theta;4)$ over the same interval (thin blue for both).  Then, the plot of $F_J^*(\theta;3)$ over $\theta \in [\theta_3,\theta_2)$  (i.e., the active interval for $n=3$, thick purple) is scaled to obtain $F_J^*(\theta;4)$ over the same interval (thin purple), and so on.  Note first that, for each $n$, at $\theta = 1$ the maximum Jain's fairness is the {\em minimum} possible, i.e., $F_J^*=1/n$, corresponding to the fairness when only one user (say $i$) contends for access (i.e., $\xbf = \pbf = \ebf_i$), as $\xbf = \ebf_i$ is the unique (up to permutation) rate vector in $\Lambda$ achieving $\theta = 1$.  Second, for each $n$, for any $\theta \leq \theta_n$ the maximum Jain's fairness is the {\em maximum} possible, i.e., $F_J^*=1$, corresponding to all $n$ users contending with equal probability, uniquely achievable by the rate vector $\xbf = \theta \ubf$.  The Jain T-F tradeoff for each $n$ up to $4$ users is shown in \figref{CJ1} (right).

\begin{algorithm}
\caption{Jain T-F tradeoff for all $n \in \{2, \ldots, n_{\rm max}\}$}
\label{alg:tfplot}
\begin{algorithmic}[1]
\For{$n=2,\ldots,n_{\rm max}$}
	\State Plot $F_J^*(\theta;n) = 1$ for $\theta \in [0,\theta_n)$
	\For{$\theta \in [\theta_n,\theta_{n-1})$}
		\State Compute $p_s^*(\theta)$ solving $T(p_s,1,n)=\theta$ (i.e., \eqnref{ps-star-inMainProp} in \thmref{mainresultChiuJain} with $t = n$) 
		\State Compute $F_J^*(\theta;n) = F_J(\xbf(\pbf(p_s^*(\theta),1,n)))$ using \eqnref{xofp}, \eqnref{Chiu-JainFairnessDefinition}, and \defref{restrictedControls}
	\EndFor
	\State Plot $F_J^*(\theta;m) = \frac{n}{m} F_J^*(\theta;n)$ for $m \in \{n,\ldots,n_{\rm max}\}$
\EndFor
\end{algorithmic}
\end{algorithm}

The following theorem gives some properties of the optimal controls, optimal rates, and the Jain T-F tradeoff.

\begin{theorem}
\label{thm:ChiuJainTFtradeoffProperties}
The Jain T-F tradeoff for $n \geq 2$ users, over $\theta \in [\theta_{n}, 1)$, has the following properties:
\begin{enumerate}
\item For fixed $n$, the small and large contention probabilities of the optimal control, $p_s^*(\theta), p_l^*(\theta)$, and the corresponding optimal rates, $x_s^*(\theta),x_l^*(\theta)$, are piecewise decreasing and increasing, respectively, in $\theta$.  More precisely, fix $t \in \{2, \ldots, n\}$ and $\theta \in [\theta_{t}, \theta_{t-1})$. Then: 
\begin{enumerate}
\item Both $p_s^*$ and $x_s^*$ are continuous and decreasing over each interval $[\theta_{t}, \theta_{t-1})$, but are not monotone over $[\theta_n,1)$.  In particular, $i)$ $\frac{\drm p_s^*(\theta)}{\drm \theta}  < 0$, $\frac{\drm x_s^*(\theta)}{\drm \theta}  < 0$, $ii)$ at $\theta = \theta_t$ they take values $p_s^*(\theta_t) = 1/t$, $x_s^*(\theta_t) = \theta_{t}/t$, and $iii)$ at $\theta = \theta_{t-1}$ they take value $p_s^*(\theta_{t-1}) = x_s^*(\theta_{t-1})= 0$.
\item Both $p_l^*$ and $x_l^*$ are continuous and increasing over $[\theta_n,1)$, but neither is differentiable at each $\theta_t$ for $t \in \{2,\ldots,n-1\}$.  In particular, $i)$ $\frac{\drm p_l^*(\theta)}{\drm \theta}  > 0$, $\frac{\drm x_l^*(\theta)}{\drm \theta} > 0$, $ii)$ at $\theta = \theta_t$ they take values $p_l^*(\theta_t) = 1/t$, $x_l^*(\theta_t) = \theta_{t}/t$, and $iii)$ at $\theta = \theta_{t-1}$ they take value $p_l^*(\theta_{t-1}) = 1/(t-1)$ and $x_l^*(\theta_{t-1})= \theta_{t-1}/(t-1)$.
\end{enumerate}
\item For fixed $n$, the T-F tradeoff curve is decreasing in $\theta$, i.e., $\frac{\drm}{\drm \theta} F_J^*(\theta;n) < 0$.
\item For fixed $\theta$, the T-F tradeoff curve is decreasing in $n$, i.e., $F_J^*(\theta;n) > F_J^*(\theta;n+1)$.
\item For fixed $n$, the T-F tradeoff curve is continuous but nondifferentiable at $\{\theta_{t}\}_{t = 2}^{n-1}$, i.e., $\left. F_J^*(\theta;n) \right\vert_{\theta \downarrow \theta_t}= \left. F_J^*(\theta;n) \right\vert_{\theta \uparrow \theta_t}$, but $\left. \frac{\drm}{\drm \theta} F_J^*(\theta;n)\right|_{\theta \downarrow \theta_t} \neq \left. \frac{\drm}{\drm \theta} F_J^*(\theta;n)\right|_{\theta \uparrow \theta_t}$ for each $t \in \{2,\ldots,n-1\}$.
\item For fixed $n$, the T-F tradeoff curve is piecewise convex in $\theta$, i.e., $\frac{\drm^2}{\drm \theta^2} F_J^*(\theta;n) > 0$, 
for $\theta \in [\theta_{t}, \theta_{t-1})$ with $t \in \{2, \ldots, n\}$.
\end{enumerate}
\end{theorem}

\begin{figure}[!ht]
\centering
\includegraphics[width=0.49 \textwidth ]{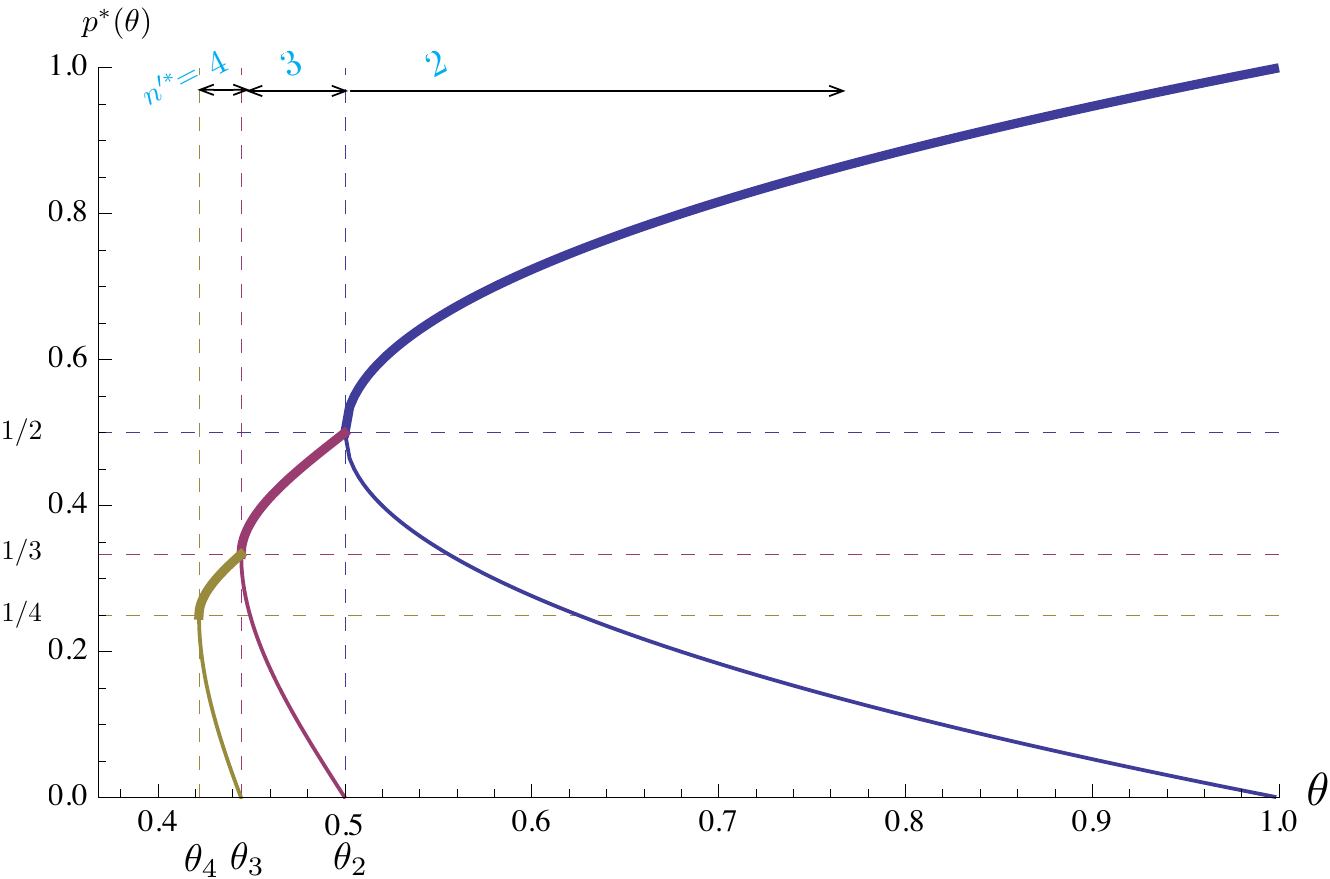}
\includegraphics[width=0.49 \textwidth ]{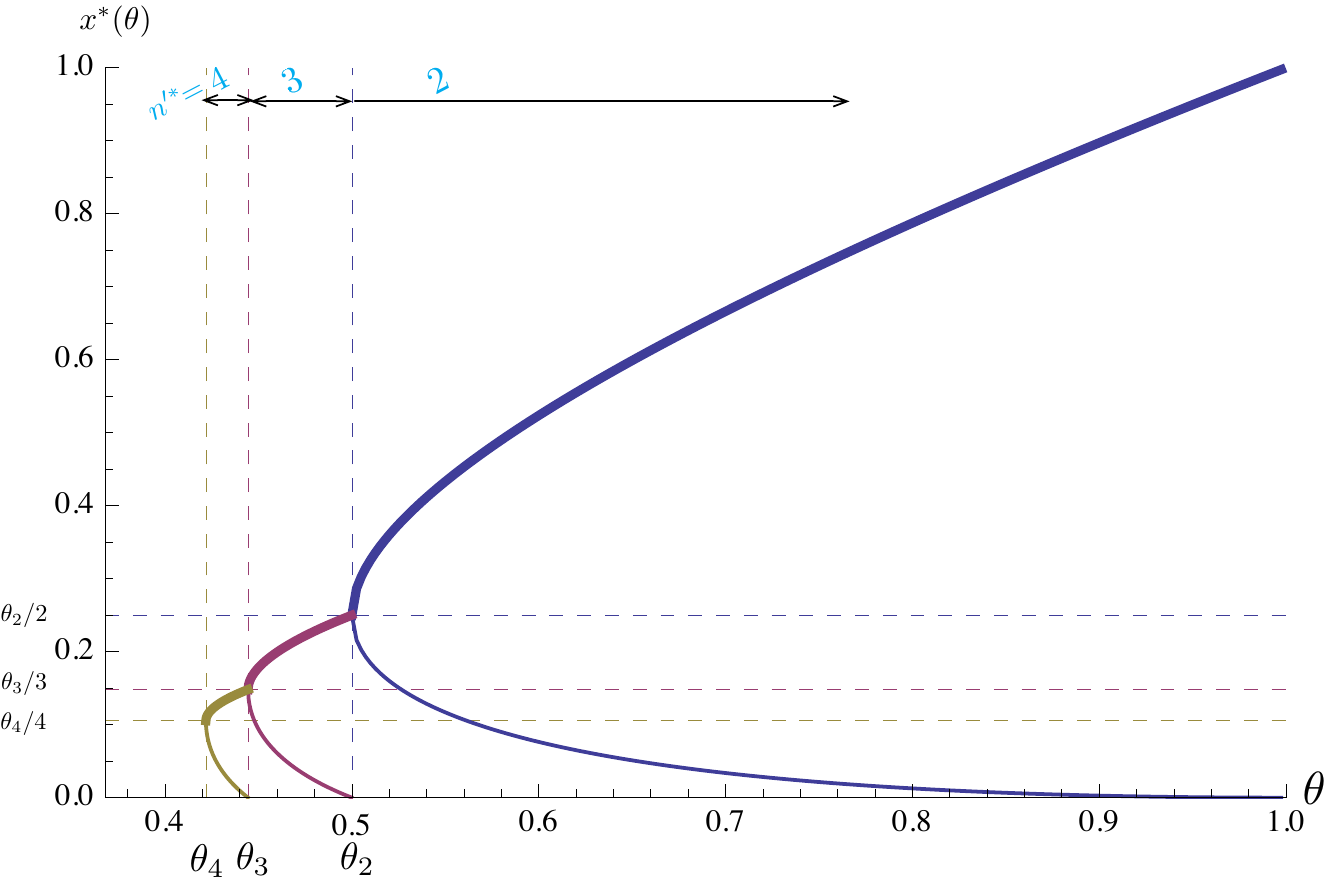}
\caption{Illustration of property $1)$ in \thmref{ChiuJainTFtradeoffProperties}: Optimal controls $p_s^*(\theta)$ (left, lower/thinner branches), $p_l^*(\theta)$ (left, upper/thicker branches) and optimal rates $x_s^*(\theta)$ (right, lower/thinner branches), $x_l^*(\theta)$ (right, upper/thicker branches) versus target throughput $\theta$, for $n = 4$ users. Vertical gridlines indicate the $\theta_{t}$'s: $(\theta_{2}, \theta_{3}, \theta_{4}) = (\frac{1}{2}, \frac{4}{9}, \frac{27}{64}) \approx (0.5, 0.4444, 0.4219)$. Horizontal gridlines indicate the corresponding optimal controls (left) and optimal rates (right) when $\theta = \theta_{t}$. Shown also are the optimal number of active users $n'^{*}$ for different ranges of $\theta$.}
\label{fig:CJ2}
\end{figure}

The proof is found in \appref{JainPropPf}.  \figref{CJ2} shows $p_s^*(\theta),p_l^*(\theta)$ (left) and $x_s^*(\theta),x_l^*(\theta)$ (right), illustrating property $1)$ in \thmref{ChiuJainTFtradeoffProperties}.  Properties $2)$ through $5)$ in \thmref{ChiuJainTFtradeoffProperties} can be seen from \figref{CJ1} (right).  Finally, we mention that a plot of the interpolated function $\tilde{T}(F)$ \eqnref{tfjinterpolation} in \thmref{mainresultChiuJain} (not shown) on the actual T-F tradeoff in \figref{CJ1} would show the interpolation lies above the true tradeoff, and is tight only at the critical throughputs $\boldsymbol\theta$.

%% file: Sec5-AlphaFair.tex
%%%%%%%%%%%%%%%%%%%%%%%%%%%%%%%%%%%%%%%%%%%%%%
% Section V. $\alpha$-fair utility functions
%%%%%%%%%%%%%%%%%%%%%%%%%%%%%%%%%%%%%%%%%%%%%%

\section{$\alpha$-fair network utility maximization}
\label{sec:TFtradeoffUsingAlphaFairUtilityFunctionsWithThroughputConstraint}
In this section we investigate the throughput-fairness tradeoff within the framework of $\alpha$-fair utility functions \cite{MoWal2000}, \cite{Atk1970}. Recall the objective $F_{\alpha}$ (for $\alpha \geq 0$), the $\alpha$-fair utility function $U_{\alpha}$, the throughput function $T$, and the mapping between a control $\pbf$ and a rate vector $\xbf(\pbf)$ given in \eqnref{falphadef}, \eqnref{AlphaFairUtilityFunction}, \eqnref{ThroughputOfxbf}, and \eqnref{xofp} respectively. The optimization under a throughput equality constraint is:
\begin{equation}
\label{eq:AlphaFairTFtradeoffProblemFormulation}
\boxed{
\max_{\pbf \in [0,1)^n} F_{\alpha}(\xbf(\pbf)) = \sum_{i=1}^{n} U_{\alpha}(x_{i} (\pbf)) ~\mbox{s.t.}~ T(\xbf(\pbf)) = \theta.}
\end{equation}
We solve this problem for $\alpha \geq 1$. In this following we give $i)$ preliminary results in \secref{AlphaFairPrelim}, $ii)$ the main results in \secref{AlphaFairMain}, and $iii)$ some additional properties of the $\alpha$-fair throughput-fairness tradeoff in \secref{AlphaFairProp}.

%%%%%%%%%%%%%%%%%%%%%%%%%%%%%%%%%%%%%%%%%%%%%%
\subsection{Preliminary results}
\label{sec:AlphaFairPrelim}

We start with the special case $n = 2$.
\begin{proposition} 
\label{prp:AlphaFair-TFtradeoffNequals2}
The throughput--fairness tradeoff under $\alpha$-fairness ($\alpha \geq 1$), for $n=2$ users, is 
\begin{eqnarray} 
\label{eq:AlphaFair-TFtradeoffNequals2}
F_{\alpha}^*(\theta) & = & \left\{ \begin{array}{lll}
-\frac{2}{\alpha - 1} \left(\frac{2}{\theta}\right)^{\alpha - 1}, \; & \theta \in  (0, \frac{1}{2}] & \alpha > 1\\
-\frac{1}{\alpha - 1} \left( \left( \frac{\theta + \sqrt{2\theta-1}}{2} \right)^{1 - \alpha} + \left( \frac{\theta - \sqrt{2\theta-1}}{2} \right)^{1 - \alpha} \right), \; & \theta \in (\frac{1}{2}, 1) & \\
-2 \log\frac{2}{\theta}, \; & \theta \in  (0, \frac{1}{2}] & \alpha = 1\\
-2 \log \frac{2}{1 - \theta}, \; & \theta \in (\frac{1}{2}, 1)
\end{array} \right.
\end{eqnarray}
\end{proposition}
\begin{IEEEproof}
The proof resembles that of \prpref{TFtradeoffNequals2} in \secref{JainPrelim}. The all-rates equal ray $\{\xbf : x_1 = x_2\}$ can still be viewed as the maximum fairness line as the maximum $\alpha$-fair objective is attained by points either on this line or closest to this line, subject to the throughput constraint $x_1 + x_2 = \theta$. This follows from the Schur-concavity of the objective (\prpref{SchurConcavityCJFairnessAndAlphaFairUtiityFunction} in \secref{MajorizationAttempt}) and (the proof of) \corref{MajorizationCorollary} in \secref{MajorizationAttempt}. Therefore, when $\theta \leq 1/2$, the maximizer is on the ray $\{\xbf : x_1 = x_2\}$ and hence $(x_{1}^{*}, x_{2}^{*}) = (\frac{\theta}{2}, \frac{\theta}{2})$; when $\theta > 1/2$, the maximizer is obtained by finding the points on the boundary of $\Lambda$ that satisfy the throughput constraint (as they are the closest to the all-rates equal ray, see \figref{ProofThm1}), which gives $(x_{1}^{*}, x_{2}^{*}) = \left( \frac{\theta \pm \sqrt{2\theta-1}}{2}, \frac{\theta \mp \sqrt{2\theta-1}}{2}\right)$. Substitution of the expressions of the maximizers into the objective yields \eqnref{AlphaFair-TFtradeoffNequals2}.
\end{IEEEproof}

The basic idea in solving the throughput-fairness tradeoff under $\alpha$-fairness (\thmref{mainresultAlphaFairWhenAlphaGreaterThanOrEqualToOne}) is to first apply \corref{MajorizationCorollary} in \secref{MajorizationAttempt} to restrict the feasible set from $\pbf \in [0,1)^n$ to $\partial\Smc$, and then apply \prpref{AtMostTwoDistinctNonzeroComponentValues} in \secref{opttputconst} to further restrict it to $\partial\Smc_{1,2}$. The optimization problem is solved with the aid of \prpref{mono2} shown below, which establishes a key monotonicity property of the objective in \eqnref{AlphaFairTFtradeoffProblemFormulation} under the throughput equality constraint over the restricted set $\pbf \in \partial\Smc_2$. It plays a similar role to that of \prpref{mono1} in proving \thmref{mainresultChiuJain} (\secref{JainMain}).

Leveraging the $(p_s,k,n')$ parameterization of $\pbf$ in \defref{restrictedControls} and the definition of $T(p_s,k,n')$ in \eqnref{signedtputgap} in \secref{opttputconst} we define
\begin{equation}
\label{eq:AlphaFairObjectiveShorthand}
F_{\alpha}(p_s,k,n') \equiv F_{\alpha}(\xbf(\pbf(p_s,k,n'))).
\end{equation}

\begin{proposition} 
\label{prp:mono2}
Under the constraints $\pbf \in \partial \Smc_{2}$ (with $\pbf = \pbf(p_s,k,n')$) and $T(p_s,k,n') = \theta$, the objective $F_{\alpha}(p_s,k,n')$ \eqnref{AlphaFairObjectiveShorthand} for $\alpha \geq 1$ is increasing in $k$ for $k \in \{1, \ldots, n'-1 \}$ when $n'$ is held fixed. Thus the maximum of $F_{\alpha}(p_{s}, k, n')$ is attained when $k^{*} = n' - 1$.
\end{proposition} 
The proof is found in \appref{AlphaFairPrelimPf}.

%%%%%%%%%%%%%%%%%%%%%%%%%%%%%%%%%%%%%%%%%%%%%%
\subsection{Main results}
\label{sec:AlphaFairMain}
For general $(n, \theta)$, where $n > 2$ and $\theta \in (0, 1)$, we will again give an {\em implicit} characterization of the T-F tradeoff under $\alpha$-fairness when $\alpha \geq 1$. The main theorem of this subsection is a characterization of the optimal control $\pbf^{*}$ for each $\theta$ (as the solution of a polynomial equation) from which we can compute $F_{\alpha}(\xbf(\pbf^{*}))$.

\begin{theorem}[Throughput-fairness tradeoff under $\alpha$-fair when $\alpha \geq 1$] \label{thm:mainresultAlphaFairWhenAlphaGreaterThanOrEqualToOne}
The throughput--fairness tradeoff for $n \geq 2$ users under $\alpha$-fairness when $\alpha \geq 1$, with a throughput equality constraint $T(\xbf) = \theta$, for $\theta \in (0,1)$, includes two regimes, parameterized by $\theta$: 
\begin{enumerate}
\item  if $\theta \leq \theta_{n}$, then the maximum fairness is 
\begin{equation} 
\label{eq:AlphaFair-TFtradeoffGeneralNThetaLeqThetaN}
F_{\alpha}^{*} (\theta) = \left\{ \begin{array}{ll}
-n \log \left( \frac{n}{\theta} \right), \; & \alpha = 1 \\
- \frac{n}{\alpha - 1} \left(\frac{n}{\theta}\right)^{\alpha-1}, \; & \alpha > 1
\end{array} \right. ,
\end{equation}
achieved when every user receives equal rate: $x_{i}(\pbf^{*}) = \theta / n$.
\item if $\theta \in (\theta_{n}, 1)$, then $\pbf^{*} = p_{s}^{*} \ebf_{1} + p_{l}^{*}\sum_{i=2}^{n} \ebf_{i}$ where $p_{l}^{*} = p_{l}(p_{s}^{*}, k^{*}, n'^{*})$ according to \eqref{eq:plofpskn} with $k^{*} = n-1$, $n'^{*} = n$, and $p_{s}^{*}$ the unique real root on $(0, 1/n)$ of the following polynomial equation
\begin{equation} 
\label{eq:ps-star-AlphaFair}
\left((n-1) p_{s}^{}\right)^{2} (1-p_{s}^{})^{n-2} + (1-(n-1) p_{s}^{}) (1 - p_{s}^{})^{n-1} = \theta.
\end{equation}
\end{enumerate}
\end{theorem}
The proof is found in \appref{AlphaFairMainPf}. The T-F tradeoff plots for $n = \{1, \ldots, 4\}$ users are illustrated in \figref{AlphaFairTFTradeoffVaryingn}.

Observe the difference between regime $1)$ in \thmref{mainresultAlphaFairWhenAlphaGreaterThanOrEqualToOne} for $\alpha$-fairness when $\alpha \geq 1$ and regime $1$ in \thmref{mainresultChiuJain} for Jain's fairness: although the maximizers are the same, the objective is increasing in $\theta$ in the former, whereas it is constant in the latter. 
Observe also the asymmetry between regime $2)$ in \thmref{mainresultAlphaFairWhenAlphaGreaterThanOrEqualToOne} and regimes $2)$ and $3)$ in \thmref{mainresultChiuJain}: $k^*=n'^{*}-1$ and $n'^* = n$ for all $\theta \in (\theta_n,1)$ in the former, while $k^*=1$ and $n'^* = t$ for $\theta \in [\theta_t,\theta_{t-1})$ in the latter. Thus, the optimal control vector $\pbf^{*}$ for $\alpha$-fairness has $n'^{*}-1$ users with ``small'' contention probability $p_s^{*}$ and one user with ``large'' contention probability $p_l^{*}$ for $n'^{*}$ always equal to $n$, while the optimal control vector $\pbf^{*}$ for Jain's fairness has one user with $p_s^{*}$ and $n'^{*}-1$ users with $p_l^{*}$, for $n'^{*}$ determined by the active throughput interval containing $\theta$.

Similar to \secref{JainMain}, we now address the case where the throughput constraint in \eqnref{AlphaFairTFtradeoffProblemFormulation} is an inequality $T(\xbf(\pbf)) \geq \theta$. 

\begin{theorem} 
\label{thm:5}
If the throughput equality constraint is changed to an inequality constraint $T(\xbf(\pbf)) \geq \theta$ then the solution in \thmref{mainresultAlphaFairWhenAlphaGreaterThanOrEqualToOne} of the $\alpha$-fair utility maximization problem \eqnref{AlphaFairTFtradeoffProblemFormulation} when $\alpha \geq 1$ is only affected in the first regime, namely when $\theta \leq \theta_{n}$. More precisely, if $\theta \leq \theta_{n}$, then the maximum fairness is independent of $\theta$ and is given by
\begin{equation} 
\label{eq:AlphaFair-TFtradeoffGeneralN-InequalityThroughputConstraint}
F_{\alpha}^{*} (\theta) = \left\{ \begin{array}{ll}
-n \log \left(\frac{n}{\theta_{n}} \right), \; & \alpha = 1 \\
- \frac{n}{\alpha - 1} \left(\frac{n}{\theta_{n}}\right)^{\alpha-1}, \; & \alpha > 1
\end{array} \right. ,
\end{equation}
where the maximizer in the control space is a uniform vector $\pbf^{*} = \ubf$.
\end{theorem}
The proof is found in \appref{AlphaFairMainPf}.

%%%%%%%%%%%%%%%%%%%%%%%%%%%%%%%%%%%%%%%%%%%%%%
\subsection{Properties of the $\alpha$-fair T-F tradeoff}
\label{sec:AlphaFairProp}

\begin{figure}[!ht]
\centering
\includegraphics[width=0.49 \textwidth ]{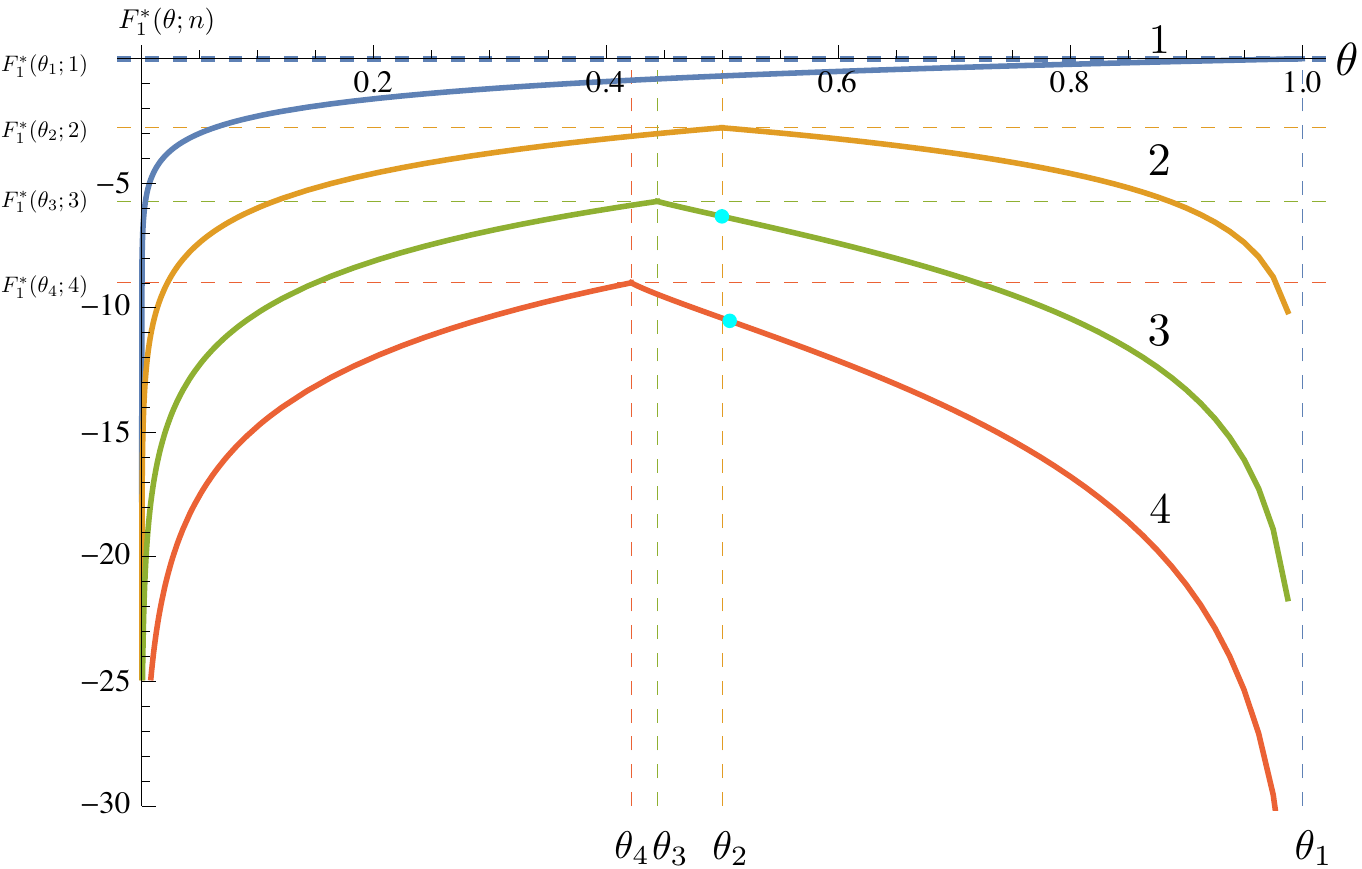}
\includegraphics[width=0.49 \textwidth ]{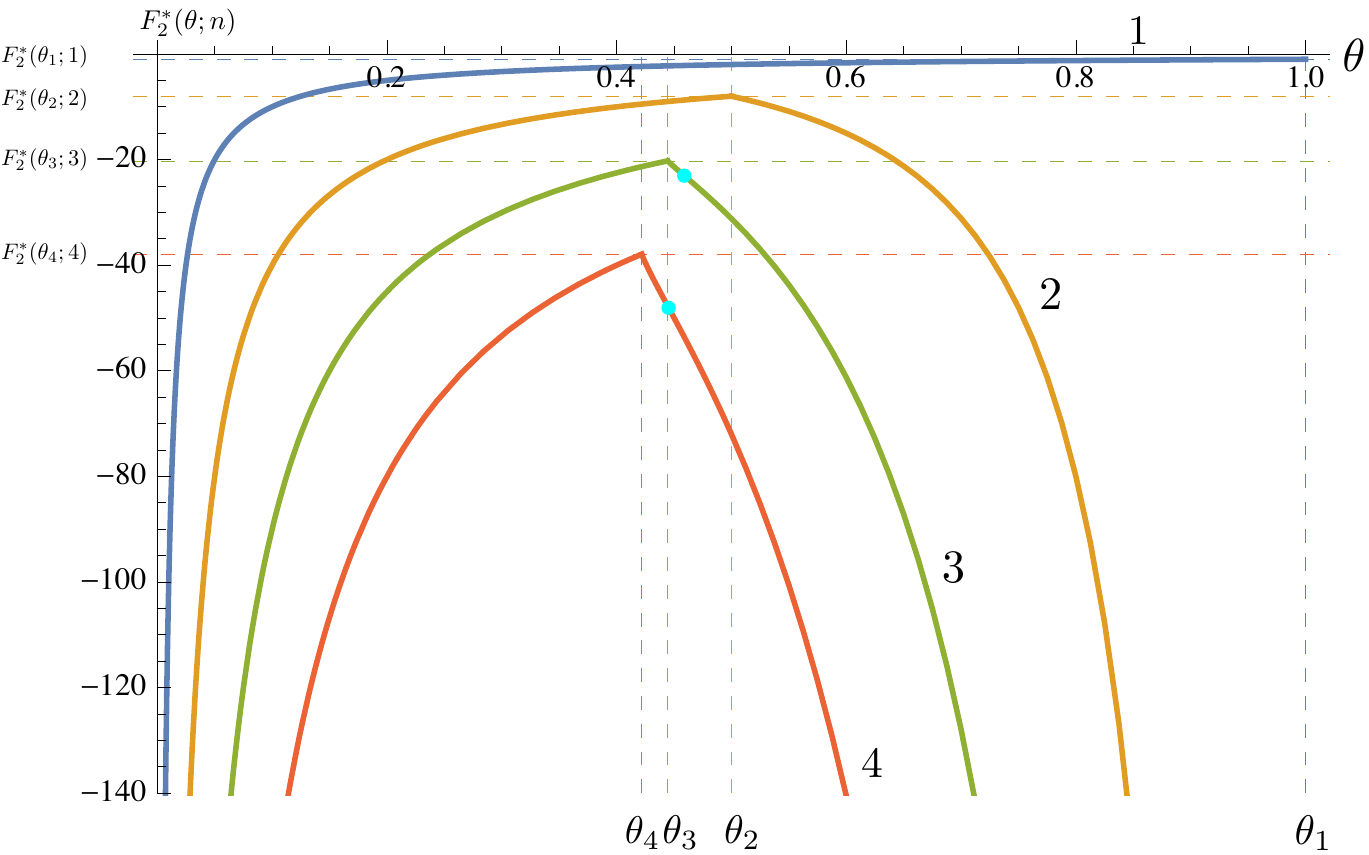}
\caption{Illustration of \thmref{mainresultAlphaFairWhenAlphaGreaterThanOrEqualToOne} and properties $2)$ and $3)$ in \thmref{AlphaFairTFtradeoffProperties}: T-F tradeoff under $\alpha$-fairness when $n = 1$ (blue), $2$ (orange), $3$ (green), and $4$ (red) users, for $\alpha = 1$ (left) and $\alpha = 2$ (right). Vertical gridlines indicate the $\theta_{t}$'s and horizontal gridlines indicate the corresponding optimal $\alpha$-fair objective for each $n$ at $\theta = \theta_{n}$ i.e., $F_{\alpha}^{*}(\theta_{n}; n)$.  Shown as cyan dots are the ``inflection'' points upon which the T-F curves transitions from convex decreasing to concave decreasing, for $n > 2$. The thresholding $\mathring{\theta}_{\alpha}(n)$ is computed using \eqnref{thresholdingtheta}.} 
\label{fig:AlphaFairTFTradeoffVaryingn}
\end{figure}

\begin{figure}[!ht]
\centering
\includegraphics[width=0.49 \textwidth ]{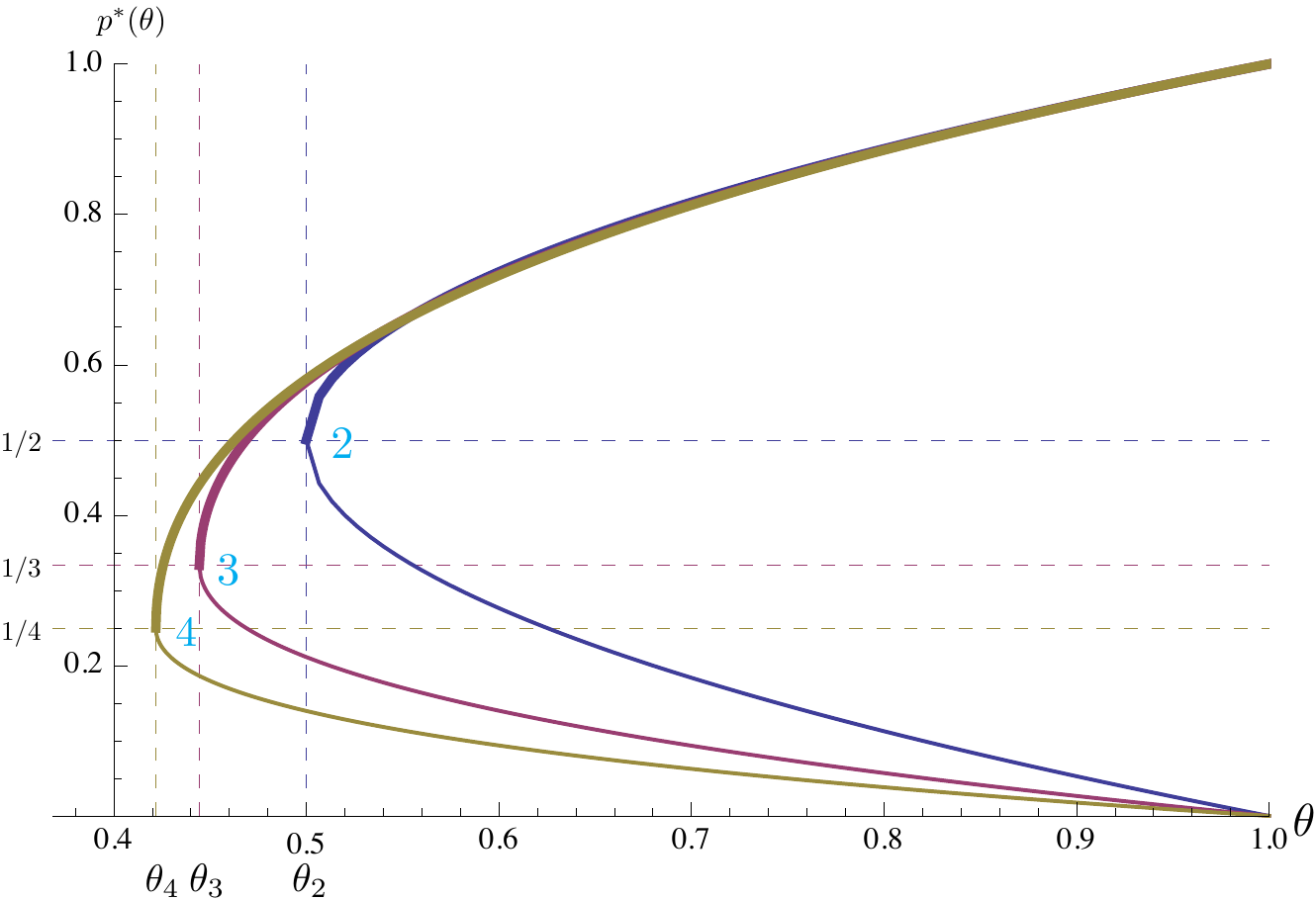}
\includegraphics[width=0.49 \textwidth ]{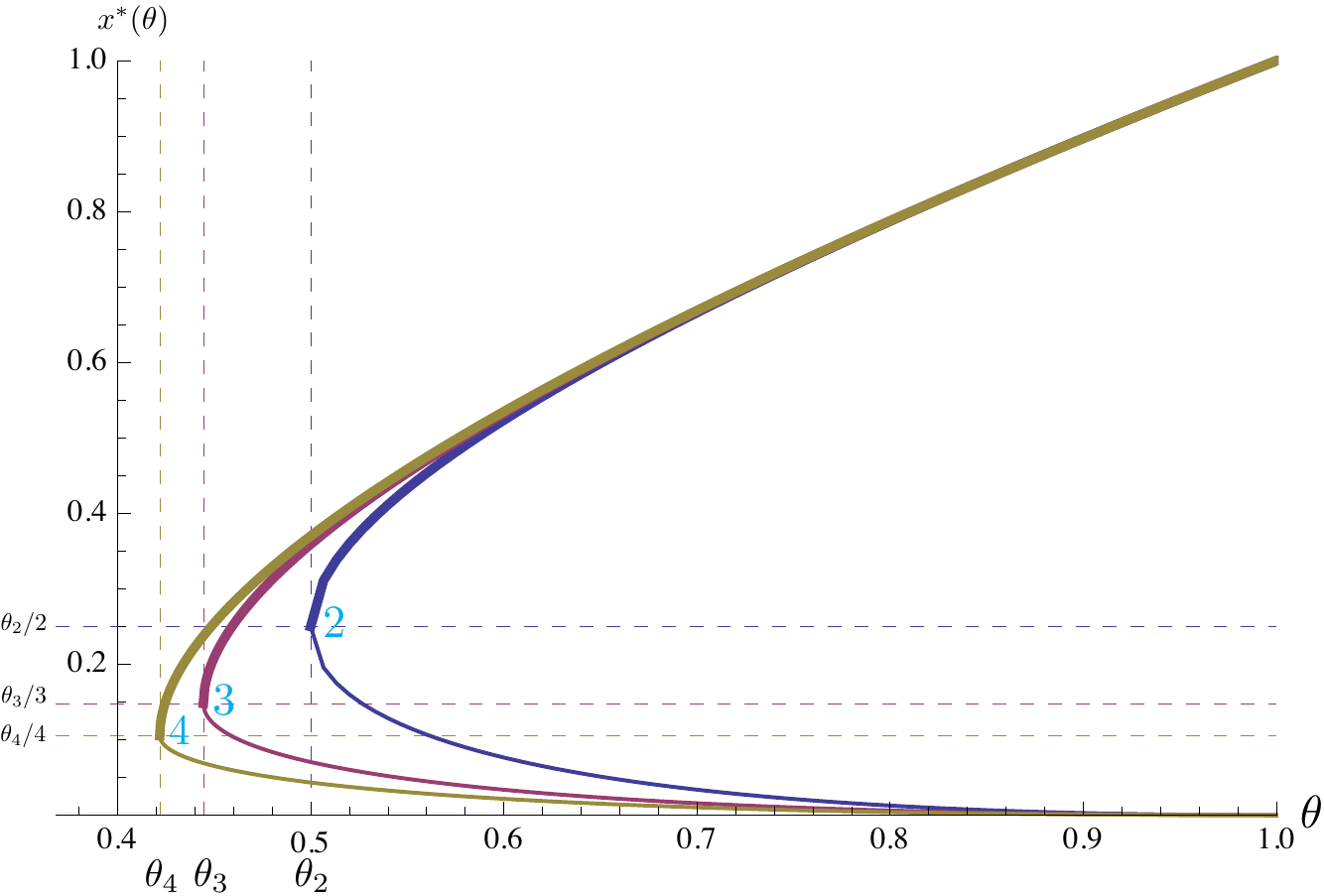}
\caption{Illustration of property $1)$ in \thmref{AlphaFairTFtradeoffProperties}: Optimal controls $p_s^*(\theta)$ (left, lower/thinner branches), $p_l^*(\theta)$ (left, upper/thicker branches) and optimal rates $x_s^*(\theta)$ (right, lower/thinner branches), $x_l^*(\theta)$ (right, upper/thicker branches) versus target throughput $\theta$, for $n = 2$ (blue), $3$ (purple), and $4$ (yellow) users. Vertical gridlines indicate the $\theta_{t}$'s: $(\theta_{2}, \theta_{3}, \theta_{4}) = (\frac{1}{2}, \frac{4}{9}, \frac{27}{64}) \approx (0.5, 0.4444, 0.4219)$. Horizontal gridlines indicate the corresponding optimal controls (left) and optimal rates (right) when $\theta = \theta_{t}$. Different from the case of Jain's fairness (see \figref{CJ2} in \secref{JainProp} where only the plots for $n = 4$ users are shown), here $n'^{*} = n$ holds irrespective of the value of $\theta$.}
\label{fig:AlphaFairOptContensionProbANDOptRateVectorVsTheta}
\end{figure}

The follow theorem gives some properties of the T-F tradeoff for the $\alpha$-fair objective. 
\begin{theorem} 
\label{thm:AlphaFairTFtradeoffProperties}
The T-F tradeoff for $n \geq 2$ users under $\alpha$-fairness for $\alpha \geq 1$, with  target throughput $\theta \in (\theta_{n}, 1)$, has the following properties:
\begin{enumerate}
\item For fixed $\alpha$ and $n$, the smaller ($p_{s}^{*}$) and larger ($p_{l}^{*}$) components of the optimal control are decreasing and increasing in $\theta$ respectively, i.e., $\frac{\drm p_s^*(\theta)}{\drm \theta} < 0$, $\frac{\drm p_l^*(\theta)}{\drm \theta} > 0$.  The smaller ($x_s^*$) and larger ($x_l^*$) components of the corresponding optimal rate vectors are likewise decreasing and increasing in $\theta$, i.e., $\frac{\drm x_s^*(\theta)}{\drm \theta} < 0$, $\frac{\drm x_l^*(\theta)}{\drm \theta} > 0$.

\item  For fixed $\alpha$ and $n$, the maximum $\alpha$-fair objective ($F_{\alpha}^{*}$) is decreasing in $\theta$ i.e., $\frac{\drm}{\drm \theta} F_{\alpha}^*(\theta; n) < 0$, and is continuous and differentiable.  For $n=2$, $F_{\alpha}^*(\theta;2)$ is concave (i.e., $\frac{\drm^2}{\drm \theta^2} F_{\alpha}^*(\theta; 2) < 0$).  For $n > 2$, there exists a throughput threshold $\mathring{\theta}_{\alpha}(n)$ such that  $F_{\alpha}^*(\theta;n)$ is convex (concave) in $\theta$ for $\theta < \mathring{\theta}_{\alpha}(n)$ ($\theta > \mathring{\theta}_{\alpha}(n)$).

\item For fixed $\alpha$ and $\theta \in (\theta_{n}, 1)$, the maximum $\alpha$-fair objective is decreasing in $n$, i.e., $F_{\alpha}^*(\theta; n) > F_{\alpha}^*(\theta; n+1)$.  
%\item For fixed $n$ and $\theta \in (\theta_{n}, 1)$, there exists a threshold $\mathring{\alpha}(\theta, n)$ such that $\theta$ lies in the interval over which $F_{\alpha}^*(\theta; n)$ is convex (concave) for $\alpha \lessgtr \mathring{\theta}_{\alpha}(n)$.
\end{enumerate}
\end{theorem}
The proof is found in \appref{AlphaFairPropPf}.
\figref{AlphaFairOptContensionProbANDOptRateVectorVsTheta} shows $p_s^*(\theta),p_l^*(\theta)$ (left) and $x_s^*(\theta),x_l^*(\theta)$ (right), illustrating property $1)$ in \thmref{AlphaFairTFtradeoffProperties}. \figref{AlphaFairTFTradeoffVaryingn} illustrates properties $2)$ and $3)$ for the cases of $\alpha = 1$ (left) and $\alpha = 2$ (right).

%% file: Sec6-Conclusion.tex
%%%%%%%%%%%%%%%%%%%%%%%%%%%%%%%%%%%%%%%%%%%%%%
% Section VI. Conclusion
%%%%%%%%%%%%%%%%%%%%%%%%%%%%%%%%%%%%%%%%%%%%%%

\section{Conclusion}
\label{sec:Conclusion}

We have presented six theorems that characterize the throughput--fairness tradeoff under slotted Aloha, using both Jain's fairness measure (Theorems 1-3), and the $\alpha$-fair measure (Theorems 4-6).  The key property enabling the analysis is \prpref{AtMostTwoDistinctNonzeroComponentValues}, which reduces the set of potential extremizers of the fairness functions from $[0,1)^n$ to $\partial \Smc_{1, 2}$, i.e., those controls taking at most two nonzero values.  Theorems 1 and 3 address the case of a throughput equality constraint, $T(\xbf) = \theta$, and Theorems 2 and 4 address the case of a throughput inequality constraint $T(\xbf) \geq \theta$.  The main point is that the throughput--fairness tradeoff is the same for both types of constraints (for $\theta \geq \theta_n$).  The key difference between the Jain and $\alpha$-fair tradeoff under a throughput constraint $\theta \in [\theta_t,\theta_{t-1})$ is in the nature of the optimal controls: to maximize the Jain fairness objective requires $n'^* = t$ active users, of which $k^*=1$ use a small contention probability and rate and $t-1$ use a large contention probability and rate, while to maximize the $\alpha$-fair objective requires {\em all} ($n'^* = n$) users be active with $k^*=n-1$ small users, and one large user.  Perhaps the most surprising result (to us) is the fact that the Jain throughput-fairness tradeoff is piecewise convex over each critical throughput interval $[\theta_t,\theta_{t-1})$ for $t \in [n]$, but not convex overall, i.e., over $[\theta_n,1)$.  

%% file: AppA-PropertiesProofs.tex
%%%%%%%%%%%%%%%%%%%%%%%%%%%%%%%%%%%%%%%%%%%%%%
% Appendix A. Proofs from Properties of Optimal Controls section
%%%%%%%%%%%%%%%%%%%%%%%%%%%%%%%%%%%%%%%%%%%%%%
\section{Proofs from \secref{PropOptCont}}
\label{app:PropOptContPf}

Proofs from \secref{MajorizationAttempt} and \secref{opttputconst} are given in \appref{MajorizationAttemptPf} and \appref{opttputconstPf} respectively.

\subsection{Proofs from \secref{MajorizationAttempt}}
\label{app:MajorizationAttemptPf}
The following lemma is used in the proof of \prpref{MajorizationProperties}, below. 

\begin{lemma}
\label{lem:StrictConvexSetBoundaryProperty}
Fix a set of $m \geq 2$ points $\Vmc \equiv \{v_{1}, \ldots, v_{m} \} \subset \Rbb^n$  such that no $v_i$ can be expressed as a convex combination of any other points in $\Vmc$, and denote by $\Cmc_{h} \equiv \text{conv}(\Vmc)$ the {\em convex hull} of $\Vmc$.  Fix a {\em strictly convex} set, denoted $\Cmc_{s}$, whose boundary also includes the set $\Vmc$, namely $\partial \Cmc_{s} \supseteq \Vmc$. Then the boundary of $\Cmc_{s}$ intersects $\Cmc_{h}$ only at the $m$ points that generate $\Cmc_{h}$, namely $\partial \Cmc_{s} \cap \Cmc_{h} =\Vmc$.
\end{lemma}

\begin{IEEEproof}
Note $\Vmc \subseteq \partial \Cmc_{s} \cap \Cmc_{h}$ by assumption. We need to show the intersection $\partial \Cmc_{s} \cap \Cmc_{h}$ can never include any other point. Recall a set $\Amc$ is strictly convex if for any $x, y \in \Amc$, every point on the {\em line segment} connecting $x$ and $y$ other than the end points is in the interior of $\Amc$.  First we observe $\Cmc_{s} \supseteq \Cmc_{h}$, by virtue of the fact that the convex hull is the smallest convex set that contains $\Vmc$. Second, we prove by contradiction that the intersection $\partial \Cmc_{s} \cap \Cmc_{h}$ can only consist of points on the boundary of $\Cmc_{h}$ (denoted $\partial \Cmc_{h}$). Assume there exists a point $v_{\rm int} \in \partial \Cmc_{s} \cap \Cmc_{h}$ that is an interior point of $\Cmc_{h}$. This means there exists a neighborhood of $v_{\rm int}$ that resides in $\Cmc_{h}$, however, as $v_{\rm int}$ is also on the boundary of $\Cmc_{s}$, every neighborhood of $v_{\rm int}$ must contain points that belong to neither $\Cmc_{s}$ nor $\Cmc_{h}$ (as $\Cmc_{s} \supseteq \Cmc_{h}$). This contradiction shows $\partial \Cmc_{s} \cap \Cmc_{h} \subseteq \partial \Cmc_{h}$. 
Observe that, since $\Cmc_h$ is a polytope, it has the property that any point on its boundary aside from the vertices, i.e., $v \in \partial\Cmc_h \setminus \Vmc$, may be expressed as a strict convex combination of two other points on the boundary, say $v',v{''} \in \partial \Cmc_h$.  Third, the previous sentence applies to any point $v \in (\partial \Cmc_{s} \cap \Cmc_{h}) \setminus \Vmc$, since such points are in $\partial\Cmc_h \setminus \Vmc$.  But the implied ability to represent $v$ as a strict convex combination of $v',v{''} \in \Cmc_s$ violates the assumed strict convexity of $\Cmc_s$, since it implies a boundary point of $\Cmc_s$ lies on the open line segment formed by two other points in $\Cmc_s$.  This establishes no such point exists, thereby proving the lemma.
\end{IEEEproof}

\begin{IEEEproof}[Proof of \prpref{MajorizationProperties}]
Write $\Pi(\xbf)$ to denote the $n!$ permutations of $\xbf$.  For item $1)$, we apply Prop.\ C.1 in Ch.\ 4 of \cite[pp.\ 162]{MarOlk2011} (Rado, 1952) which says $\abf \prec \bbf$ if and only if $\abf$ lies in the convex hull of the $n!$ permutations of $\bbf$, denoted $\text{conv}(\Pi(\bbf))$. Let $\xbf \in \Lambda_{\theta}^{\rm int}$; it suffices to establish $\xbf' \in \partial \Lambda_{\theta}$ with $\xbf' \prec \xbf$.  The geometric argument below is illustrated in \figref{ProofThm1} by replacing $\xbf^{*}$ in the figure with $\xbf'$.  Define $\cbf \equiv \theta \ubf$.  First: it follows from \lemref{AllRatesEqualRay} that $\cbf \not\in \Lambda$ (since $\theta > \theta_n$), but that $\cbf \in \text{conv}(\Pi(\xbf))$ (using the convex combination of $\Pi(\xbf)$ with all weights equal to $1/n!$).  Second: it follows from the convexity\footnote{The complement of $\Lambda$, i.e., $\Lambda^c \equiv \Rbb^n_+ \setminus \Lambda$ is shown to be convex by Post in \cite{Pos1985}.} of $\Lambda^c$ that there exists a unique point $\xbf' \in \partial\Lambda$ on the line segment connecting $\xbf$ with $\cbf$.  Third: it follows from the convexity of $\Hmc_{\theta}$ that $\xbf' \in \Hmc_{\theta}$ (which contains both $\xbf,\cbf$), and therefore, $\xbf' \in \partial\Lambda_{\theta}$ (as it lies in both $\partial\Lambda$ and $\Hmc_{\theta}$).  Fourth: this point $\xbf' \in \text{conv}(\Pi(\xbf))$ by the convexity of $\text{conv}(\Pi(\xbf))$ (which contains both $\xbf,\cbf$).  Fifth: by Rado's result, $\xbf' \prec \xbf$, which concludes the proof of item $1)$.

For item $2)$, we again apply Rado's result and prove by contradiction. Assume there exist distinct (up to permutation) $\xbf, \xbf'$ both in $\partial \Lambda_{\theta}$ satisfying $\xbf' \prec \xbf$, equivalently, $\xbf' \in \text{conv}(\Pi(\xbf)) \equiv \Cmc_{h}$. The contradiction will establish $\partial\Lambda_{\theta} \cap \Cmc_{h} = \Pi(\xbf)$, meaning the only feasible points (i.e., in $\partial\Lambda_{\theta}$) that are majorized by $\xbf$ (i.e., in $\Cmc_{h}$) are permutations of the original point $\xbf$.  This provides the desired contradiction since permutations of a point do not majorize each other.  Our approach to establishing $\partial\Lambda_{\theta} \cap \Cmc_{h} = \Pi(\xbf)$ is to apply \lemref{StrictConvexSetBoundaryProperty}, with $\Vmc = \Pi(\xbf)$ and $\Cmc_s = \Lambda^c_{\theta} = \Lambda^c \cap \Hmc_{\theta}$.  To apply \lemref{StrictConvexSetBoundaryProperty} we must show $i)$ $\Cmc_s$ is strictly convex, and $ii)$ $\partial \Cmc_s \supseteq \Vmc$, i.e., $\partial \Lambda_{\theta} \supseteq \Pi(\xbf)$ (since $\partial\Lambda^c = \partial\Lambda$).  The lemma establishes the desired result, $\partial\Cmc_s \cap \Cmc_h =\Pi(\xbf)$.  It remains to show $i)$ and $ii)$.  $i)$ Subramanian and Leith \cite[Lem.\ 1 and Remark $1$ in \S II-A]{SubLei2013} have shown that $\Lambda^{c}$ is strictly convex\footnote{Post \cite{Pos1985} establishes the tangent hyperplane equation of every point on $\partial \Lambda$.} in $\Rbb_{+}^{n}$. As strict convexity is preserved under intersection with affine spaces, it follows that $\Cmc_{s}$ is strictly convex.  $ii)$ By assumption $x \in \partial\Lambda_{\theta}$, which ensures $\Pi(\xbf) \subset \partial\Lambda_{\theta}$ since $\Lambda$ and $\Hmc_{\theta}$ are permutation invariant.  This establishes item $2)$.

\end{IEEEproof}

\begin{IEEEproof}[Proof of \corref{MajorizationCorollary} for the case of Jain's fairness \eqnref{Chiu-JainFairnessDefinition} (formulated as \eqnref{TFtradeoffProblemFormulation})]

Given that $\xbf^{*}$ satisfies the throughput constraint $\xbf^{*} \in \Hmc_{\theta}$, we need to show $\xbf^{*} \in \partial \Lambda$, i.e., the optimal rate vector $\xbf^*$ is Pareto efficient.  Refer to \figref{ProofThm1} for geometric intuition.  Recall $\mathbf{0}$ denotes the origin and $\mbf \equiv \frac{1}{n} \theta_n \mathbf{1} \in \partial\Lambda$.  Define the following: $i)$ $\cbf = c \mathbf{1}$ with $c \equiv \theta / n$, $ii)$ $\mathrm{ray}(\mathbf{0},\mathbf{1})$ as the ray emanating from $\mathbf{0}$ in the direction $\mathbf{1}$ (holding $\mathbf{0}$, $\mbf$, and $\cbf$).
Recall $i)$ $\Hmc_{\theta} = \{ \xbf : \sum_i x_i = \theta\}$ is the hyperplane with normal $\mathbf{1}$ (and thereby orthogonal to $\mathrm{ray}(\mathbf{0},\mathbf{1})$), and $ii)$ $\xbf = \xbf(\pbf) \in \Hmc_{\theta} \cap \Lambda$ is a feasible rate vector under the throughput constraint for feasible control $\pbf$.  Observe $\Hmc_{\theta}$ intersects with $\mathrm{ray}(\mathbf{0},\mathbf{1})$ at $\cbf$.  Finally, note that the objective in \eqnref{TFtradeoffProblemFormulation} is $\frac{1}{2}d (\xbf,\mathbf{0})^{2}$.  

Since $\Hmc_{\theta}$ is orthogonal to $\mathrm{ray}(\mathbf{0},\mathbf{1})$, it follows that $(\mathbf{0},\cbf,\xbf)$ form a right triangle with the right angle at $\cbf$, and therefore, by the Pythagorean theorem, $d (\xbf, \mathbf{0})^{2} = d (\xbf, \cbf)^{2} + d (\cbf, \mathbf{0})^{2}$.  It follows that the objective $\frac{1}{2} d (\xbf, \mathbf{0})^{2}$ is minimized iff $d (\xbf, \cbf)^{2}$ is minimized (over $\xbf \in \Hmc_{\theta} \cap \Lambda$).  Observe the assumption $\theta \geq \theta_n$ ensures $\cbf \not\in \partial \Lambda$ for $\theta > \theta_n$, and $\cbf = \mbf \in \partial\Lambda$ for $\theta = \theta_n$ (in which case the unique global minimizer is $\xbf^* = \mbf$).  Fix a candidate feasible point $\xbf \in \Hmc_{\theta} \cap \Lambda$ and consider the line segment connecting $\xbf$ with $\cbf$: it must intersect $\partial\Lambda$, and this point is denoted $\xbf^*(\xbf)$.  It is clear that any feasible $\xbf'$ on the line segment $(\xbf,\cbf)$ not equal to $\xbf^*(\xbf)$ is suboptimal to $\xbf^*(\xbf)$ in that $d(\xbf',\cbf) > d(\xbf^*(\xbf),\cbf)$.  This shows the desired minimizer $\xbf^{*} \in \partial \Lambda$. Equivalently (\cite{Pos1985}, recall \eqnref{LambdaBoundary}) this means the corresponding optimal control $\pbf^{*}$ (in the sense of \eqnref{xofp}) is in $\partial \Smc$.
\begin{figure}[!h]
\centering
\includegraphics[width=0.49 \textwidth ]{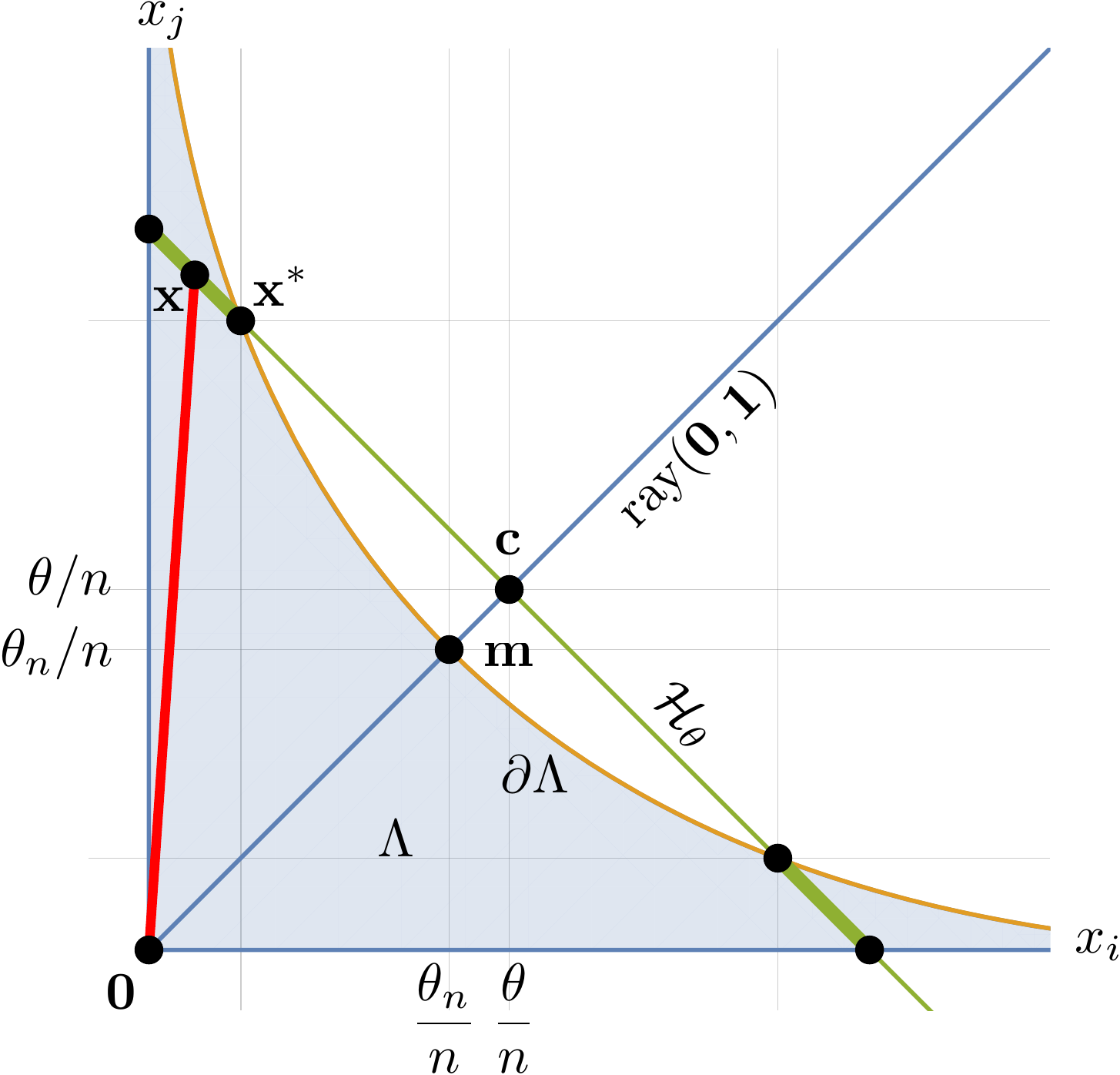}
\caption{Illustration of the proof of \corref{MajorizationCorollary} for the case of Jain's fairness: $\xbf(\pbf^{*}) \in \partial \Lambda$.  The feasible set under a throughput constraint $T(\xbf) = \theta$ is the intersection of $\Lambda$ with the throughput hyperplane $\Hmc_{\theta} = \{ \xbf : \sum_i x_i = \theta\}$, indicated by the two bold green line segments.}
\label{fig:ProofThm1}
\end{figure}
\end{IEEEproof}

\subsection{Proofs from \secref{opttputconst}}
\label{app:opttputconstPf}

\begin{IEEEproof}[Proof of \prpref{tputeqconst}]
We prove the three statements in the order they are given.

{\em Proof of 1).}  Observe by definition of $\pbf \in \overline{\partial\Smc_2}$ and $T(\xbf(\pbf))$, we may write $T(p_s,k,n') = k x_{s} + (n'-k) x_{l}$.  Substituting the expressions for $x_s,x_l$ in \eqnref{xsxl} in \defref{restrictedControls} yields:
\begin{equation}
\label{eq:gpsasbdf}
T(p_{s}, k,n') =  -\frac{(1-p_{s})^{k-1} \left(-k n' p_{s}^2+ n' p_{s} + k - n\right)(1- p_{l})^{n'-k}}{ k p_{s} + n' - k - 1}.
\end{equation}
The partial derivative w.r.t. $p_{s}$ is
\begin{equation} 
\label{eq:dgdps}
\frac{\partial}{\partial p_{s}} T(p_s,k,n') = -\frac{k (1-p_{s})^{k-2} (n' p_{s}-1) (1- p_{l})^{n'-k} ( k p_s(n' p_s - 2) - (n'-1-k))}{(k p_{s} + n'  - k - 1)^2}.
\end{equation}
One can easily verify this derivative is nonpositive on the regime of interest, and thus $T(p_s,k,n')$ is monotone decreasing in $p_s$ on $(0,1/n']$, and as such there can exist at most one value of $p_s$ solving $T(p_s,k,n') = \theta$.

{\em Proof of 2).}  Since $n'$ is fixed, we write $T(p_{s}, k, n')$ defined in \eqnref{signedtputgap} as $T(p_{s}, k)$. Observe that the monotonicity of $T(p_s,k)$ ensures $\Rmc(k,n') = [T(1/n',k,n'),T(0,k,n'))$.   
Observe from \eqnref{plofpskn} that $p_l(1/n',k,n') = p_s$ and $p_l(0,k,n') = 1/(n'-k)$.  Substitution of $(p_s,p_l) = (1/n',1/n')$ and $(p_s,p_l) = (0,1/(n'-k))$ in \eqnref{gpsasbdf} yields $\Rmc(k,n') = [\theta_{n'},\theta_{n'-k})$.  As $\theta_{n'}$ is constant in $k$, while $\theta_{n'-k}$ is increasing in $k$, it follows that the intervals forming each $\Rmc(k,n')$ are nested and increasing in $k$. 

{\em Proof of 3).} Recall $i)$ $\theta_n \leq \cdots \leq \theta_1$ \eqnref{criticalthroughputs}, $ii)$ $\Rmc(k,n') = [\theta_{n'},\theta_{n'-k})$ (Item $2)$), and $iii)$ by assumption, the target $\theta$ lies in $[\theta_t,\theta_{t-1})$, for some $t \in \{2, \ldots, n\}$.  First, observe $n' \geq t$ needs to hold, since for $n' \leq t - 1$ we have 
\begin{equation}
\label{eq:RangeOfAchievableThroughputFirstStep}
\Rmc(k, n') \cap [\theta_{t}, \theta_{t-1}) = [\theta_{n'},\theta_{n'-k}) \cap [\theta_{t}, \theta_{t-1}) =  \emptyset.
\end{equation}
Second, refer to \figref{DtnpRegion} (right).  As evident from the figure, $\theta \in \Rmc(k,n')$ if and only if $k \in \{n'-t+1, \ldots, n'-1\}$. 
\end{IEEEproof}

\begin{IEEEproof}[Proof of \prpref{AtMostTwoDistinctNonzeroComponentValues}]

We prove the two statements in the order they are given.

{\em Proof of $i)$.} The main idea of the proof is to establish the impossibility of any $\pbf \in [0,1)^n$ simultaneously being an extremizer and having $|\Vmc(\pbf)| > 2$. Observe we may partition the feasible set $[0,1)^n$ into $\{ \pbf \in [0,1)^n : |\Vmc(\pbf)| \leq 2\}$ and $\{ \pbf \in [0,1)^n : |\Vmc(\pbf)| > 2\}$.  We now show any $\pbf$ with $|\Vmc(\pbf)| > 2$ cannot satisfy the KKT conditions, given below, necessary for $\pbf$ to be an extremizer.

We first consider the case of a throughput inequality constraint, $T(\xbf(\pbf)) \geq \theta$.  
Introducing Lagrange multipliers $\mu_{\theta}$ for $T(\xbf(\pbf)) \geq \theta$, $\boldsymbol\lambda = (\lambda_{i},i \in [n])$ for $\pbf \geq \mathbf{0}$, and $\boldsymbol\nu = (\nu_i, i \in [n])$ for $\pbf < \mathbf{1}$, the Lagrangian is:
\begin{equation}
\label{eq:JainLagrangian1}
\Lmc (\pbf, \mu_{\theta}, \boldsymbol\lambda, \boldsymbol\nu) = F_{\alpha}(\xbf(\pbf)) + \mu_{\theta} (\theta - T(\xbf(\pbf))) + \sum_{i=1}^{n} \lambda_{i} \left(-p_{i}\right) + \sum_{i=1}^{n} \nu_{i} \left(p_{i} - 1 \right).
\end{equation}

The first-order Karush-Kuhn-Tucker (KKT) necessary conditions for a {\em maximizer} are, for each $i \in [n]$:
\begin{eqnarray*}
\text{stationarity} & & \frac{\partial \Lmc}{\partial p_{i}} = 0 \\
\text{primal feasibility} & & \theta - T(\xbf(\pbf)) \leq 0 \\
& & -p_{i} \leq 0, ~ p_{i} - 1 < 0 \\
\text{dual feasibility} & & \mu_{\theta} \leq  0, \lambda_{i} \leq  0, \nu_{i} \leq 0 \\
\text{comp.\ slackness} & & \mu_{\theta} (\theta - T(\xbf(\pbf))) = 0 \\
& & \lambda_{i} \left(-p_{i}\right) = 0, ~ \nu_{i} \left(p_{i} - 1\right) = 0.
\end{eqnarray*}
The KKT conditions for a {\em minimizer} are the same, with the signs on each Lagrange multiplier on each inequality constraint reversed.  As is evident from the proof below, the sign of the multipliers is inessential to establishing the result, and therefore the result holds for {\em both} minimization and maximization.  

The first step of the proof is to derive the condition $g_k = 0$ in \eqnref{giJain} below from the KKT stationarity condition $\frac{\partial \Lmc}{\partial p_k} = 0$ when $0 < p_k < 1$.  Towards that goal, we make the following definitions, where the dependence of these quantities upon $\pbf$ is omitted for brevity:
\begin{eqnarray}
F_{\alpha} \equiv F_{\alpha}(\xbf(\pbf)), & &  
T \equiv T(\xbf(\pbf)) \nonumber \\
\pi = \pi(\pbf) \equiv \prod_j (1-p_{j}), & & 
\pi_{i} = \pi_i(\pbf) \equiv \frac{\pi}{1-p_i}.
\label{eq:FTpi}
\end{eqnarray}
Observe $x_i = x_i(\pbf)$ in \eqnref{xofp} may be written in terms of $\pi$ as $x_i = \frac{p_i}{1-p_i} \pi$. Differentiation of  \eqnref{JainLagrangian1} yields:
\begin{equation}
\label{eq:JainLagrangian23}
\frac{\partial \Lmc}{\partial p_{i}} = \frac{\partial F_{\alpha}}{\partial p_{i}} - \mu_{\theta} \frac{\partial T}{\partial p_{i}} - \lambda_i + \nu_i.
\end{equation}
The following partial derivatives may be established after some algebra:
\begin{eqnarray}
\label{eq:UsefulPartialDerivativesXTF}
\frac{\partial x_j(\pbf)}{\partial p_i} &=& \left\{ \begin{array}{cl} 
\frac{\pi}{1-p_j}, \; & i = j \\
-\frac{\pi p_j}{(1-p_i)(1-p_j)}, \; & i \neq j 
\end{array} \right. \nonumber \\
\frac{\partial T}{\partial p_{i}} & = & \frac{\pi_{i} - T}{1-p_{i}} \nonumber \\
\frac{\partial F_{\alpha}}{\partial p_{i}} &=& -\frac{1-\alpha}{1-p_{i}} F_{\alpha} + \frac{\pi}{\left( 1 - p_{i} \right)^{2}} x_{i}^{-\alpha} 
\end{eqnarray}
Substitution of the above into \eqnref{JainLagrangian23} yields
\begin{equation}
\frac{\partial \Lmc}{\partial p_{i}} = \frac{g_i}{1-p_{i}}  - \lambda_{i} + \nu_{i},
\end{equation}
where $\gbf = (g_i, i \in [n])$ has components
\begin{equation}
\label{eq:giJain}
g_i \equiv -(1-\alpha) F_{\alpha} + \pi_{i} x_{i}^{-\alpha} + \mu_{\theta} (T - \pi_{i}).
\end{equation}
The quantity $g_i$ has the following important property: if $k \in [n]$ is such that $0 < p_k < 1$ then stationarity and complementary slackness require $\frac{\partial \Lmc}{\partial p_{k}} = \lambda_{k} = \nu_{k} = 0$, which in turn requires $g_k = 0$.  Next fix two distinct indices, $i_1$ and $i_2$, such that $0<p_{i_1}, p_{i_2}<1$, which by the above argument, requires $g_{i_1} = g_{i_2} = 0$.  Substituting \eqnref{giJain} into this equation, substituting the earlier expressions for $\pi_i$ and $x_i$, and solving for $\mu_{\theta}$ yields:
\begin{equation}
\label{eq:muthetaJainFair}
\mu_{\theta}(i_1,i_2) \equiv \frac{(1-p_{i_1}) f_1(p_{i_2};\alpha) - (1-p_{i_2}) f_1(p_{i_1};\alpha)}{\pi^{\alpha} (p_{i_2} - p_{i_1})},
\end{equation}
where
\begin{equation}
\label{eq:JainFairfyalpha}
f_1(y;\alpha) \equiv \left( \frac{1}{y} - 1 \right)^{\alpha} \text{ for } y \in (0,1).
\end{equation}
Here $\mu_{\theta}(i_1,i_2)$ denotes the unique value of the Lagrange multiplier $\mu_{\theta}$ enforced by the KKT conditions for indices $i_1,i_2$.  

As, by assumption, $\vert \Vmc(\pbf) \vert > 2$, there exist at least three distinct indices $\{j,k,l\}$ with $0 < p_j < p_k < p_l < 1$.  As there can only be one value for $\mu_{\theta}$, it follows that $\mu_{\theta}(j,k) = \mu_{\theta}(j,l) = \mu_{\theta}(k,l)$.  Equating $\mu_{\theta}(j,k) = \mu_{\theta}(j,l)$ and simplifying gives
\begin{equation} 
\label{eq:KeyEqInvolvingjklUsingLambdaTheta}
(p_{k} - p_{j}) f_1(p_l;\alpha) + (p_{l} - p_{k}) f_1(p_j;\alpha) = (p_{l} - p_{j}) f_1(p_k;\alpha).
\end{equation}
The assumed ordering of $p_j,p_k,p_l$ ensures that $p_k$ may be written as a convex combination of $(p_j,p_l)$, i.e., $p_{k} = t \cdot p_{l} + (1-t) \cdot p_{j}$ for
\begin{equation}
\label{eq:t1mtJainFair}
t = t(p_j,p_k,p_l) \equiv \frac{p_{k} - p_{j}}{p_{l} - p_{j}},~~
1-t = \frac{p_l - p_k}{p_l - p_j}.
\end{equation}
By the assumptions on $p_j$, $p_k$, and $p_l$, both $t$ and $1-t$ are in $ (0,1)$. Subtitution of the above into \eqnref{KeyEqInvolvingjklUsingLambdaTheta} yields:
\begin{equation} \label{eq:KeyEqInvolvingjklRewrittenUsingLambdaTheta}
t \cdot f_1(p_l;\alpha) + (1 - t) \cdot f_1(p_j;\alpha) = f_1(t \cdot p_{l} + (1-t) \cdot p_{j};\alpha).
\end{equation}
To summarize thus far, the KKT conditions applied to these three distinct nonzero values require each of the three pairs of indices to agree on the value of the Lagrange multiplier $\mu_{\theta}$ \eqnref{muthetaJainFair}, and this is equivalent to the condition that \eqnref{KeyEqInvolvingjklRewrittenUsingLambdaTheta} holds for $t$ in \eqnref{t1mtJainFair}.  The natural interpretation of \eqnref{KeyEqInvolvingjklRewrittenUsingLambdaTheta} is that the function $f_1(y;\alpha)$ has the property that the convex combination, with parameter $t$, of the values $f_1(p_l)$ and $f_1(p_j)$ equals the value of $f_{1}$ at the convex combination of the arguments $p_j$ and $p_l$ with the same parameter $t$.  Geometrically, this requires the point $(p_k,f_1(p_k))$ to lie on the chord connecting $(p_j,f_1(p_j))$ with $(p_l,f_1(p_l))$, as illustrated in \figref{ProofPropVmcLeq2}  (left).

\begin{figure}[!ht]
\centering
\includegraphics[width=\textwidth]{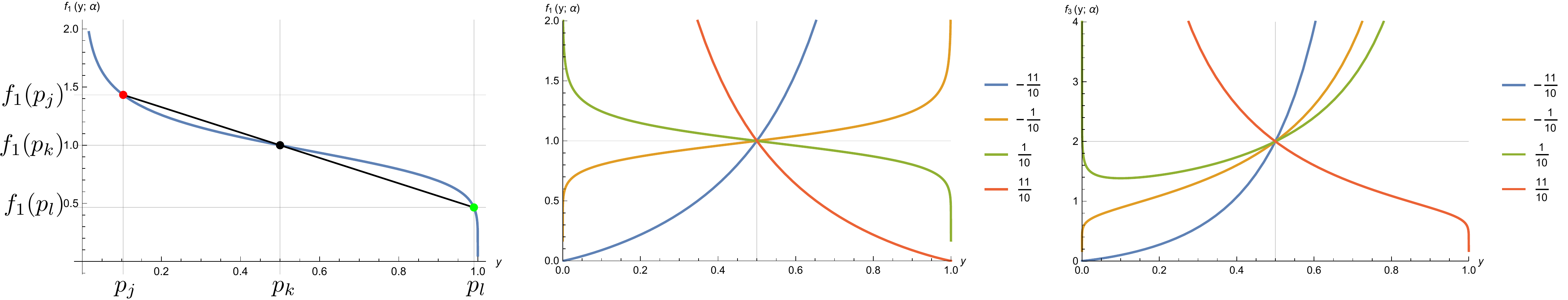}
\caption{Proof of \prpref{AtMostTwoDistinctNonzeroComponentValues}.  {\bf Left:} an optimal $\pbf$ with three distinct nonzero values $0 < p_j < p_k < p_l < 1$ must satisfy \eqnref{KeyEqInvolvingjklRewrittenUsingLambdaTheta}, which states the point $(p_k,f_1(p_k; \alpha))$ must lie on the chord connecting $(p_j,f_{1}(p_j; \alpha))$ with $(p_l,f_1(p_l; \alpha))$, for $f_1(y;\alpha)$ in \eqnref{JainFairfyalpha}. Here $\alpha = 1/6$ and the coordinates of the three points are (approximately) $(0.1036, 1.4329)$ (red), $(1/2, 1)$ (black), and $(0.99, 0.4649)$ (green).  {\bf Middle:} the function $f_1(y;\alpha)$ for various $\alpha$; we established $f_1$ to be strictly convex for $\alpha \in (-\infty, -1] \cup [1, \infty)$, precluding $\pbf$ to be optimal for all such $\alpha$.  {\bf Right:} the function $f_3(y;\alpha)$ for various $\alpha$; we established $f_3$ to be strictly monotone for $\alpha \in (-\infty, -1] \cup [1, \infty)$, precluding the throughput inequality to be loose for all such $\alpha$.}
\label{fig:ProofPropVmcLeq2}
\end{figure}

Recall a univariate function $f$ is strictly convex if its domain $\mathrm{dom} f$ is convex and
\begin{equation}
f(s y_{1} + (1-s) y_{2}) < s f(y_{1}) + (1-s) f(y_{2}), ~ \forall y_{1}, y_{2} \in \mathrm{dom} f, ~ \forall s \in (0,1),
\end{equation}
and is strictly concave if the inequality is reversed.  In particular, the above strict inequality, for both strictly convex and strictly concave functions, ensures \eqnref{KeyEqInvolvingjklRewrittenUsingLambdaTheta} cannot hold for any $t$, and thus a contradiction is reached in the assumed optimality of the $\pbf$ with three or more distinct values, for any $\alpha$ for which $f_1(y;\alpha)$ is strictly convex or strictly concave.  Our analysis is inconclusive in the regime where $f_1(y;\alpha)$ is neither strictly convex nor strictly concave: it may or may not be possible to satisfy  \eqnref{KeyEqInvolvingjklRewrittenUsingLambdaTheta}. 

This motivates us to investigate the convexity / concavity of the function $f_1(y;\alpha)$ in $y$.  The second derivative (w.r.t. $y$) is
\begin{equation} 
f_1^{(2)}(y;\alpha) = y^{-4} \left(y^{-1} - 1\right)^{\alpha - 2} f_2(\alpha;y)
\end{equation}
for
\begin{equation}
f_2(\alpha;y) \equiv \alpha (1+ \alpha - 2y).
\end{equation}
Since the domain of $y$ is $(0,1)$, the sign of $f_1^{(2)}(y;\alpha)$ is determined by $f_2(\alpha;y)$, which we view as a quadratic in $\alpha$ with parameter $y$.  Recall $f_1^{(2)}(y;\alpha) \gtrless 0$ is a sufficient condition for $f_1(y;\alpha)$ to be strictly convex (concave) in $y$. 
Define the sets 
\begin{eqnarray}
\Amc_{f_2} &=& \{ \alpha : f_2(\alpha;y) > 0 ~ \forall y \in (0,1) \} \nonumber \\
\Amc_{f_2}^+ &=& \{ \alpha > 0: f_2(\alpha;y) > 0 ~ \forall y \in (0,1) \} \nonumber \\
\Amc_{f_2}^- &=& \{ \alpha < 0: f_2(\alpha;y) > 0 ~ \forall y \in (0,1) \}
\end{eqnarray}
and note $f_1(y;\alpha)$ is strictly convex in $y$ for $\alpha \in \Amc_{f_2}$.  Next, observe $\Amc_{f_2} = \Amc_{f_2}^+ \cup \Amc_{f_2}^-$, since ${f_2}(0;y) = 0$.  Furthermore, it is evident that $\Amc_{f_2}^+ = [1,\infty)$ and $\Amc_{f_2}^- = (-\infty,-1]$, and so $\Amc_{f_2} = (-\infty,-1] \cup [1,\infty)$.

Similarly it can be verified there is no value of $\alpha \in \Rbb$ for which $f_2(\alpha;y) < 0$ {\em for all} $y \in (0,1)$, meaning $f_1(y;\alpha)$ is not strictly concave on $(0,1)$ for any $\alpha$.  In summary, we've established the impossibility of an optimal $\pbf$ having $|\Vmc(\pbf)| > 2$ for $\alpha \in (-\infty, -1] \cup [1, \infty)$, as illustrated in \figref{ProofPropVmcLeq2} (middle).  

We next consider the case of a throughput equality constraint, $T(\xbf(\pbf)) = \theta$.  The only change in the KKT conditions from the inequality constraint case is that now the sign of the Lagrange multiplier $\mu_{\theta}$ is unrestricted.  However, observe that the above proof for the inequality constraint case does not rely upon the dual feasibility condition of $\mu_{\theta}$.  As such, the above proof holds in this case as well.

{\em Proof of $ii)$.} By assumption that the optimizer $\pbf^*$ has $|\Vmc(\pbf^*)| = 2$, we denote the two nonzero component values by $0 < p_k < p_l < 1$.  We prove by contradiction. Assuming the throughput constraint does not hold with  equality namely $T(\xbf(\pbf^*)) > \theta$, it follows that the corresponding Lagrange multiplier $\mu_{\theta}$ is zero, and in particular we must have $\mu_{\theta}(k,l) = 0$ in \eqnref{muthetaJainFair}.  This expression may be rearranged as $f_3(p_k;\alpha) = f_3(p_l;\alpha)$, for
\begin{equation}
f_3(y;\alpha) \equiv \frac{f_1(y;\alpha)}{1-y}.
\end{equation}
We next establish that $f_3(y;\alpha)$ is strictly monotone in $y \in (0,1)$ for all $\alpha \in (-\infty, -1] \cup [1, \infty)$, as illustrated in \figref{ProofPropVmcLeq2}.  This strict monotonicity means it is impossible to have $0 < p_k < p_l < 1$ and $f_3(p_k;\alpha) = f_3(p_l;\alpha)$.  The first derivative of $f_3$ (w.r.t. $y$) is
\begin{equation}
f_3^{(1)}(y;\alpha) = \frac{y - \alpha}{y (1-y)^{2}} f_1(y; \alpha).
\end{equation}
And thus $f_3$ is either always strictly monotone increasing (when $\alpha \in (-\infty, -1]$) or always strictly monotone decreasing in $y$ (when $\alpha \in [1, \infty)$), for all $y \in (0,1)$. This implies $\mu_{\theta}(k,l) = 0$ cannot hold, which in turn implies, as a consequence of complementary slackness, at an optimizer $\pbf^{*}$ that has the property that $\vert \Vmc(\pbf^{*}) \vert = 2$, the throughput inequality constraint must be tight i.e., $T(\xbf(\pbf^{*})) = \theta$.

Note that in all the above analysis, the expression for the $\alpha \neq 1$ case of $F_{\alpha}$, defined in \eqnref{falphadef}, is used. As the claimed regime of $\alpha$ (i.e., $(-\infty, -1] \cup [1, \infty)$) to which the assertion of this proposition applies includes $\alpha = 1$, it is necessary to verify it also holds for this case. 
This is done separately below.
\end{IEEEproof}

\begin{IEEEproof}[Proof of \prpref{AtMostTwoDistinctNonzeroComponentValues} for the $\alpha = 1$ case]
We prove the two parts in the order they are given.

{\em Proof of $i)$.} 
The domain $\pbf \in [0,1)^{n}$ allows us to rule out the possibility of any component $p_{i} = 1$. We will further dismiss the case when there exists some component $p_{i} = 0$, because if any such zero component exists in $\pbf$, then the corresponding rate $x_{i} = 0$, which gives the objective $F_{1}(\xbf(\pbf)) = -\infty$ meaning it is uninteresting/infeasible if we were to minimize/maximize $F_{1}(\xbf)$.  Let $\pbf$ obey $|\Vmc(\pbf)| > 2$; we will show any such point cannot satisfy the KKT conditions.

We first consider the case of a throughput inequality constraint, $T(\xbf(\pbf)) \geq \theta$. Since $F_{1}(\xbf)$ is maximized iff $\tilde{F}_{1} (\xbf) \equiv \prod_{j=1}^{n} x_{j}$ is maximized (for $\xbf > \mathbf{0}$), we work with $\tilde{F}_{1}$. Introduce Lagrange multipliers $\mu_{\theta}$, $\boldsymbol\lambda$, and $\boldsymbol\nu$, and form exactly the same Lagrangian \eqnref{JainLagrangian1}, with the objective replaced by $\tilde{F}_{1}$.

As $0 < p_i < 1$ it follows that $\lambda_{i} = \nu_{i}= 0$. As $0 < p_i < 1$ holds for all $i \in [n]$, it follows that $\pi (\pbf) \neq 0$ (defined in \eqnref{FTpi}) , and as such the stationarity equation $\frac{\partial \Lmc}{\partial p_{i}} = 0$ of \eqnref{JainLagrangian1} may be solved for $\mu_{\theta}$:
\begin{equation} 
\label{eq:PropFairmutheta}
\mu_{\theta} = - \frac{ n - \frac{1}{p_{i}}}{\frac{1}{1-p_{i}} - A(\pbf)}    \prod_{j=1}^{n} p_{j} (1-p_{j})^{n-2},
\end{equation}
where $A(\pbf) \equiv \sum_{j=1}^{n} \frac{p_{j}}{1-p_{j}}$.

Fixing indices $i_1,i_2$ with $0 < p_{i_1} < p_{i_2} < 1$, the two equations $\frac{\partial \Lmc}{\partial p_{i_1}}= 0$ and $\frac{\partial \Lmc}{\partial p_{i_2}} = 0$ may each be solved for $\mu_{\theta}$ in \eqnref{PropFairmutheta}, equated with each other, and the resulting equation may be solved for $A(\pbf)$:
\begin{equation} 
\label{eq:KeyequationpkplRevised}
%\frac{n - \frac{1}{p_{k}}}{\frac{1}{1-p_{k}} - A(\pbf)}  =  \frac{n - \frac{1}{p_{l}}}{\frac{1}{1-p_{l}} - A(\pbf)} \Rightarrow  
A(\pbf) = A(i_1,i_2) = \frac{1 - n p_{i_1} p_{i_2}}{(1-p_{i_1}) (1-p_{i_2})}.
\end{equation}
Here $A(i_{1}, i_{2})$ denotes the value of $A(\pbf)$ obtained from the KKT stationarity condition for indices $i_{1}, i_{2}$.

Now consider three distinct indices $\{j,k,l\}$ with $0 < p_j < p_k < p_l < 1$.  As there can only be one value for $A$, it follows that $A(j,k) = A(j,l) = A(k,l)$.  Equating any pair out of these three and simplifying yields $p_{s} = 1 / n$ where $s$ is the common index in the two pairs of indices. Collectively this implies $p_{j} = p_{k} = p_{l} = 1 / n$, which is a contradiction. This shows $\vert \Vmc (\pbf) \vert \leq 2$.

We now consider the case of a throughput equality constraint, $T(\xbf(\pbf)) = \theta$. Since in this case there is no restriction on the sign of the corresponding Lagrange multiplier $\mu_{\theta}$, the above proof holds as well.

{\em Proof of $ii)$.} 
For the second part of the proposition, we prove by contradiction. Given $\vert \Vmc(\pbf^{*}) \vert = 2$, meaning $\pbf^{*}$ has components $p_{k}$, $p_{l}$ satisfying $0 < p_{k} < p_{l} < 1$, if the throughput inequality constraint is not tight at $\pbf^{*}$, then due to complementary slackness it follows $\mu_{\theta} = 0$, which would imply $p_{k} = p_{l} = 1/n$, a contradiction.
\end{IEEEproof}

%% file: AppB-JainProofs.tex
%%%%%%%%%%%%%%%%%%%%%%%%%%%%%%%%%%%%%%%%%%%%%%
% Appendix B. proofs from Jain section
%%%%%%%%%%%%%%%%%%%%%%%%%%%%%%%%%%%%%%%%%%%%%%
\section{Proofs from \secref{TFtradeoffUsingJain'sFairness}}
\label{app:ProofsFromJainSection}

Proofs from \secref{JainPrelim}, \secref{JainMain}, and \secref{JainProp} are given in \appref{JainPrelimPf}, \appref{JainMainPf}, and \appref{JainPropPf}, respectively.

%%%%%%%%%%%%%%%%%%%%%%%%%%%%%%%%%%%%%%%%%%%%%%
\subsection{Proofs from \secref{JainPrelim}}
\label{app:JainPrelimPf}

\begin{IEEEproof}[Proof of \prpref{mono1}]
We establish the two statements in the order they are given.

{\em Proof of $1)$.} 
Recall the implicit definition of $p_s(k,n',\theta)$ in \eqnref{psfunctheta} in \prpref{tputeqconst} enables us to write $T(p_s(k,n',\theta),k,n') = \theta$.  Note first that $\theta$ is held constant in \prpref{mono1}.  Moreover, in the proof of $1)$ we furthermore hold $n'$ constant, while in the proof of $2)$ we instead hold $n_l = n'-k$ constant.  Because of this, we suppress in the proof of $1)$ the dependence on both $\theta$ and $n'$, and in particular, $p_s(k)$ is defined as the unique solution, when it exists, to the equation $T(p_s(k),k) = \theta$, and $F_{-1}(p_{s}(k, n', \theta), k, n')$ (defined in \eqnref{JainObjectiveShorthand}) is written as $F_{-1}(p_{s}(k), k)$. It is convenient to treat $k$ as a continuous variable in what follows, i.e., to replace $k \in \{1,\ldots,n'-1\}$ with $k \in [1,n'-1]$.  Note here we write $p_{s}(k)$ because the throughput equality constraint (implicitly) determines $p_{s}$ as a function of $k$ under the $(p_{s}, k, n')$ parameterization. It is straightforward to establish $\frac{\partial}{\partial p_s}T(p_s,k) \neq 0$ over the domain of $(p_s,k)$, and as such we can apply the implicit function theorem:
\begin{equation} \label{eq:dpsdk}
\frac{\drm}{\drm k}p_s(k) = - \frac{\frac{\partial }{\partial k}T(p_{s}, k)}
{\frac{\partial }{\partial p_{s}}T(p_{s}, k)}.
\end{equation}
The \textit{total derivative}\footnote{In this case, some authors such as Chiang and Wainwright \cite{ChiWai2005} may call this \textit{partial total derivative} and use a different notation (see discussion toward the end of Section $8.4$). It is ``partial'' because the function ($F_{-1}$) by definition still depends on another exogenous variable ($n'$); it is ``total'' in that it fully captures both the direct and indirect influence of $k$.} of $F_{-1}$ w.r.t.\ $k$ is
\begin{equation}
\label{eq:PartialTotalDerivativeJainsFairnessNprimeHeldConstant}
\frac{\drm }{\drm k}F_{-1}(p_s(k),k) = \frac{\partial }{\partial k}F_{-1}(p_s,k) + \frac{\partial }{\partial p_s}F_{-1}(p_s,k) \frac{\drm }{\drm k}p_s(k).
\end{equation}
Computing and substituting the three derviatives in the above expression yields:  
\begin{equation}
\label{eq:dF1psk}
\frac{\drm }{\drm k} F_{-1}(p_{s}(k), k)
= \frac{(1-p_{s})^{2 (k-1)} (1-p_{l})^{2 (n'-k)} }{2 (k p_{s}+n'-k-1)^2} f_1(p_{s},k),
\end{equation}
where
\begin{eqnarray}
f_1(p_{s},k) 
&\equiv& (n' p_{s}-1) (-2 k (p_{s}-1)+n' (p_{s}-2)+1) + \nonumber \\
& & 2 (p_{s}-1) (k-n') (k (p_{s}-1)+n'-1) \log \frac{1-p_{l}}{1-p_{s}}.
\end{eqnarray}
It is evident from \eqnref{dF1psk} that showing $F_{-1}(p_s(k),k)$ to be increasing in $k$ is equivalent to showing $f_1(p_s,k) > 0$.  After rearrangement, it may be seen that showing $f_1(p_s,k) > 0$ is equivalent to showing
\begin{equation} 
\label{eq:JainsFairnessLHSlessthanRHS}
f_3(p_s,k) <  f_2(p_s, k),
\end{equation}
where
\begin{equation}
\label{eq:f3psklog1pz}
f_3(p_s,k) \equiv \log\left(1 + \frac{p_{l} - p_{s}}{1-p_{l}}\right),
\end{equation}
and 
\begin{equation}
f_2(p_s, k) \equiv \frac{(n' p_{s}-1) (-2 k (p_{s}-1)+n' (p_{s}-2)+1)}{2 (p_{s}-1) (k-n') (k p_{s} + n'  - k - 1)}.
\end{equation}
In $F_{-1},f_1,f_2,f_3$ above the variable $p_s$ is not in fact free, but instead is determined by $T(p_s(k),k) = \theta$.  Below, we show a stronger result that in fact \eqnref{JainsFairnessLHSlessthanRHS} holds for {\em all} $k \geq 0$ and for {\em all} $p_s \in (0,1/n')$.  Our approach to showing \eqnref{JainsFairnessLHSlessthanRHS} is as follows: to show two univariate functions $g_1(x),g_2(x)$ with domain $\Rbb_+$ are ordered as $g_1(x) < g_2(x)$ for all $x$, it suffices to show $i)$ $g_1'(x) \leq g_2'(x)$ and $ii)$ $g_1(0) < g_2(0)$ (which can be easily verified by working with a new function $g_{2}(x) - g_{1}(x)$).  The first step towards \eqnref{JainsFairnessLHSlessthanRHS} is to establish the ordering of the derivatives. 
Recalling $p_{l} = p_{l} (p_{s}, k, n')$ \eqnref{plofpskn}, define 
\begin{equation}
z = z(p_s,k) \equiv \frac{p_l - p_s}{1-p_l} = \frac{1-n' p_s}{n'-1-k + k p_{s}} > 0,
\end{equation}
substitute $z$ into \eqnref{f3psklog1pz}, and observe:
\begin{equation}
\label{eq:df2dkmdldk}
\Delta(p_s,k) = \frac{(n' p_{s}-1)^3}{2 (p_{s}-1) (k-n')^2 (k p_{s} + n'  - k - 1)^2} > 0,
\end{equation}
for $\Delta(p_s,k) \equiv \frac{\partial}{\partial k}f_2(p_{s}, k) - \frac{\partial}{\partial k}f_3(p_s, k)$.  The second step towards \eqnref{JainsFairnessLHSlessthanRHS} is to establish $f_3(p_s,0) < f_2(p_s,0)$.  In fact we show 
\begin{equation}
\label{eq:log1pzts3rdord}
f_3(p_s,0) < f_4(z(p_s,0)) < f_2(p_s,0),
\end{equation}
for $f_4(z) \equiv z - \frac{1}{2} z^{2} + \frac{1}{3} z^{3}$.  The first inequality in \eqnref{log1pzts3rdord} follows from the series expansion of $\log(1+z)$ and valid for all $z > 0$.  The second inequality in \eqnref{log1pzts3rdord} is established by computing
\begin{equation}
f_2(p_{s}, 0) - f_4(z(p_s,0)) = \frac{(n' p_{s}-1)^3 (2 n' p_{s}+n'-3)}{6 (n'-1)^3 n' (p_{s}-1)},
\end{equation}
which is positive for all $n' \geq 3$.  Note $n' \geq 2$ since $\pbf \in \partial \Smc_{2}$, and the $n' = 2$ case can be skipped as $k = 1$ always holds.
This concludes the proof of the first part of the proposition.

{\em Proof of $2)$.}  In the second statement of \prpref{mono1} we again hold $\theta$ constant, but instead of also holding $n'$ constant (as in the first statement), we now hold $n_l$ constant, where $n_l$ is the number of components in $\pbf \in \partial\Smc_2$ taking (the larger) value $p_l$.  It is clear that we can just as easily parameterize $\pbf \in \partial\Smc_2$ using the three free parameters $[p_s,k,n_l]$ as with $(p_s,k,n')$ (the change in parameterization emphasized by the change from parentheses to square braces) using the mapping $k+n_l = n'$ (with $p_s$ and $k$ still defined as before).  The new parameters must take values such that $p_s \in (0,1/(k+n_l))$, and $(k,n_l) \in \Dmc_n$, where 
\begin{equation}
\Dmc_n \equiv \{ (k,n_l) \in \Nbb^2 : k \geq 1, n_l \geq 1, k+n_l \leq n\}.
\end{equation}

We now define the functions $T[p_s,k,n_l] = T(\xbf(\pbf[p_s,k,n_l]))$ and $F_{-1}[p_s,k,n_l] = F_{-1}(\xbf(\pbf[p_s,k,n_l]))$ under this new parameterization.  The throughput constraint $T[p_s,k,n_l] = \theta$ again implicitly defines a function $p_s[k,n_l,\theta]$ satisfying $T[p_s[k,n_l,\theta],k,n_l] = \theta$.  Analogous to part $1)$ of the proof, we suppress the dependence upon $n_l$ and $\theta$, and again because the throughput equality constraint determines $p_{s}$ as a function of $k$, we write $p_s[k,n_l,\theta]$ as $p_s[k]$, the throughput constraint function as $T[p_s[k],k] = \theta$, and the objective $F_{-1}[p_{s}[k, n_{l}, \theta], k, n_{l}]$ as $F_{-1}[p_{s}[k], k]$.  

It is straightforward to establish $\frac{\partial}{\partial p_s}T[p_s,k]  \neq 0$ over the domain of $(p_s,k)$, and as such we can apply the implicit function theorem (which again treats $k$ as a continuous variable):
\begin{equation} 
\label{eq:dpsdk2}
\frac{\drm }{\drm k}p_s[k] = - \frac{\frac{\partial }{\partial k}T[p_{s}, k]}
{\frac{\partial }{\partial p_{s}}T[p_{s}, k]}.
\end{equation}
The \textit{total derivative} of $F_{-1}$ w.r.t.\ $k$ is
\begin{equation}
\label{eq:totder2}
\frac{\drm }{\drm k}F_{-1}[p_s[k],k] = \frac{\partial }{\partial k}F_{-1}[p_s,k] + \frac{\partial }{\partial p_s}F_{-1}[p_s,k] \frac{\drm }{\drm k}p_s[k].
\end{equation}
Computing and substituting the above derivatives yields
\begin{equation}
\frac{\drm }{\drm k}F_{-1}[p_s[k],k] = \frac{1}{2 n_l}(1-p_{s})^{2 (k-1)} \left(\frac{k p_{s}+n_l-1}{n_l}\right)^{2 n_l-1} f_5[p_s,k],
\end{equation}
where
\begin{equation}
f_5[p_s,k] \equiv - k p_{s}^{3} + (n_{l} + 1) p_{s}^{2} - 2 n_{l} p_{s}   - 2 n_l (1 - p_{s}) \log (1-p_{s}),
\end{equation}
and the sign of the derivative is easily seen to equal the sign of the above function.  Thus part $2)$ of the proposition is established by showing $f_5[p_s,k] > 0$ for $k \in [n-n_l]$ and $p_s \in (0,1/(k+n_l))$.  Using the upper bound $\log(1-p_{s}) \leq (-p_{s}) - \frac{1}{2} \left(-p_{s}\right)^{2}$ we obtain
\begin{equation}
f_5[p_s,k] \geq p_s^2 (1 - p_s (k+n_l)) > 0.
\end{equation}
This concludes the proof of the second part of the proposition.
\end{IEEEproof}

%%%%%%%%%%%%%%%%%%%%%%%%%%%%%%%%%%%%%%%%%%%%%%
\subsection{Proofs from \secref{JainMain}}
\label{app:JainMainPf}

\begin{IEEEproof}[Proof of \thmref{mainresultChiuJain}]
There are three regimes for $\theta$ given in \thmref{mainresultChiuJain}.  The proof consists of two parts: part $i)$ addresses regime $1$, while part $ii)$ addresses regimes $2)$ and $3)$.  

{\em Part $i)$} (Regime $1)$).
The claim here is that the maximum fairness of $1$ is achievable, attained when all the $x_{i}$'s are equal to $\theta / n$. It is not hard to see all the $x_{i}$'s are equal iff all the associated controls $p_{i}$'s (i.e., satisfying \eqnref{xofp}) are equal, in which case $\theta / n = x_{i} = p (1-p)^{n-1}$ for each $i \in [n]$, for some $p \in [0,1]$ to be determined. The existence of such a $p$ follows from \lemref{AllRatesEqualRay} and thus the claim is proved.

{\em Part $ii)$} (Regimes $2)$ and $3)$).  This part of the proof is divided into three steps.  Recall $\pbf^{*}$ denotes the optimal control.

\textbf{Step 1:} $\pbf^* \in \partial\Smc$.  That $\pbf^*$ must be a probability vector follows from \corref{MajorizationCorollary} in \secref{MajorizationAttempt}.

\textbf{Step 2:} $\pbf^* \in \partial\Smc_{1,2}$. By \prpref{AtMostTwoDistinctNonzeroComponentValues} in \secref{opttputconst}, $|\Vmc(\pbf^*)| \leq 2$, as the minimization problem \eqnref{TFtradeoffProblemFormulation} is a special case of the extremization problem \eqnref{extremization} in \prpref{AtMostTwoDistinctNonzeroComponentValues} with $\alpha = -1$.  Then together with $\pbf^* \in \partial\Smc$, it gives $\pbf^* \in \partial\Smc_{1,2}$.

\textbf{Step 3:} Following  Remark \ref{rem:JainMergingRegimes2And3}, regimes $2)$ and $3)$ are grouped together meaning the target throughput $\theta \in [\theta_{t}, \theta_{t-1})$. 
By item $3)$ in \prpref{tputeqconst}, the set of feasible $(k,n')$ pairs for which there exists a $\pbf \in \partial \Smc_{1,2}$ satisfying $T(\xbf(\pbf)) = \theta$ is the set $\Dmc_{t,n}$ in \eqnref{DtnpRegion}, illustrated in \figref{DtnpRegion}.

Case $1:$ \textit{assuming} $\pbf^{*} \in \partial \Smc_{2}$, we can then apply the two monotonicity properties stated in \prpref{mono1} to the set $\Dmc_{t,n}$, which shows the optimal $(k^{*}, n'^{*}) = (1, t)$. Applying $(k^{}, n'^{}) = (1, t)$ to the throughput constraint equation \eqnref{psfunctheta} yields \eqnref{ps-star-inMainProp}. Furthermore, as $p_{s}^{*} \in (0, 1/n'^{*})$ this in turn shows (due to the monotonicity established in item $1)$ of \prpref{tputeqconst}) the achievable throughput range by varying $p_{s}$ is the open interval $(\theta_{t}, \theta_{t-1})$. 

Case $2:$ \textit{assuming} $\pbf^{*} \in \partial \Smc_{1}$, we let such a $\pbf^{*}$ be parameterized by $n'$ (\defref{restrictedControls}). The corresponding extremizer in the rate space is $\xbf^{*} = \xbf^{*}(n') \equiv \frac{\theta_{n'}}{n'} \sum_{i=1}^{n'} \ebf_{i}$. Satisfying the feasibility constraint for $\theta \in [\theta_{t}, \theta_{t-1})$ requires $n'^{} \leq t$, and in fact $n'^{}$ can only equal $t$ due to its integer support. This shows the optimal $n'^{*} = t$ (thus $\pbf^{*} = (1 / t) \sum_{i=1}^{t} \ebf_{i}$ and $F_{J}^{*} = t/n$). Furthermore, this in turn shows if $\theta = \theta_{t}$ then the corresponding $\pbf^{*} \in \partial \Smc_{1}$.

Clearly the target throughput range $[\theta_{t}, \theta_{t-1})$ is partitioned as $(\theta_{t}, \theta_{t-1}) \cup \{\theta_{t}\}$ where the extremizers for the former (regime $3)$) and latter (regime $2)$) are found in cases $1$ and $2$ respectively.

Finally, $\tilde{T}(F)$ in \eqnref{tfjinterpolation} is obtained by observing the above results for regime $2$ as $n$ points $\{(T_t,F_t)\}_{t \in [n]}$ on the throughput--fairness tradeoff plot, with $T_t = \theta_t$ and $F_t = t/n$.  Thus, to interpolate the $n$ points via a function $\tilde{T}(F)$ it suffices to use $\tilde{T}(F) = T_{n F}$ and treat $F$ as a continuous variable. 
\end{IEEEproof}

\begin{IEEEproof}[Proof of \thmref{2}]
Part $i)$ (Regime $1)$). In the proof of Thm.\ \ref{thm:mainresultChiuJain} it is shown that for this regime, the maximum fairness $1$ can be attained with the throughput constraint satisfied with equality. This continues to hold here.

Part $ii)$ (Regimes $2)$ and $3)$). The second and third regimes namely the case when $\theta \geq \theta_{n}$.

\textbf{Step 1:} $\pbf^* \in \partial\Smc$. This is because the global minimizer must lie on a hyperplane $\Hmc_{\theta^{*}} = \{ \xbf : \sum_i x_i = \theta^{*}\}$ for \textit{some} $\theta^{*} \geq \theta$. Then the same step in the proof of \thmref{mainresultChiuJain} applies.

\textbf{Step 2:} $\pbf^* \in \partial\Smc_{1,2}$. The same step in the proof of \thmref{mainresultChiuJain} applies, as the extremization problem \eqnref{extremization} in \prpref{AtMostTwoDistinctNonzeroComponentValues} includes the case of throughput inequality constraint.

\textbf{Step 3} is divided into two sub-steps, one for each regime.  Recall $\partial\Smc_{1,2}$ is the disjoint union of $\partial\Smc_{1}$ and $\partial\Smc_{2}$, and $n'$ denotes the number of nonzero component(s) of $\pbf^{*}$.

\underline{Regime $2)$}: when $\theta = \theta_{t}$ for some $t \in [n]$. 

Case $1$: \textit{assuming} $\pbf^{*} \in \partial \Smc_{2}$, since Item $ii)$ of \prpref{AtMostTwoDistinctNonzeroComponentValues} says under the assumption $\vert \Vmc(\pbf^{*}) \vert = 2$, an extremizer has to satisfy the throughput constraint with equality, this justifies we can apply \thmref{mainresultChiuJain} (regime $2)$). Doing so gives the extremizer as $\pbf^{*} = (1/t) \sum_{i=1}^{t} \ebf_{i}$. But this contradicts our assumption that $\pbf^{*} \in \partial \Smc_{2}$.

Case $2$: \textit{assuming} $\pbf^{*} \in \partial \Smc_{1}$, it follows that $\xbf (\pbf^{*}) = \frac{\theta_{n'}}{n'} \sum_{i=1}^{n'} \ebf_{i}$. On one hand, $T(\xbf(\pbf^{*})) \geq \theta$ for such an $\xbf$ requires $n' \leq t$; on the other hand, the objective $F_{-1}(\xbf(\pbf^{*}))$, to be minimized, is decreasing in $n'$. Together they imply the optimal $n'^{*} = t$, with the corresponding fairness $F_{J}^{*} = t/n$. 

Therefore, the extremizer for $\theta = \theta_{t}$ actually comes from $\partial \Smc_{1}$ and is given by $\pbf^{*} = (1/t) \sum_{i=1}^{t} \ebf_{i}$ with $F_{J}^{*} = t/n$.

\underline{Regime $3)$}: when $\theta \in (\theta_{t}, \theta_{t-1})$ for some $t \in \{2, \ldots, n \}$. 

Case $1$: \textit{assuming}  $\pbf^{*} \in \partial \Smc_{1}$: similar to what we have done above, satisfying the feasibility constraint $T(\xbf) \geq \theta$ requires $n' \leq t-1$, while the objective function $F_{-1}(\xbf)$ is decreasing in $n'$ which means $n'$ is desired to be as large as possible. Together they imply the optimal $n'^{*} = t-1$, with the corresponding fairness $F_{J}^{*} = (t-1)/n$.

Case $2$: \textit{assuming} $\pbf^{*} \in \partial \Smc_{2}$: again item $ii)$ of \prpref{AtMostTwoDistinctNonzeroComponentValues} justifies \thmref{mainresultChiuJain} (regime $3)$) is applicable. Furthermore, in this case, the optimal solution $\pbf^{*}$ from $\partial \Smc_{2}$ is such that $F_{J}^{*} \in ((t-1)/n, t/n)$ due to the monotonicity and continuity of the T-F tradeoff curve (\thmref{ChiuJainTFtradeoffProperties}, items $2, 4$) and the just proved result for regime $2)$. 

As the optimal solution from $\partial \Smc_{2}$ outperforms that from $\partial \Smc_{1}$, this shows the desired extremizer is indeed from $\partial \Smc_{2}$ and is as stated for regime $3)$ in \thmref{mainresultChiuJain}.

In summary, the solution to the Jain throughput--fairness tradeoff \eqnref{TFtradeoffProblemFormulation} remains unchanged.
\end{IEEEproof}

%%%%%%%%%%%%%%%%%%%%%%%%%%%%%%%%%%%%%%%%%%%%%%
\subsection{Proofs from \secref{JainProp}}
\label{app:JainPropPf}

The following lemma is essential to the proof of item $5)$ of \thmref{ChiuJainTFtradeoffProperties}. 

\begin{lemma} 
\label{lem:KeyLemmaChiuJainConvexDecreasing}
Given an integer $n \geq 3$, the following two polynomials in $n$ are both positive for $p_{s} \in (0, 1/n)$.
\begin{eqnarray} \label{eqn:fdeAndfnuBothOfn}
f_{\rm de}(n; p_{s}) & = & n^2 p_s^2+n \left( p_s^{4} - 2 p_s^{3} + 2 p_s^{2} - 4 p_s + 1 \right) - 2 p_s^{2} + 4 p_s -1 \nonumber \\
f_{\rm nu}(n; p_{s}) & = & n^4 \left(p_s^5-p_s^4-2 p_s^3+5 p_s^2-4 p_s\right)+ \nonumber \\
& & n^3 \left(p_s^7+p_s^6-12 p_s^5+19 p_s^4-10 p_s^3-6 p_s^2+6 p_s+5\right)+ \nonumber \\
& & n^2 \left(2 p_s^7-16 p_s^6+41 p_s^5-55 p_s^4+45 p_s^3-20 p_s^2+17 p_s-20\right)+ \nonumber \\
& & n \left(7 p_s^4-18 p_s^3+24 p_s^2-35 p_s+26\right) \nonumber \\
& & -6 p_s^2+16 p_s-11.
\end{eqnarray}
\end{lemma}

\begin{figure}[!ht]
\centering
\includegraphics[width= 0.4\textwidth ]{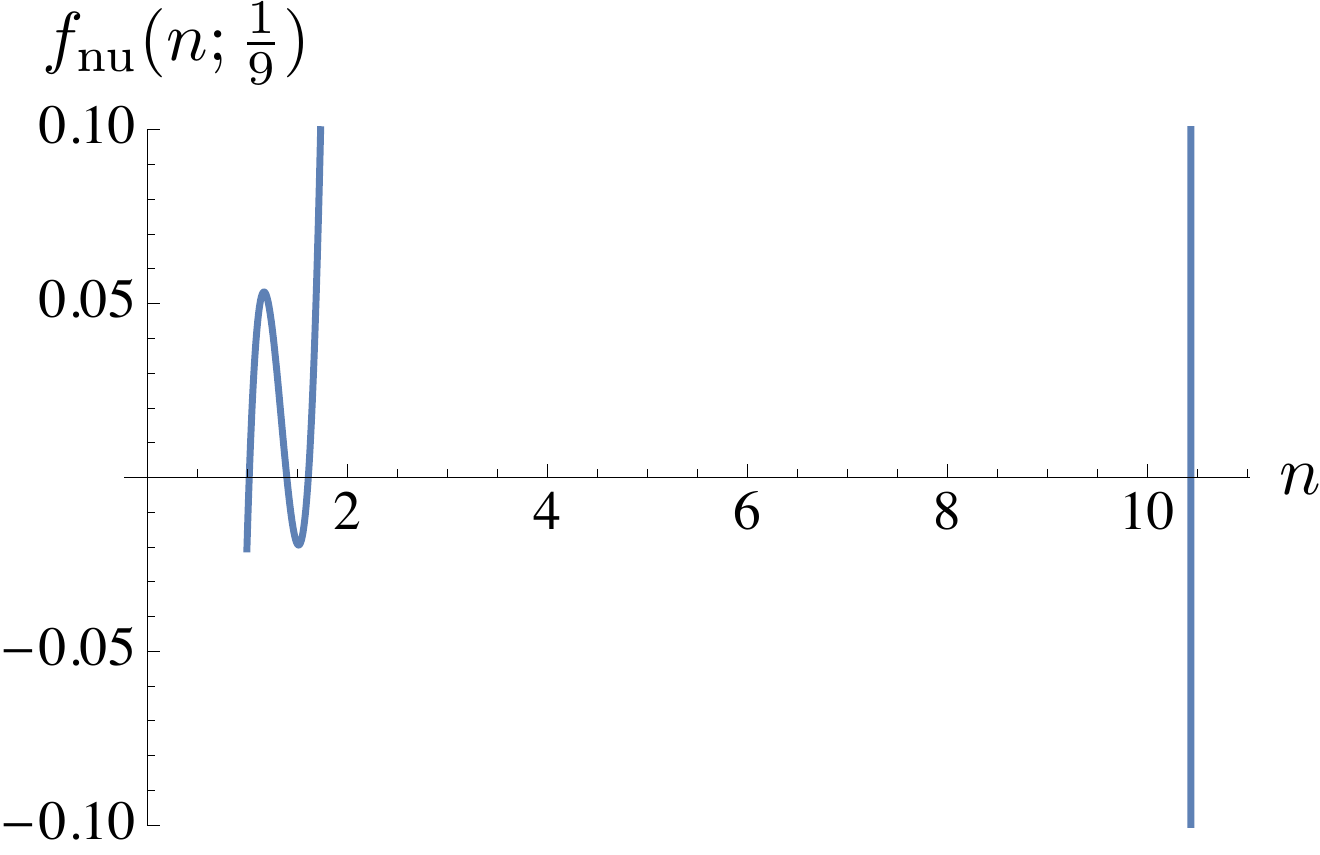}
\caption{$f_{\rm nu}(n; p_{s})$ when $p_{s} = 1/9$; the top part is not shown, in order to better view all the roots.  Three of them are between $1$ and $2$ and the remaining one is in $(1/p_{s}, \infty)$. As $f_{\rm nu}(n; p_{s})$ is a $4^{\rm th}$ order polynomial in $n$, it has a total of four roots and hence no root exists in the interval $(2, 1/p_{s})$.}
\label{fig:fnuOfn-psEqualsOneNinth}
\end{figure}

\begin{IEEEproof}[Proof of \lemref{KeyLemmaChiuJainConvexDecreasing}]
In both parts of the proof we treat $n$ as a continuous variable and view $p_{s}$ as fixed. 

{\em Part $i)$} ($f_{\rm de}(n; p_{s}) > 0$).
We prove this by showing $f_{\rm de}(3; p_{s}) > 0$ and $\frac{\drm}{\drm n} f_{\rm de}(n; p_{s})  > 0$ for all $n \geq 3$. First, $f_{\rm de}(3; p_{s}) = 3 p_{s}^4-6 p_{s}^3+13 p_{s}^2-8 p_{s}+2$. Since this quartic (in $p_{s}$) has all its four roots being complex, this means this polynomial (in $p_{s}$) is either always positive or always negative for all $p_{s} \in \Rbb$.  We can test this by setting $p_{s} = 0$ and this shows its positiveness. Second, $\frac{\drm}{\drm n} f_{\rm de}(n; p_{s})  = 2 n p_{s}^2+p_{s}^4-2 p_{s}^3+2 p_{s}^2-4 p_{s}+1$, which is lower bounded by $p_{s}^4-2 p_{s}^3+8 p_{s}^2-4 p_{s}+1$ since $n \geq 3$.  Again this quartic (in $p_{s}$) can be shown to have all its four roots being complex and we can use any specific value of $p_{s} \in \Rbb$ to verify its positiveness.

{\em Part $ii)$} ($f_{\rm nu}(n; p_{s}) > 0$). The condition $p_{s} \in (0, 1/n)$ for $n \geq 3$ then translates to $n \in [3, 1/p_{s})$. We will focus on showing $f_{\rm nu}(n; p_{s})$ as a polynomial in $n$ does not have any real root on $n \in [3, 1/p_{s})$, which suggests $f_{\rm nu}(n; p_{s})$ is either always positive or always negative on this interval and we then only need to test this out using any specific point in the interval. A plot of $f_{\rm nu}(n; p_{s})$ versus $n$ for fixed $p_{s}$ is shown in \figref{fnuOfn-psEqualsOneNinth}. In the following we will show a slightly stronger result, namely to extend the domain of interest to $(2, 1/p_{s})$. For notational simplicity we let $p_{s} = 1/m$ for $m > n$ and express the coefficients of the polynomial $f_{\rm nu}(n; p_{s})$ using $m$, and we will also use the shorter notation $f_{\rm nu}(n)$.  The $j^{\rm th}$ derivative (w.r.t.\ $n$) of $f_{\rm nu}$ is denoted $f^{(j)}_{\rm nu}(n) \equiv \frac{\drm^{j}}{\drm n^{j}} f_{\rm nu}(n)$.

We use the Budan-Fourier theorem, which (partially) characterizes the number of real roots of a polynomial in any given interval. Specifically, let $v(a)$ and $v(b)$ denote the number of sign changes (i.e., sign variation) of the Fourier sequence $\{ f_{\rm nu}(n), f_{\rm nu}^{(1)}(n), \ldots, f_{\rm nu}^{(4)}(n)  \}$ when $n  = a$ and $b$ respectively, for $a < b$.  This theorem says the number of real roots in $(a, b)$,  each root counted with proper multiplicity, equals $v(a) - v(b)$ minus an even nonnegative integer.

We can verify $v(2) = 1$ since the signs of the Fourier sequence are $+ ~ + ~ + ~ \mp ~ -$ (note the sign of $f_{\rm nu}^{(3)}(2)$ is undetermined, if we only know $m > 3$).  We can further verify $v(m) = 1$ since the signs of the Fourier sequence are $+ ~ - ~ - ~ - ~ -$). Since $v(2) - v(m)$ already equals $0$, applying Budan-Fourier theorem, we see the polynomial $f_{\rm nu}(n)$ has no real root on $(2, m)$.

The Fourier sequence at $a=2$ and $b=m$ are given below in a form that facilitates checking their sign.  Namely:
\begin{eqnarray} 
\label{eq:Budan-FourierSequence}
m^7 f_{\rm nu}(2) &=& (m-2) \left( m^4 (m^2-6) + m^2 (20 m-30) + (24 m-8)\right) \nonumber \\
m^7 f_{\rm nu}^{(1)}(2) &=& \left( m^{3}(6 m^4-23 m^3+ 32 m^2-22 m-17) + m (52 m-52) + 20\right) \nonumber \\
m^7 f_{\rm nu}^{(2)}(2) &=& 2 \left( m^{4} (10 m^3-43 m^2 + 64 m-63 ) + m (35 m^2-7 m-10 ) + 8  \right) \nonumber \\
m^7 f_{\rm nu}^{(3)}(2) &=& 6 \left( 5 m^7-26 m^6+34 m^5-26 m^4+11 m^3-4 m^2+m+1 \right) \nonumber \\
m^5 f_{\rm nu}^{(4)}(2) &=& -24 \left( m^3 (4 m-5) + (2 m^2+m-1) \right) 
\end{eqnarray}
and 
\begin{eqnarray}
m^5 f_{\rm nu}(m) &=& (m - 2) (m - 1)^7 \nonumber \\
m^6 f_{\rm nu}^{(1)}(m) &=& -(m - 1)^5 \left( m^3 + m (7 m-9) + 4 \right) \nonumber \\
m^7 f_{\rm nu}^{(2)}(m) &=& -2 (m - 1)^3 \left( m^3 (9 m^2-m-17) + m (17 m-7) +2 \right) \nonumber \\
m^7 f_{\rm nu}^{(3)}(m) &=& -6 \left( m^6 (11 m-26) + 14 m^5 + m^3 (14 m-23) + (12 m^2-m-1) \right) \nonumber \\
m^5 f_{\rm nu}^{(4)}(m) &=& -24 \left( m^3 (4 m-5) + (2 m^2+m-1) \right).
\end{eqnarray}
Since $m = 1/p_{s} > n \geq 3$, it is not hard to verify the sign of the terms grouped by inner parentheses to be positive (and hence determine $v(2)$ and $v(m)$), except for $f_{\rm nu}^{(3)}(2)$, but, as mentioned, this sign does not affect the value of $v(2)$.

It remains to use any specific point on $(2, m)$ to determine the sign of $f_{\rm nu}(n)$ over the entire $(2, m)$, e.g.,
\begin{equation}
f_{\rm nu}(3) = \frac{1}{m^{7}} \left( m^{3} (22 m^4-98 m^3+129 m^2-81 m-42)  + (126 m^2-117 m+45) \right).
\end{equation}
It can be verified that the quartic and quadratic enclosed by the two pairs of inner parentheses in the above expression are both positive. This shows $f_{\rm nu}$ is positive at $n = 3$, and as argued above, this proves $f_{\rm nu}$ is positive over $n \in (2, 1/p_{s})$.

\end{IEEEproof}

\begin{IEEEproof}[Proof of \thmref{ChiuJainTFtradeoffProperties}]
We write $F_{J}^{*}(\theta; n)$ to denote the optimized Jain's fairness under a throughput constraint $T(\xbf) = \theta$, where $n$ serves as a parameter but not a free variable in the optimization. 

The feasible set $\Lambda$ is parameterized by $\pbf$ via \eqnref{xofp}, and when $\theta \in [\theta_{t}, \theta_{t-1})$ for $t \in \{2, \ldots, n\}$, we know from \thmref{mainresultChiuJain} the unique extremizer is characterized by the tuple $(p_{s}^{*}, k^{*}, n'^{*})$, with $k^* = 1$ and $n'^*=t$, defined in \defref{restrictedControls}. It is clear from \thmref{mainresultChiuJain} that this tuple is a function of $\theta$, and may be written as $(p_{s}^{*}(\theta), 1,t)$. Therefore the notation $F_{J}^{*} (\theta; n) $ should be understood as
\begin{equation} 
\label{eq:FJStarThetaN}
F_{J}^{*} (\theta; n) 
\equiv F_{J}(\xbf(\pbf(p_{s}^{*}(\theta), 1,t))),
\end{equation}
with $F_{J}$ defined in \eqnref{Chiu-JainFairnessDefinition}.  Observe also the identity 
\begin{equation}
\label{eq:TpsknprimeStar}
\theta \equiv T(p_{s}^{*}(\theta), 1,t),
\end{equation}
for $T(p_{s}, k, n')$ defined in \eqnref{signedtputgap} and $p_{s}^{*}$ the solution of  \eqnref{ps-star-inMainProp}. This is used to compute the dependence of $p_{s}^{*}$ on $\theta$. 

Item $1)$. That $p_{s}^{*}(\theta)$ is piecewise decreasing in $\theta$ follows from \eqnref{TpsknprimeStar} and \prpref{tputeqconst} (item $1)$):
\begin{equation}
\label{eq:psStarDecInTheta}
\frac{\drm p_{s}^{*}(\theta)}{\drm \theta} 
= \left( \frac{\drm \theta(p_{s}^{*})}{\drm p_{s}^{*}}\right)^{-1} 
=\left( \frac{\drm }{\drm p_{s}^{*}} T (p_{s}^{*}, 1, t) \right)^{-1}
< 0,
\end{equation}
That $p_{l}^{*}(\theta)$ is piecewise increasing in $\theta$ follows from $p_{l}$ in \defref{restrictedControls} and \eqnref{psStarDecInTheta}.

In fact, a stronger statement is that $p_{l}^{*}(\theta)$ is increasing in $\theta$, albeit not everywhere differentiable. To see this, let us look at two adjacent active throughput intervals on the T-F plot: $[\theta_{t}, \theta_{t-1})$, $[\theta_{t-1}, \theta_{t-2})$. When $\theta$ sweeps over the first interval (for which $n'^{*} = t$), $p_{s}^{*}$ decreases from $1/t$ (when $\theta = \theta_{t}$) to $0$ (when $\theta = \theta_{t-1}$) and correspondingly $p_{l}^{*}$ increases from $1/t$ (when $\theta = \theta_{t}$) to $1/(t-1)$ (when $\theta = \theta_{t-1}$). Moving onto the second interval (for which $n'^{*} = t-1$), similarly, $p_{s}^{*}$ ($p_{l}^{*}$) decreases (increases) from $1/(t-1)$ ($1/(t-1)$) to $0$ ($1/(t-2)$). Clearly $p_s^*$ is not  monotonic over the entire $\theta \in (\theta_{n}, 1)$ whereas $p_{l}^{*}$ is monotonic.

Next we show $p_{l}^{*}(\theta)$ is not differentiable at the boundary of active throughput intervals. More precisely, at the boundary of the two intervals $[\theta_{t}, \theta_{t-1})$ and $[\theta_{t-1}, \theta_{t-2})$, i.e., $\theta = \theta_{t-1}$, we compute the left- and right- derivative respectively and show they are not equal. That is, nondifferentiability at $\theta_{t-1}$ is established by showing
\begin{equation}
\label{eq:dpldTnonsmooth}
\left. \frac{\drm }{\drm \theta} p_{l}(p_{s}^{*}(\theta), 1,t) \right \vert_{p_{s}^{*}(\theta) = 0}     \neq     \left. \frac{\drm }{\drm \theta} p_{l}(p_{s}^{*}(\theta), 1,t-1) \right \vert_{p_{s}^{*}(\theta) = \frac{1}{t-1}},
\end{equation}
where
\begin{equation}
\frac{\drm p_{l}^{*} (\theta)}{\drm \theta}
 = \frac{\drm p_{l}^{*}(\theta)}{\drm p_{s}^{*}} \frac{\drm p_{s}^{*}}{\drm \theta} 
 = \frac{ \frac{\drm p_{l}^{*}(\theta)}{\drm p_{s}^{*}}  }{  \frac{\drm \theta(p_{s}^{*})}{\drm p_{s}^{*}}  }
 =   \frac{ - \frac{k^{*}}{n'^{*} - k^{*}}  }{ \frac{\drm}{\drm p_{s}^{*}}  T (p_{s}^{*}, 1, t)  },
\end{equation}
with $T (p_{s}^{*}, 1, t)$ again coming from \eqnref{TpsknprimeStar}. Since the LHS of \eqref{eq:dpldTnonsmooth} equals $\left(\frac{t-2}{t-1}\right)^{2-t}$ whereas the RHS equals infinity, this establishes \eqref{eq:dpldTnonsmooth}.

Next, we look at the dependence of $x_{s}^{*}(\theta)$, $x_{l}^{*}(\theta)$ upon $\theta$. Let $\theta \in [\theta_{t}, \theta_{t-1})$. As $(k^{*}, n'^{*}) = (1, t)$, we have
\begin{equation}
\label{eq:xsAndxlStarVersusTheta}
x_{s}^{*} = p_{s}^{*} (1 - p_{l}^{*})^{t-1}, ~ x_{l}^{*} = p_{l}^{*} (1 - p_{s}^{*}) (1 - p_{l}^{*})^{t-2}.
\end{equation}
That $\frac{\drm x_{s}^{*}(\theta)}{\drm \theta} < 0$ follows easily from $\frac{\drm p_{s}^{*}(\theta)}{\drm \theta} < 0$ and $\frac{\drm p_{l}^{*}(\theta)}{\drm \theta} > 0$. To show $\frac{\drm x_{l}^{*}(\theta)}{\drm \theta} > 0$, it can be seen from \eqnref{xsAndxlStarVersusTheta} that it suffices to show $p_{l}^{*} (1 - p_{l}^{*})^{t-2}$ is increasing in $p_{l}^{*}$: we can verify the function $p(1 - p)^{t-2}$ is increasing in $p$ when $p \in (0, 1/(t-1))$, which includes the range of $p_{l}^{*}$ when $\theta \in [\theta_{t}, \theta_{t-1})$ namely $[1/t, 1/(t-1))$. This proves $\frac{\drm x_{l}^{*}(\theta)}{\drm \theta} > 0$.

Finally, we want to show at the boundary of active throughput intervals, $x_{l}^{*}(\theta)$ is not differentiable. First, let $\pbf^{*}$ be parameterized by $(p_{s}^{*}, k^{*}, n'^{*})$. Applying the chain rule, we have
\begin{equation}
\label{eq:dxldTnonsmoothFirstStep}
\frac{\drm x_{l}^{*}}{\drm \theta} 
= \frac{\drm x_{l}^{*}}{\drm p_{s}^{*}} \frac{\drm p_{s}^{*}}{\drm \theta} 
= \frac{ \frac{\drm }{\drm p_{s}^{*}} p_{l}^{*} (1 - p_{s}^{*})^{k^{*}} (1 - p_{l}^{*})^{n'^{*} - k^{*} - 1} }  { \frac{\drm}{\drm p_{s}^{*}} T (p_{s}^{*}, k^{*}, n'^{*})  }
= \frac{1 - p_{s}^{*}}{1 - n'^{*} p_{s}^{*}}.
\end{equation}
Second, we need to show the derivative $\frac{\drm x_{l}^{*}}{\drm \theta} $ in \eqnref{dxldTnonsmoothFirstStep} when $\theta$ is in $[\theta_{t}, \theta_{t-1})$ and approaches $\theta_{t-1}$ from below does not equal to this derivative when $\theta$ is in $[\theta_{t-1}, \theta_{t-2})$ and approaches $\theta_{t-1}$ from above. Therefore, similar to \eqnref{dpldTnonsmooth}, we need to verify
\begin{equation}
\label{eq:dxldTnonsmoothSecondStep}
\left. \frac{\drm x_{l}^{*}}{\drm \theta}  \right \vert_{(0, 1, t)} \neq \left. \frac{\drm x_{l}^{*}}{\drm \theta}  \right \vert_{(\frac{1}{t-1}, 1, t-1)}.
\end{equation}
Applying the computed result in \eqnref{dxldTnonsmoothFirstStep}, we see the LHS of \eqnref{dxldTnonsmoothSecondStep} equals $1$ while its RHS equals infinity: this shows the nondifferentiability of $x_{l}^{*}(\theta)$ at the critical throughputs.

Item $2)$. We claim that it suffices to show the monotone decreasing property when $\theta \in [\theta_{n}, \theta_{n-1})$ for each $n \geq 2$. To see this, we prove by mathematical induction. For the base case, namely when $n = 2$, there is only one active throughput interval $[\theta_{2}, \theta_{1})$ and the monotonicity follows from the assumption. Now assuming the monotonicity holds for $n = n_{0} \geq 2$ i.e., $\frac{\drm}{\drm \theta}  F_{J}^{*}(\theta; n_{0}) < 0$ over $\theta \in [\theta_{n_{0}}, 1)$, we need to show it continues to hold when $n = n_{0} + 1$ i.e., $\frac{\drm}{\drm \theta} F_{J}^{*}(\theta; n_{0}+1) < 0$ over $\theta \in [\theta_{n_{0}+1}, 1)$. There are two cases: when $\theta \in [\theta_{n_{0}+1}, \theta_{n_{0}})$ the monotonicity follows from the assumption; when $\theta \in [\theta_{n_{0}}, 1)$, specializing \eqnref{RecusionJain} with $l = 1$, $n = n_{0} + 1$ gives $F_{J}^{*}(\theta; n_{0}+1) = \frac{n_{0}}{n_{0} + 1} F_{J}^{*}(\theta; n_{0})$: the monotonicity then follows from the induction hypothesis. This proves the claim. 

Now, let the number of users be $n$ and $\theta \in [\theta_{n}, \theta_{n-1})$. \thmref{mainresultChiuJain} says $k^{*} = 1$, $n'^{*} = t = n$ and we can compute
\begin{equation}
\label{eq:dFStardThetaOriginalExpression}
\frac{\drm}{\drm \theta}  F_{J}^{*}(\theta; n)
= \frac{\drm}{\drm p_{s}^{*}(\theta)} F_{J}(p_{s}^{*}(\theta); n)     \left( \frac{\drm \theta(p_{s}^{*})}{\drm p_{s}^{*}}\right)^{-1} 
= \frac{\drm}{\drm p_{s}^{*}(\theta)} F_{J}(p_{s}^{*}(\theta); n)    \left( \frac{\drm }{\drm p_{s}^{*}} T (p_{s}^{*}, 1, n) \right)^{-1},
\end{equation}
where the second equality comes from \eqnref{TpsknprimeStar}. Substituting the definition of $F_{J}$ and $T$ in \eqnref{Chiu-JainFairnessDefinition} and \eqnref{signedtputgap}, we get
\begin{equation} 
\label{eq:dFdTExpression}
\frac{\drm}{\drm \theta} F_{J}^{*}(\theta; n)  = - \frac{2 (1 - p_{s}^{*}) (n+p_{s}^{*}-2)^3 \left(\frac{n+p_{s}^{*}-2}{n-1}\right)^{-n} (n ( - p_{s}^{*} (1- p_{s}^{*}) +1)-1)}{n \left[n^2 p_{s}^{*2}+n \left((p_{s}^{*}-2) \left(p_{s}^{*2}+2\right) p_{s}^{*}+1\right)-2 (p_{s}^{*}-2) p_{s}^{*}-1\right]^2},
\end{equation}
which can be verified to be negative for all $n \geq 2$ and $p_{s}^{*} \in (0, 1)$. 
Finally the monotone decreasing property over $[\theta_{n}, 1)$ (namely not just piecewise) follows from continuity of the T-F curve, shown in item $4)$.

Item $3)$. Again we decompose the interval $[\theta_{n}, 1)$ into $[\theta_{n}, \theta_{n-1}) \cup [\theta_{n-1}, 1)$. When $\theta \in [\theta_{n}, \theta_{n-1})$ this property automatically holds because for all $n_{s} < n$ we have $F_{J}^{*}(\theta; n_{s}) \equiv 1$ since $\theta < \theta_{n-1} \leq \theta_{n_{s}}$.  When $\theta \in [\theta_{n-1}, 1)$, specializing \eqnref{RecusionJain} with $l = 1$ gives $F_{J}^{*}(\theta; n) = \frac{n-1}{n} F_{J}^{*}(\theta; n-1) < F_{J}^{*}(\theta; n-1)$, which proves the desired monotone decreasing in $n$ property. Graphically, this corresponds to the observation that as $n$ increases, the T-F tradeoff curve will tend closer to the $\theta$-axis. Furthermore, since the sequence $\{\theta_{n}\}$ is decreasing in $n$, the range of $\theta$ for which the maximum achievable fairness is less than $1$ (namely $(\theta_{n}, 1)$) always extends toward the lower bound $1/\erm$, and thus the full curve for any given $n$ will tend closer to the $F_{J}^{*}$-axis, too.

Item $4)$. We first prove continuity in three steps. $a)$ The extremizers in regime $2$ can be viewed as limiting cases of those in regime $3$.  $b)$ Within regime $3$, since the root (on the complex plane) of a polynomial equation is continuous in its coefficients \cite[\S 3.9]{Tyr1997}, and since the polynomial equation \eqref{eq:ps-star-inMainProp} only has a single real root ($p_{s}^{*}$) it must also be continuous. $c)$ The function $F_{J}^{*}$ in \eqnref{FJStarThetaN} is continuous in $p_{s}^{*}$. We next prove nondifferentiability occurs when $\theta = \theta_{n_{s}}$ for all $n_{s}$ smaller than $n$. We claim it suffices to only verify this when $n_{s} = n - 1$ but for \textit{all} $n \geq 3$. To see this, specializing \eqnref{RecusionJain} with $l = 1$ and taking the derivative w.r.t.\ $\theta$ gives
\begin{equation}
\frac{\drm}{\drm \theta} F_{J}^{*}(\theta; n)  = \frac{n-1}{n} \frac{\drm}{\drm \theta} F_{J}^{*}(\theta; n-1), ~ \forall \theta \geq \theta_{n-1}.
\end{equation}
This implies the non-differentiability will be ``inherited'' as $n$ increases (by $1$), and hence one can prove this claim using mathematical induction similar to what is done in the proof of item $2)$.
Mathematically we compare the following two (scaled) derivatives and show they are not equal at the throughput boundary $\theta_{n-1}$.
\begin{eqnarray}
& & \left. \frac{\drm}{\drm \theta} F_{J}^{*}(\theta; n) \right\vert_{\theta \uparrow \theta_{n-1}} \neq \frac{n-1}{n} \left. \frac{\drm}{\drm \theta} F_{J}^{*} (\theta; n-1) \right\vert_{\theta \downarrow \theta_{n-1}}  \label{eq:non-smooth} 
\end{eqnarray}
Note when the number of users is $n$, $\theta_{n-1}$ is the right-end of its active interval $[\theta_{n}, \theta_{n-1})$ and is attained when $\lim_{\theta \uparrow \theta_{n-1}} p_{s}^{*}(\theta) = 0$, whereas when the number of users is $n-1$, $\theta_{n-1}$ is the left-end of its active interval $[\theta_{n-1}, \theta_{n-2})$ and is attained when $\lim_{\theta \downarrow \theta_{n-1}} p_{s}^{*}(\theta) = 1/(n-1)$. Therefore the LHS of \eqnref{non-smooth} is given by \eqnref{dFdTExpression} with $p_{s}^{*}$ set to $0$ while the derivative in the RHS of \eqnref{non-smooth} is given by \eqnref{dFdTExpression} with $n$ reparameterized as $n-1$ and $p_{s}^{*}$ set to $1/(n-1)$. We can verify their ratio is $(n-2)/(n-1)$ which does not equal $1$, although it approaches $1$ as $n \to \infty$:
\begin{eqnarray}
& & \lim_{n \to \infty} \frac{   \left. \frac{\drm}{\drm \theta} F_{J}^{*}(\theta; n) \right\vert_{\theta \uparrow \theta_{n-1}}   }{   \frac{n-1}{n} \left. \frac{\drm}{\drm \theta} F_{J}^{*} (\theta; n-1) \right\vert_{\theta \downarrow \theta_{n-1}}    }  = 1  \label{eq:asym-smooth}
\end{eqnarray}

Item $5)$. We claim again that it suffices to show convexity when $\theta \in [\theta_{n}, \theta_{n-1})$ but for \textit{all} $n \geq 2$; the proof of this claim is similar to the one given in proving item $2)$: essentially \eqnref{RecusionJain} implies the T-F curve for $\theta$ in a non-active throughput interval may be obtained by linear scaling of some appropriate curve section for which $\theta$ lies in its active throughput interval.

We establish convexity by showing the second derivative is positive:
\begin{eqnarray} 
\label{eq:d2FdT2ComputingApproach}
\frac{\drm^{2}}{\drm \theta^{2}} F_{J}^{*}(\theta; n) 
&=& \frac{\drm}{\drm \theta} \left( \frac{\drm}{\drm \theta} F_{J}(p_{s}^{*}(\theta); n) \right) \nonumber \\
&=& \frac{  \frac{\drm}{\drm p_{s}^{*}(\theta)} \frac{\drm}{\drm \theta} F_{J}(p_{s}^{*}(\theta); n)  }{\frac{\drm}{\drm p_{s}^{*}} T(p_{s}^{*}, 1, n)} \nonumber \\
& \stackrel{(a)}{=} & \frac{  \frac{\drm}{\drm p_{s}^{*}(\theta)} \left(  \frac{\frac{\drm }{\drm p_{s}^{*}(\theta)} F_{J}(p_{s}^{*}(\theta); n)}{ \frac{\drm }{\drm p_{s}^{*}} T (p_{s}^{*}, 1, n)}  \right)   }{\frac{\drm}{\drm p_{s}^{*}} T(p_{s}^{*}, 1, n)}  \nonumber \\
&=& \frac{  \frac{\drm^{2} }{\drm p_{s}^{*}(\theta)^{2}} F_{J}(p_{s}^{*}(\theta); n)  \frac{\drm }{\drm p_{s}^{*}} T (p_{s}^{*}, 1, n)  -     \frac{\drm }{\drm p_{s}^{*}(\theta)} F_{J}(p_{s}^{*}(\theta); n)     \frac{\drm^{2} }{\drm p_{s}^{*2}} T (p_{s}^{*}, 1, n)  }{ \left( \frac{\drm}{\drm p_{s}^{*}} T(p_{s}^{*}, 1, n) \right)^{3}},
\end{eqnarray}
where $(a)$ is from \eqnref{dFStardThetaOriginalExpression}. Since we know $\frac{\drm}{\drm p_{s}^{*}} T(p_{s}^{*}, 1, n) < 0$ for $p_{s}^{*} \in (0, 1/n)$ (applying \prpref{tputeqconst}, item $1)$), showing $\frac{\drm^{2}}{\drm \theta^{2}} F_{J}^{*}(\theta; n)  > 0$ is equivalent to showing the numerator in \eqnref{d2FdT2ComputingApproach} is negative. Thus we compute
\begin{eqnarray}  
\label{eq:d2FdT2ApplyingkStarEqOne-ChiuJain}
& & \frac{\drm^{2}}{\drm \theta^{2}} F_{J}^{*}(\theta; n) \cdot \left( \frac{\drm}{\drm p_{s}^{*}} T(p_{s}^{*}, 1, n) \right)^{3} \nonumber \\
& & = -\frac{2 (n-1)^2  (n p_s^{*}-1)^2 (n p_s^{*}+n-2)^2}{n (n+p_s^{*}-2)^4 \left(\frac{n+p_s^{*}-2}{n-1}\right)^{-n}} \cdot \frac{f_{\rm nu}(n; p_{s}^{*})}{f_{\rm de}(n; p_{s}^{*})^{3}},
\end{eqnarray}
where the functions $f_{\rm nu}$ and $f_{\rm de}$ are defined in \eqref{eqn:fdeAndfnuBothOfn} in Lem.\ \ref{lem:KeyLemmaChiuJainConvexDecreasing} 
%(note they can be viewed as a function of $n$ only, although conceptually they are written as $f_{\rm nu}(n; p_{s}^{*})$ and $f_{\rm de}(n; p_{s}^{*})$, yet for our problem $p_{s}^{*}$ is not a function of $n$ (but the target throughput $\theta$), see \eqref{eq:ps-star-inMainProp}, and even if it is, the conclusion of Lem.\ \ref{lem:KeyLemmaChiuJainConvexDecreasing} still holds). 
Hence we need to show $\frac{f_{\rm nu}(n; p_{s}^{*})}{f_{\rm de}(n; p_{s}^{*})^{3}}$ is positive. This follows from Lem.\ \ref{lem:KeyLemmaChiuJainConvexDecreasing} which assumes $n \geq 3$. For $n = 2$ we can actually prove the convexity directly, leveraging the closed-form expression shown in \prpref{TFtradeoffNequals2}. Specifically, the second derivative can be computed as
\begin{equation}
 \frac{\drm^{2}}{\drm \theta^{2}} F_{J}^{*}(\theta; 2) = \frac{2 \theta^{2} ( - 2 \theta +3) + 2}{(\theta^{2} + 2 \theta - 1)^{3}},
\end{equation}
which can be shown to be positive for $\theta \in [1/2, 1)$. This completes the proof.
\end{IEEEproof}

%% file: AppC-AlphaFairProofs.tex
%%%%%%%%%%%%%%%%%%%%%%%%%%%%%%%%%%%%%%%%%%%%%%
% Appendix C. Proofs from AlphaFair section
%%%%%%%%%%%%%%%%%%%%%%%%%%%%%%%%%%%%%%%%%%%%%%
\section{Proofs from \secref{TFtradeoffUsingAlphaFairUtilityFunctionsWithThroughputConstraint}}
\label{app:ProofsFromAlphaFairSection}

Proofs from \secref{AlphaFairPrelim}, \secref{AlphaFairMain}, and \secref{AlphaFairProp} are given in \appref{AlphaFairPrelimPf}, \appref{AlphaFairMainPf}, and \appref{AlphaFairPropPf}, respectively.

%%%%%%%%%%%%%%%%%%%%%%%%%%%%%%%%%%%%%%%%%%%%%%
\subsection{Proofs from \secref{AlphaFairPrelim}}
\label{app:AlphaFairPrelimPf}

The following lemma is used in the proof of \prpref{mono2} for the $\alpha > 1$ case.
\begin{lemma} 
\label{lem:Z3decreasingInAlpha}
Given $p_{s} \in (0, 1/n)$, $k \in [n-1]$ and $\alpha \geq 1$, the function $f_{2}(p_{s}, k; \alpha)$ defined in \eqref{eq:NewLHSGreaterRHS20151217} is decreasing in $\alpha$.
\end{lemma}
\begin{IEEEproof}
Recall the notation shorthand $r_{x}$ defined in \eqnref{ratioShortHands} in \secref{opttputconst} and observe $r_{x} > 1$. A scaled version of the partial derivative of $f_{2}$ w.r.t.\ $\alpha$ is
\begin{eqnarray} 
\label{eq:partialT3partialAlpha}
(\alpha -1)^{2} \frac{\partial}{\partial \alpha} f_{2}(p_{s}, k; \alpha) &=& g_{1} (p_{s}, k; \alpha)
\end{eqnarray}
where 
\begin{eqnarray}
g_{1}(p_{s}, k; \alpha) & \equiv & \frac{p_{l} - p_{s}}{r_x^{\alpha} - 1} + \frac{\alpha (\alpha -1) (p_{l} - p_{s}) r_x^{\alpha}  \log r_x}{\left(r_x^{\alpha} - 1 \right)^{2}} - p_{s} (1- p_{l}).
\end{eqnarray}
We must show $g_{1} \leq 0$ for all $\alpha \geq 1$.  Towards that goal, the first derivative of $g_1$ with respect to $\alpha$ is
\begin{eqnarray} 
\label{eq:T3DerivativeNumeratorScaled}
\frac{\partial }{\partial \alpha} g_{1}(p_{s}, k; \alpha) &=& - \frac{(\alpha -1) (p_{l} - p_{s})  r_x^{\alpha} \log  r_x }{\left(  r_x^{\alpha} - 1 \right)^{3}} \cdot g_{2}(p_{s}, k; \alpha)
\end{eqnarray}
where 
\begin{eqnarray}
g_{2}(p_{s}, k; \alpha) & \equiv & \tilde{g}_{2}(r_{x}; \alpha) = -2 r_x^{\alpha} + \alpha \left( r_x^{\alpha} + 1 \right) \log r_x + 2,
\end{eqnarray}
and $\tilde{g}_{2}(r_{x}; \alpha)$ is a reparameterization of $g_{2}(p_{s}, k; \alpha)$. The derivative of $\tilde{g}_{2}(r_{x}; \alpha)$ with respect to $\alpha$ is
\begin{eqnarray}
\frac{\partial}{\partial \alpha} \tilde{g}_{2}(r_{x}; \alpha) & = & \tilde{g}_{3} (r_{x}; \alpha) r_{x}^{\alpha} \log r_{x}
\end{eqnarray}
where
\begin{eqnarray}
\tilde{g}_{3} (r_{x}; \alpha) & \equiv &  -1 + r_{x}^{-\alpha} + \alpha \log r_{x}.
\end{eqnarray}
Thus $\tilde{g}_{3}$ determines the sign of $\frac{\partial}{\partial \alpha} \tilde{g}_{2}(r_{x}; \alpha)$. We can verify $ \frac{\partial \tilde{g}_{3}}{\partial \alpha} = (\alpha + 1) \log r_{x} > 0$, and furthermore
\begin{eqnarray}
\tilde{g}_{3}(r_x;1) & = & -1+ \frac{1}{r_{x}} + \log r_{x} \nonumber \\
& \geq &- \frac{r_{x}-1}{r_{x}} + 2 \left( \frac{r_{x}-1}{r_{x}+1}\right) \nonumber \\
&=& \frac{(r_{x}-1)^{2}}{r_{x} (r_{x}+1)} > 0.
\end{eqnarray}
The inequality comes from a series expansion of the natural logarithm based on the inverse hyperbolic tangent function 
\begin{equation}
\log y = 2 \tanh^{-1} \frac{y-1}{y+1}  = 2 \sum_{n=0}^{\infty} \frac{1}{2 n + 1} \left(\frac{y-1}{y+1}\right)^{2 n + 1},
\end{equation}
valid for any $y > 0$. This shows $\frac{\partial}{\partial \alpha} \tilde{g}_{2}(r_{x}; \alpha) \geq 0$ meaning $\tilde{g}_{2}$ is nondecreasing in $\alpha$.  Next, 
\begin{eqnarray}
\tilde{g}_{2}(r_{x}; 1) & = & -2 r_{x} + (r_{x}+1) \log r_{x} + 2 \nonumber \\
& > & -2 r_{x} + (r_{x}+1) 2 \left( \frac{r_{x}-1}{r_{x}+1}\right) + 2 = 0,
\end{eqnarray}
where the bounding comes again from the series expansion of logarithm based on the inverse hyperbolic tangent function. This means $\tilde{g}_{2} \geq 0$ (and in particular, $g_{2} \geq 0$) for all $\alpha \geq 1$, which, according to \eqref{eq:T3DerivativeNumeratorScaled}, implies $g_{1}$ is decreasing in $\alpha$ for $\alpha \geq 1$. Since we can verify $g_{1}(p_s,k;1) = 0$, this means $g_{1}(p_s,k;\alpha) \leq 0$ for all $\alpha \geq 1$. It then follows from \eqnref{partialT3partialAlpha} that $\frac{\partial f_{2}}{\partial \alpha} \leq 0$ for all $\alpha \geq 1$. This concludes the proof of the lemma.
\end{IEEEproof}

\begin{IEEEproof}[Proof of \prpref{mono2}]
The cases $\alpha > 1$ and $\alpha = 1$ are proved separately. Recall the notation shorthands $r_{x}$, $r_{\bar{p}}$ defined in \eqnref{ratioShortHands} in \secref{opttputconst} and observe $r_{x}, r_{\bar{p}} > 1$.

{\em Proof for the case $\alpha > 1$.}

Fix $n'$. We will write the objective $F_{\alpha}(p_{s}, k, n')$, defined in \eqnref{AlphaFairObjectiveShorthand}, as $F_{\alpha}(p_{s}, k)$ to suppress the dependence on $n'$. Similar to the proof of \prpref{mono1}, we treat $k$ as a continuous variable and compute the \textit{total derivative} to take into account the throughput constraint. More precisely, we apply \eqnref{dpsdk} and \eqnref{PartialTotalDerivativeJainsFairnessNprimeHeldConstant} (with $F_{-1}$ replaced by $F_{\alpha}$), which yields
\begin{eqnarray} 
\label{eq:TotalDerivativeAlphaFairOriginalObjWRTk}
\frac{\drm}{\drm k} F_{\alpha}(p_s(k),k) 
&=&  \frac{ x_{l}^{-\alpha} x_{s}^{1-\alpha} }{p_{s} (1 - n' p_{s}) (\alpha - 1) B} f_{1}(p_{s}, k)
\end{eqnarray}
where 
\begin{eqnarray}
f_{1}(p_{s}, k) & \equiv & (\alpha-1) (1-p_{s}) B (n'-k) \left(x_{l}^{\alpha} - x_{s}^{\alpha}\right) \log r_{\bar{p}} \nonumber \\
& &  - (1 - n' p_{s}) \left((\alpha + p_{s} -1) B x_{l}^{\alpha} - (1-p_{s}) (-B+\alpha (n'-k)) x_{s}^{\alpha} \right)
\end{eqnarray}
and $B = (n'-k)(1-p_{l})$.  To show $f_{1}(p_{s}, k)$ in \eqnref{TotalDerivativeAlphaFairOriginalObjWRTk} is nonnegative, we show an equivalent inequality which is less ``coupled''. More precisely, showing $f_{1}(p_{s}, k)$ is nonnegative is equivalent to showing 
\begin{eqnarray} 
\label{eq:NewLHSGreaterRHS20151217}
\log r_{\bar{p}} & \geq & \frac{1 - n' p_{s}}{(n'-k) (1-p_{s}) (1-p_{l})} f_{2} (p_{s}, k; \alpha)
\end{eqnarray}
where
\begin{eqnarray}
f_{2} (p_{s}, k; \alpha) & \equiv & \frac{(\alpha-1+p_{s}) (1-p_{l}) r_{x}^{\alpha} - (\alpha-1+p_{l}) (1-p_{s})}{(\alpha -1) \left(r_{x}^{\alpha} - 1\right)}.
\end{eqnarray}
Observe in \eqref{eq:NewLHSGreaterRHS20151217}, only one side of the inequality involves logarithm and only one side has terms involving $\alpha$ (c.f., showing the positiveness of \eqnref{dF1psk} via \eqnref{JainsFairnessLHSlessthanRHS}, in the proof of \prpref{mono1}). In particular, only $f_{2}$ depends on $\alpha$.  Since Lem.\ \ref{lem:Z3decreasingInAlpha} asserts $f_{2}$ is decreasing in $\alpha$ for the regime of interest, this means to prove \eqref{eq:NewLHSGreaterRHS20151217} we only need to prove it for the $\alpha = 1$ case.  Applying L'H\^opital's rule, we have 
\begin{equation}
\lim_{\alpha \to 1}f_{2} (p_{s}, k; \alpha) = \frac{(1-p_{s}) \left(p_{l} \left(1 + p_{s} \log r_{x} \right) - p_{s}\right) (1-p_{l})}{p_{l} - p_{s}}. 
\end{equation}
Observing $r_{x} = \frac{p_{l}}{p_{s}} + r_{\bar{p}} $ and $\frac{1-n' p_{s}}{(n'-k) (p_{l}-p_{s})} = 1$, showing \eqref{eq:NewLHSGreaterRHS20151217} amounts to showing
\begin{equation}
f_{3} (p_{s}, k) \equiv (1 - p_{s} p_{l}) \log r_{\bar{p}} - p_{l} + p_{s} - p_{s} p_{l} \log \frac{p_{l}}{p_{s}} \geq 0.
\end{equation}
The partial derivative of $f_{3}$ w.r.t. $k$ is 
\begin{equation}
\frac{\partial }{\partial k} f_{3}(p_s,k)= - \frac{1 - n' p_{s}}{(n'-k)^{3} (1-p_{l})}  f_{4}(p_s,k)
\end{equation}
where 
\begin{equation}
f_{4}(p_{s}, k) \equiv p_{s} (1-p_{l}) (n'-k) \log r_{x} - (1-n' p_{s}).
\end{equation}
By applying the logarithm inequality \eqnref{InequalityOnLog}, $f_4(p_s,k)$ may be shown to be upper bounded by 0. This implies $f_{3}(p_s,k)$ is increasing in $k$, and thus it suffices to verify $f_{3}(p_s,0) \geq 0$.  Note that although the $k = 0$ case is not included in the $\vert \Vmc(\pbf) \vert = 2$ scenario, all the relevant functions are nonetheless well-defined.  We must show
\begin{equation}
n' f_{3}(p_s,0) = n' p_{s} + p_{s} \log (n' p_{s}) + (n' - p_{s}) \log \frac{n'(1-p_{s})}{n'-1} - 1 \geq 0.
\end{equation}
By taking the partial derivative of $f_{3}(p_s,0)$ w.r.t.\ $p_{s}$ it is easily seen that it is nonpositive (by checking the monotonicity of $\frac{\drm}{\drm p_s} f_3(p_s,0)$ w.r.t.\ $p_{s}$). Therefore, to show $f_{3}(p_s,0) \geq 0$ over $p_s \in [0,1/n']$ it suffices to show $f_3(1/n',0) \geq 0$.  As $f_3(1/n',0) = 0$, the above arguments collectively imply $f_{3}(p_s,k) \geq 0$ for all $0 \leq k \leq n'-1$ and $p_{s} \in (1, 1/n')$. Therefore inequality \eqnref{NewLHSGreaterRHS20151217} is proved, which means the total derivative \eqnref{TotalDerivativeAlphaFairOriginalObjWRTk} is positive and thus the optimal $k^{*} = n'-1$.

{\em Proof for the case $\alpha = 1$.}

We again fix $n'$ and suppress the dependence of $F_1(p_{s}, k, n')$ on $n'$:
\begin{equation}
F_{1}(p_{s}, k) = \log \left( p_{s}^{k} p_{l}^{n'-k} \left( \left( 1 - p_{s}\right)^{k} \left( 1 - p_{l} \right)^{n'-k} \right)^{n'-1} \right).
\end{equation}
We also write $F_{1}(p_{s}(k), k)$ to take into account the throughput constraint. Applying \eqnref{dpsdk} and \eqnref{PartialTotalDerivativeJainsFairnessNprimeHeldConstant} (with $F_{-1}$ replaced by $F_{1}$) yields (where $\tanh^{-1}$ is the inverse hyperbolic tangent function) the \textit{total derivative}
\begin{eqnarray} 
\label{eq:TotalDerivativeProportionalFairOriginalObjWRTk}
\frac{\drm}{\drm k} F_{1}(p_s(k),k) 
& \equiv & h_{1}(p_{s}, k) \nonumber \\
&=&  \frac{1}{p_s (k p_s-1)} \left( p_{s} (1-k p_{s}) \log \left(p_{l}/(1-p_{l})\right) - (n'-k) \log r_{\bar{p}}  \right. - \nonumber \\
& & \left. 2 p_s (k p_s-1) \tanh ^{-1}(1-2 p_s)-n' p_s+1 \right) .
\end{eqnarray}
We further compute the partial derivative of $h_{1}(p_{s}, k)$ w.r.t.\ $k$ and get
\begin{eqnarray} 
\label{eqn:TotalDerivativeProportionalFairOriginalObjWRTk-itsOwnPartialDerivativeWRTk}
\frac{\partial}{\partial k}  h_{1}(p_{s}, k)
&=& \frac{-(1 - n' p_{s})}{p_{s} (k p_{s}-1)^2 (n' - k - 1 + k p_{s})} h_{2} (p_{s}, k), \end{eqnarray}
where 
\begin{eqnarray}
h_{2} (p_{s}, k) & \equiv & (n'-k-1+ k p_{s}) \log r_{\bar{p}}  + n' p_{s} - 1.
\end{eqnarray}
Applying inequality \eqnref{InequalityOnLog}, $h_{2} (p_{s}, k)$ may be bounded as
\begin{equation}
h_{2} (p_{s}, k)
\leq \frac{2 k p_{s} (n' p_{s} -1)}{n'-k-1+kp_{s}}
< 0,
\end{equation}
which shows $\frac{\partial}{\partial k} h_{1}(p_{s}, k) > 0$.  Therefore, to show $h_{1}(p_{s}, k) > 0$ for all $p_{s} \in (0, 1/n')$ and $k \in [n'-1]$ we only need to show $h_{1}(p_{s}, 1)  > 0$, or equivalently,
\begin{eqnarray}
p_{s} (1-p_{s}) h_{1}(p_{s}, 1) & = & (n'-1) \log r_{\bar{p}} - p_{s} (1-p_{s}) \log r_{x} - (1 - n' p_{s}) > 0.
\end{eqnarray}
We rearrange terms and seek to prove the equivalent condition $h_{3}(p_{s}, k) > 0$, for\begin{equation} 
\label{eq:ProportionalFairLHS-ProportionalFairRHS}
h_{3}(p_{s}, k) \equiv \frac{(n'-1) \log r_{\bar{p}} - (1- n' p_{s})}{p_{s} (1 - p_{s})} - \log r_{x}.
\end{equation}
Computing the partial derivative w.r.t. $p_{s}$ and applying inequality \eqnref{InequalityOnLog}, we have the upper bound
\begin{equation}
\frac{\partial}{\partial p_{s}}  h_{3}(p_{s}, k) \leq -\frac{p_{s} (1-n' p_{s})^{2}}{1-p_{s}} < 0.
\end{equation}
This means $h_{3}(p_{s}, k)$ is decreasing in $p_{s}$ and therefore to show $h_{3}(p_{s}, k) > 0$ holds for all $p_{s} \in (0, 1/n')$ it suffices to show $h_{3}(1/n', k) \geq 0$. We can verify this indeed holds with equality. This completes the proof that $h_{1}(p_{s}, k)$, namely the total derivative \eqnref{TotalDerivativeProportionalFairOriginalObjWRTk}, is positive, implying the optimality of $k^{*} = n'-1$.
\end{IEEEproof}

%%%%%%%%%%%%%%%%%%%%%%%%%%%%%%%%%%%%%%%%%%%%%%
\subsection{Proofs from \secref{AlphaFairMain}}
\label{app:AlphaFairMainPf}

\begin{IEEEproof}[Proof of \thmref{mainresultAlphaFairWhenAlphaGreaterThanOrEqualToOne}]
Regime $i)$: $\theta \leq \theta_{n}$. That the maximizer is a uniform vector follows from the Schur-concavity of $F_{\alpha}$ w.r.t.\ $\xbf$ (\prpref{SchurConcavityCJFairnessAndAlphaFairUtiityFunction}, or by applying Thm.\  A.\ 4 in Ch.\ 3 of \cite{MarOlk2011}), and the fact that when $\theta \leq \theta_{n}$ the ``all-rates equal'' vector is always feasible, as the uniform vector is majorized by all the other vectors that have the same component sums.  To establish this feasibility, we assume the optimal rate vector $\xbf^{*}$ is such that $x_{i}^{*} = p^{*} (1-p^{*})^{n-1} = \theta / n$, $i \in [n]$, and attempt to solve for $p^{*} \in [0,1]$. The existence of such a $p^{*}$ follows from \lemref{AllRatesEqualRay} and hence the feasibility is proved.

When $\alpha = 1$, an alternative way to show the ``all-rates equal'' vector is optimal is by using the AM-GM inequality $\tilde{F}_{1} (\xbf)= \prod_{i} x_{i} \leq \left(\sum_{i} x_{i} / n \right)^{n} = (\theta / n)^{n}$, where $\tilde{F}_{1} \equiv \erm^{F_{1}(\xbf)}$. As this inequality is tight when all the $x_{i}$'s ($p_{i}$'s) are equal, the maximum $\tilde{F}_{1}^{*} = (\theta / n)^{n}$ will be attained if there exists a vector $\pbf^{*} = p^{*} \mathbf{1}$ that satisfies the throughput constraint namely $x_{i}^{*} = p^{*} (1-p^{*})^{n-1} = \theta / n$, $i \in [n]$. 

Regime $ii)$: $\theta \in (\theta_{n}, 1)$. First, $\pbf^* \in \partial\Smc$ follows from \corref{MajorizationCorollary} in \secref{MajorizationAttempt}. Second, observe $n'^{*} = n$ as otherwise any inactive user (i.e., one with zero contention probability) will make the objective $F_{\alpha}$ go to $-\infty$. Third, we claim $\pbf^{*} \in \partial \Smc_{2}$. To see this, we apply \prpref{AtMostTwoDistinctNonzeroComponentValues} (item $i)$), which, together with the fact $\pbf^{*} \in \partial \Smc$ and $n'^{*} = n$, implies that there is no feasible point if $\vert \Vmc (\pbf^{*}) \vert  = 1$, and hence $\vert \Vmc (\pbf^{*}) \vert  = 2$, meaning $\pbf^{*} \in \partial \Smc_{2}$.  Fourth, when $n'^{*} = n$ and $\theta$ are fixed, the throughput constraint \eqnref{psfunctheta} implicitly defines $p_{s}^{}$ as a function of $k^{}$ and thus we write $F_{\alpha}(p_{s}(k), k)$ (with $n'^{*}$ suppressed). It then follows from the analysis based on the total derivative, shown in \prpref{mono2}, that the optimal $k^{*} = n - 1$. Finally, the existence and uniqueness of $p_{s}^{*}$ follows from \prpref{tputeqconst} and recognizing that \eqnref{ps-star-AlphaFair} is \eqnref{psfunctheta} specialized with $k = n -1$ and $n' = n$.
\end{IEEEproof}

\begin{IEEEproof}[Proof of \thmref{5}]
Regime $i)$: $\theta \leq \theta_{n}$. Denote the original optimization problem \eqnref{AlphaFairTFtradeoffProblemFormulation} with a throughput equality constraint $T(\xbf) = \hat{\theta}$ by $\Psf_=(\hat{\theta})$, and denote the current optimization problem with a throughput inequality constraint $T(\xbf) \geq \theta$ by $\Psf_{\geq}(\theta)$. The current problem, $\Psf_{\geq}(\theta)$, may be viewed as a two-layer optimization problem where the inner layer is $\Psf_=(\hat{\theta})$, i.e., $\Psf_{\geq}(\theta) = \max_{\hat{\theta} \in [\theta, 1]} \Psf_=(\hat{\theta})$. This can be further decomposed as the following 
\begin{equation} 
\label{eq:NestedOptimizationProportionFairness}
\Psf_{\geq}(\theta) 
= \max\left( \max_{\hat{\theta} \in [\theta, \theta_{n}]} \Psf_=(\hat{\theta}), ~ \max_{\hat{\theta} \in (\theta_{n}, 1)} \Psf_=(\hat{\theta}),  ~ \Psf_=(1) \right).
\end{equation}
For the first term in \eqref{eq:NestedOptimizationProportionFairness}, since we can verify that $F_{\alpha}^{*}(\theta)$ in \eqnref{AlphaFair-TFtradeoffGeneralNThetaLeqThetaN} of Thm.\ \ref{thm:mainresultAlphaFairWhenAlphaGreaterThanOrEqualToOne} is increasing in $\theta$, at least for regime $1$, this shows $\max_{\hat{\theta} \in [\theta, \theta_{n}]} \Psf_=(\hat{\theta}) = \Psf_=(\theta_{n})$ with the maximizer $\pbf = (1/n)\mathbf{1}$. For the second term, based on the fact that there exists a tradeoff between target throughput and the $\alpha$-fair objective for regime $2$ (\thmref{AlphaFairTFtradeoffProperties}, item $2)$), it follows that $\max_{\hat{\theta} \in (\theta_{n}, 1)} \Psf_=(\hat{\theta}) \leq \Psf_=(\theta_n)$. For the third term, it can be seen that $\Psf_=(1) = -\infty$ because the only feasible point achieving a target throughput of $1$ is $\ebf_{i}$. Therefore, the solution of \eqref{eq:NestedOptimizationProportionFairness} when $\theta \leq \theta_{n}$ is given in \eqnref{AlphaFair-TFtradeoffGeneralN-InequalityThroughputConstraint}, attained when $\pbf^{*} = (1/n)\mathbf{1} = \ubf$.

Regime $ii)$: $\theta \in (\theta_{n}, 1)$. Observe the maximum of the objective will be attained when there exists some $\theta^{*} \in [\theta, 1)$ for which the throughput constraint holds with equality namely $T(\xbf(\pbf^{*})) = \theta^{*}$, then similar to what was shown in the proof of \thmref{mainresultAlphaFairWhenAlphaGreaterThanOrEqualToOne} (regime $2$), we can show $\pbf^{*} \in \partial \Smc_{2}$. Consequently, \prpref{AtMostTwoDistinctNonzeroComponentValues} (item $ii)$) says the throughput inequality constraint is tight, namely $\theta^{*} = \theta$, and thus the rest part in the proof of \thmref{mainresultAlphaFairWhenAlphaGreaterThanOrEqualToOne} (regime $2$) applies here. Therefore the assertion in regime $2$ of \thmref{mainresultAlphaFairWhenAlphaGreaterThanOrEqualToOne} continues to hold.
\end{IEEEproof}

%%%%%%%%%%%%%%%%%%%%%%%%%%%%%%%%%%%%%%%%%%%%%%
\subsection{Proofs from \secref{AlphaFairProp}}
\label{app:AlphaFairPropPf}

The following lemma is used in the proof of item $2)$ in \thmref{AlphaFairTFtradeoffProperties}.

\begin{lemma} 
\label{lem:cubicEquationHasOneValidRoot}
Assume $\alpha > 1$ and $n > 2$. The cubic polynomial $f_{\rm cubic}(p_s)$ in \eqnref{cubicEquationForThresholdingps} has only one root over $p_{s} \in (0, 1/n)$.
\end{lemma}
\begin{IEEEproof}
First, observe Descartes's rule of sign is not sufficient, as it can only assert their are either one or three positive roots.  We instead use the Budan-Fourier Theorem, as was done in the proof of \lemref{KeyLemmaChiuJainConvexDecreasing}.  Specifically, let $v(p)$ denote the number of sign changes (i.e., sign variation) of the Fourier sequence  $\{ f_{\rm cubic}(p_{}), f_{\rm cubic}^{(1)}(p_{}),  f_{\rm cubic}^{(2)}(p_{}), f_{\rm cubic}^{(3)}(p_{})  \}$ when $p_{s}  = p$. We can show $v(0) = 3$ (the signs of the Fourier sequence are: $- ~ + ~ - ~ +$) and $v(1/n) = 2$ (the signs of the Fourier sequence are: $+ ~ + ~ - ~ +$). Since $v(0) - v(1/n) = 1$, this means the polynomial $f_{\rm cubic}(p_{s})$ only has one root on $(0, 1/n)$.

The following expressions are used to establish the signs of the computed Fourier sequence:
\begin{eqnarray} 
\label{eqn:Budan-FourierSequence-cubicEquationHasOneValidRoot}
f_{\rm cubic}(0) &=& 1 - 4 \alpha^{2} \nonumber \\
f_{\rm cubic}^{(1)}(0) &=& 2 n (4 \alpha^{2} - 1) \nonumber \\
f_{\rm cubic}^{(2)}(0) &=& \left(-10 \alpha ^2+4 \alpha +4\right) n^2+(-8 \alpha -6) n+6 \nonumber \\
f_{\rm cubic}^{(3)}(0) &=& 6 n ((\alpha -1) n+1) (\alpha  n+1),
\end{eqnarray}
and 
\begin{eqnarray} 
\label{eqn:Budan-FourierSequence-cubicEquationHasOneValidRoot2}
f_{\rm cubic}(1/n) &=& (1- 2 p_{s}) (1 - 2 p_{s} + \alpha) \nonumber \\
f_{\rm cubic}^{(1)}(1/n) &=& -2 \alpha +\left(\alpha ^2+\alpha +2\right) n - 9(1 - p_{s}) \nonumber \\
f_{\rm cubic}^{(2)}(1/n) &=& \left(-4 \alpha ^2-2 \alpha +4\right) n^2+(4 \alpha -12) n+12 \nonumber \\
f_{\rm cubic}^{(3)}(1/n) &=& 6 n ((\alpha -1) n+1) (\alpha  n+1).
\end{eqnarray}
\end{IEEEproof}

\begin{IEEEproof}[Proof of \thmref{AlphaFairTFtradeoffProperties}]
We write $F_{\alpha}^{*}(\theta; n)$ to emphasize $\theta$ is the free variable and $(\alpha,n)$ are viewed as parameters. The feasible set $\Lambda$ is parameterized by $\pbf$ via \eqnref{xofp} and when $\theta \in (\theta_{n}, 1)$ we know from \thmref{mainresultAlphaFairWhenAlphaGreaterThanOrEqualToOne} that the unique extremizer can be characterized by the tuple $(p_{s}^{*}, k^{*}, n'^{*})$ as functions of $\theta$. Therefore the notation $F_{\alpha}^{*}(\theta; n)$ should be understood as
\begin{eqnarray}
\label{eq:FAlphaStarThetaAlphaN}
F_{\alpha}^{*}(\theta; n) 
& \equiv & F_{\alpha}(\xbf(\pbf (p_{s}^{*}(\theta), k^{*}(\theta), n'^{*}(\theta))); n) \nonumber \\
& = & F_{\alpha}(\xbf(\pbf (p_{s}^{*}(\theta), n-1, n)); n) \nonumber \\
& \equiv & F_{\alpha}(p_{s}^{*}(\theta); n),
\end{eqnarray}
where the second equality follows from \thmref{mainresultAlphaFairWhenAlphaGreaterThanOrEqualToOne}, and the third equivalence is a shorthand notation.

Observe also the identity (c.f., \eqnref{TpsknprimeStar})
\begin{equation}
\label{eq:AlphaFair-TpsknprimeStar}
\theta \equiv T (p_{s}^{*}(\theta), n-1, n),
\end{equation}
for $T(p_{s}, k, n')$ defined in \eqnref{signedtputgap} and $p_{s}^{*}$ solved from \eqnref{ps-star-AlphaFair}. This is useful for computing the dependence of $p_{s}^{*}$ on $\theta$. 

Item $1)$. We first look at the monotonicity of $p_{s}^{*}(\theta)$, $p_{l}^{*}(\theta)$ in $\theta$. That $p_{s}^{*}(\theta)$ is decreasing in $\theta$ follows from
\begin{equation}
\label{eq:AlphaFair-psStarDecInTheta}
\frac{\drm p_{s}^{*}(\theta)}{\drm \theta} 
= \left( \frac{\drm \theta(p_{s}^{*})}{\drm p_{s}^{*}}\right)^{-1} 
=\left( \frac{\drm }{\drm p_{s}^{*}} T (p_{s}^{*}, n-1, n) \right)^{-1}
< 0,
\end{equation}
where we apply \eqnref{AlphaFair-TpsknprimeStar} and the negativity follows from \prpref{tputeqconst} (item $1)$) in \secref{opttputconst}.
That $p_{l}^{*}(\theta) = p_{l}(p_{s}^{*}(\theta), n-1,n)$ is increasing in $\theta$ follows easily from the definition of $p_{l}$ in \defref{restrictedControls} and \eqnref{AlphaFair-psStarDecInTheta}.

Next, we look at the dependence of $x_{s}^{*}(\theta)$, $x_{l}^{*}(\theta)$ upon $\theta$. We have
\begin{eqnarray}
x_{s}^{*} &=& (n-1) p_{s}^{*2} (1 - p_{s}^{*})^{n-2} \nonumber \\
x_{l}^{*} &=& p_{l}^{*} (1 - p_{s}^{*})^{n-1}.
\label{eq:xsAndxlStarVersusThetaAlphaFair}
\end{eqnarray}
That $\frac{\drm x_{l}^{*}(\theta)}{\drm \theta} > 0$ follows from $\frac{\drm p_{s}^{*}(\theta)}{\drm \theta} < 0$ and $\frac{\drm p_{l}^{*}(\theta)}{\drm \theta} > 0$. To show $\frac{\drm x_{s}^{*}(\theta)}{\drm \theta} < 0$, applying the chain rule and recalling \eqnref{AlphaFair-psStarDecInTheta}, we need to show $p_{s}^{*2} (1 - p_{s}^{*})^{n-2}$ is increasing in $p_{s}^{*}$. Since we can verify the function $p^{2} (1-p)^{n-2}$ is increasing for all $p \in (0, 2/n) \supseteq (0, 1/n)$, this proves $\frac{\drm x_{s}^{*}(\theta)}{\drm \theta} < 0$.

Item $2)$. First, we prove $i)$ monotonicity, $ii)$ continuity, and $iii)$ differentiability. Towards $i)$ monotonicity, we compute:
\begin{eqnarray} 
\label{eq:AlphaFair-dFStardThetaOriginalExpression}
\frac{\drm}{\drm \theta}  F_{\alpha}^{*}(\theta; n)
& = & \frac{\drm}{\drm p_{s}^{*}(\theta)}  F_{\alpha}(p_{s}^{*}(\theta); n)    \left( \frac{\drm \theta(p_{s}^{*})}{\drm p_{s}^{*}}\right)^{-1} \nonumber \\
& = & \frac{\drm}{\drm p_{s}^{*}(\theta)}  F_{\alpha}(p_{s}^{*}(\theta); n)    \left( \frac{\drm }{\drm p_{s}^{*}} T (p_{s}^{*}, k^{*}, n'^{*}) \right)^{-1},
\end{eqnarray}
where the second equality comes from \eqnref{AlphaFair-TpsknprimeStar}. When $\alpha = 1$ and $\alpha > 1$, we get
\begin{eqnarray}
\label{eq:AlphaFair-dFdT}
& & \frac{\drm}{\drm \theta}  F_{1}^{*}(\theta; n)
= -\frac{1-p_{s}^{*}}{p_{s}^{*}} \frac{1}{x_{l}^{*}}  < 0, \nonumber \\
& & \frac{\drm}{\drm \theta}  F_{\alpha > 1}^{*}(\theta; n)
=  - \frac{(1-p_{l}^{*}) x_{s}^{*-\alpha} - (1-p_{s}^{*}) x_{l}^{*-\alpha}}{1-n'^{*} p_{s}^{*}} (n'^{*}-k^{*})
\end{eqnarray}
We will show \eqnref{AlphaFair-dFdT} is negative for all $p_{s}^{*} \in (0, 1/n')$. Namely we want to show $(1-p_{l}^{*}) x_{s}^{*-\alpha} - (1-p_{s}^{*}) x_{l}^{*-\alpha} > 0$, which is equivalent to $\frac{p_{s}^{*-\alpha}}{(1-p_{s}^{*})^{1-\alpha}} > \frac{p_{l}^{*-\alpha}}{(1-p_{l}^{*})^{1-\alpha}}$. Thus it suffices to show 
\begin{equation}
h(z) \equiv \frac{z^{-\alpha}}{(1-z)^{1-\alpha}}
\end{equation}
with $\alpha > 1$ is decreasing in $z$ for $z \in (0, 1)$. Since 
\begin{equation}
\frac{\drm h(z)}{\drm z} = z^{-\alpha-1} (1-z)^{\alpha -2} (z - \alpha) < 0
\end{equation}
this shows the desired monotonicity of $h(z)$, establishing that the optimal $\alpha$-fair objective is decreasing in $\theta$. 

Next, we prove $ii)$ continuity.  $a)$ since the roots (on the complex plane) of a polynomial equation are continuous in its coefficients \cite[\S 3.9]{Tyr1997} and since the polynomial equation \eqnref{ps-star-AlphaFair} only has a single real root ($p_{s}^{*}$) it must be continuous also.  $b)$ the function $F_{\alpha}^{*}$ in \eqnref{FAlphaStarThetaAlphaN} is continuous in $p_{s}^{*}$. 

Third, we prove $iii)$ differentiability.  This follows from the fact that the derivatives given in \eqnref{AlphaFair-dFdT} are continuous in $p_{s}^{*}$, which are themselves continuous in $\theta$, c.f., \eqnref{AlphaFair-psStarDecInTheta}.

Next, we investigate convexity (concavity).

Similar to what was done in the proof of \thmref{ChiuJainTFtradeoffProperties} (item $5)$), we compute the second derivative and investigate its sign:
\begin{eqnarray} 
\label{eq:AlphaFair-d2FdT2ComputingApproach} 
\frac{\drm^{2}}{\drm \theta^{2}} F_{\alpha}^{*}(\theta; n) 
&=& \frac{\drm}{\drm \theta} \left( \frac{\drm}{\drm \theta} F_{\alpha}(p_{s}^{*}(\theta); n) \right) \nonumber \\
& \stackrel{(b)}{=} & \frac{  \frac{\drm}{\drm p_{s}^{*}(\theta)} \left( \frac{\drm}{\drm \theta} F_{\alpha}(p_{s}^{*}(\theta); n) \right)  }{\frac{\drm}{\drm p_{s}^{*}} T(p_{s}^{*}, n-1, n)} \nonumber \\
& \stackrel{(c)}{=} & \frac{  \frac{\drm}{\drm p_{s}^{*}(\theta)} \left(  \frac{\frac{\drm }{\drm p_{s}^{*}(\theta)} F_{\alpha}(p_{s}^{*}(\theta); n)}{ \frac{\drm }{\drm p_{s}^{*}} T (p_{s}^{*}, n-1, n)}  \right)   }{\frac{\drm}{\drm p_{s}^{*}} T(p_{s}^{*}, n-1, n)}  \nonumber \\
& = & \frac{  \frac{\drm^{2} }{\drm p_{s}^{*}(\theta)^{2}} F_{\alpha}(p_{s}^{*}(\theta); n)  \frac{\drm }{\drm p_{s}^{*}} T (p_{s}^{*}, n-1, n)  -     \frac{\drm }{\drm p_{s}^{*}(\theta)} F_{\alpha}(p_{s}^{*}(\theta); n)     \frac{\drm^{2} }{\drm p_{s}^{*2}} T (p_{s}^{*}, n-1, n)  }{ \left( \frac{\drm}{\drm p_{s}^{*}} T(p_{s}^{*}, n-1, n) \right)^{3}},
\end{eqnarray}
where $(b)$ is due to the chain rule and \eqnref{AlphaFair-TpsknprimeStar} and $(c)$ is from \eqnref{AlphaFair-dFStardThetaOriginalExpression}. Since we know from \prpref{tputeqconst} (item $1)$) $\frac{\drm}{\drm p_{s}^{*}} T(p_{s}^{*}, n-1, n) < 0$ for $p_{s}^{*} \in (0, 1/n)$, showing $\frac{\drm^{2}}{\drm \theta^{2}} F_{\alpha}^{*}(\theta; n)  \gtrless 0$ is equivalent to showing the numerator in \eqnref{AlphaFair-d2FdT2ComputingApproach} is negative / positive. 

We consider the $\alpha = 1$ and $\alpha \geq 0$ but $\alpha \neq 1$ cases separately.  First, for $\alpha =1$: 
\begin{equation} 
\label{eq:d2FdT2ApplyingkStarEqNminusOne-AlphaFair}
\frac{\drm^{2}}{\drm \theta^{2}} F_{1}^{*}(\theta; n) \left( \frac{\drm}{\drm p_{s}^{*}} T(p_{s}^{*}, n-1, n) \right)^{3} = \frac{(n-1)^2 (1-p_s^{*})^{n-5} (n p_s^{*}-2)^2 (n p_s^{*}-1)^2}{p_s^{*2} (-n p_s^{*}+p_s^{*}+1)^2} f_{\rm quad}(p_{s}^{*}, n)  
\end{equation}
where 
\begin{eqnarray} 
\label{eq:KeyTerm-AlphaFair}
f_{\rm quad}(p_{s}^{*}, n) & \equiv &  \left(n^2-n\right) p_s^{*2}+(3-3 n) p_s^{*}+1
\end{eqnarray}
Next, for $\alpha \geq 0$ but $\alpha \neq 1$:
\begin{equation} 
\label{eq:d2FdT2ApplyingkStarEqNminusOne-AlphaFair2}
\frac{\drm^{2}}{\drm \theta^{2}} F_{\alpha}^{*}(\theta; n) \left( \frac{\drm}{\drm p_{s}^{*}} T(p_{s}^{*}, n-1, n) \right)^{3} 
= -(n-1)^3 (1-p_s^{*})^{2 n-7} (n p_s^{*}-2)^2 x_{s}^{* - \alpha} f_{2}(p_{s}^{*}, \alpha, n) 
\end{equation}
where 
\begin{eqnarray} 
\label{eq:KeyTerm-AlphaFair2}
f_{2}(p_{s}^{*}, \alpha, n) & \equiv &  -Z (p_{s}^{*}, \alpha, n ) + (1-p_{s}^{*}) - \left(\frac{x_{s}^{*}}{x_{l}^{*}}\right)^{\alpha} (1-p_{s}^{*}) \left(\frac{Z (p_{s}^{*}, \alpha, n )}{1-(n-1)p_{s}^{*}} + 1\right), 
\end{eqnarray}
and
\begin{eqnarray}
Z (p_{s}^{*}, \alpha, n )& \equiv & \alpha (n p_{s}^{*} - 2) (n p_{s}^{*} - 1).
\end{eqnarray}

Recall $\alpha$ and $n$ are assumed to be fixed. We may sometimes drop them from the parameter list by e.g., writing $f_{2}(p_{s}^{*}, \alpha, n)$ as $f_{2}(p_{s}^{*})$.

\textbf{Case 1: $\alpha > 1$.}  In this case, since the sign of $\frac{\drm^{2}}{\drm \theta^{2}} F_{\alpha}^{*}(\theta; n)$ equals the sign of $f_{2}(p_{s}^{*}, \alpha, n)$, our goal is to show, as $p_{s}^{*}$ increases from $0$ to $1/n$:
\begin{itemize}
\item when $n > 2$, there exists a thresholding $\mathring{p}_{s}^{*} = \mathring{p}_{s}^{*}(\alpha, n)$ below (above) which $f_{2} < (>) ~0$, corresponding to the T-F curve being concave when $\theta$ is large (convex when $\theta$ is small);
\item when $n = 2$, it always holds that $f_{2} < 0$, meaning the T-F curve is always concave.
\end{itemize}

\begin{figure}[!ht]
\centering
\includegraphics[width=6.5in]{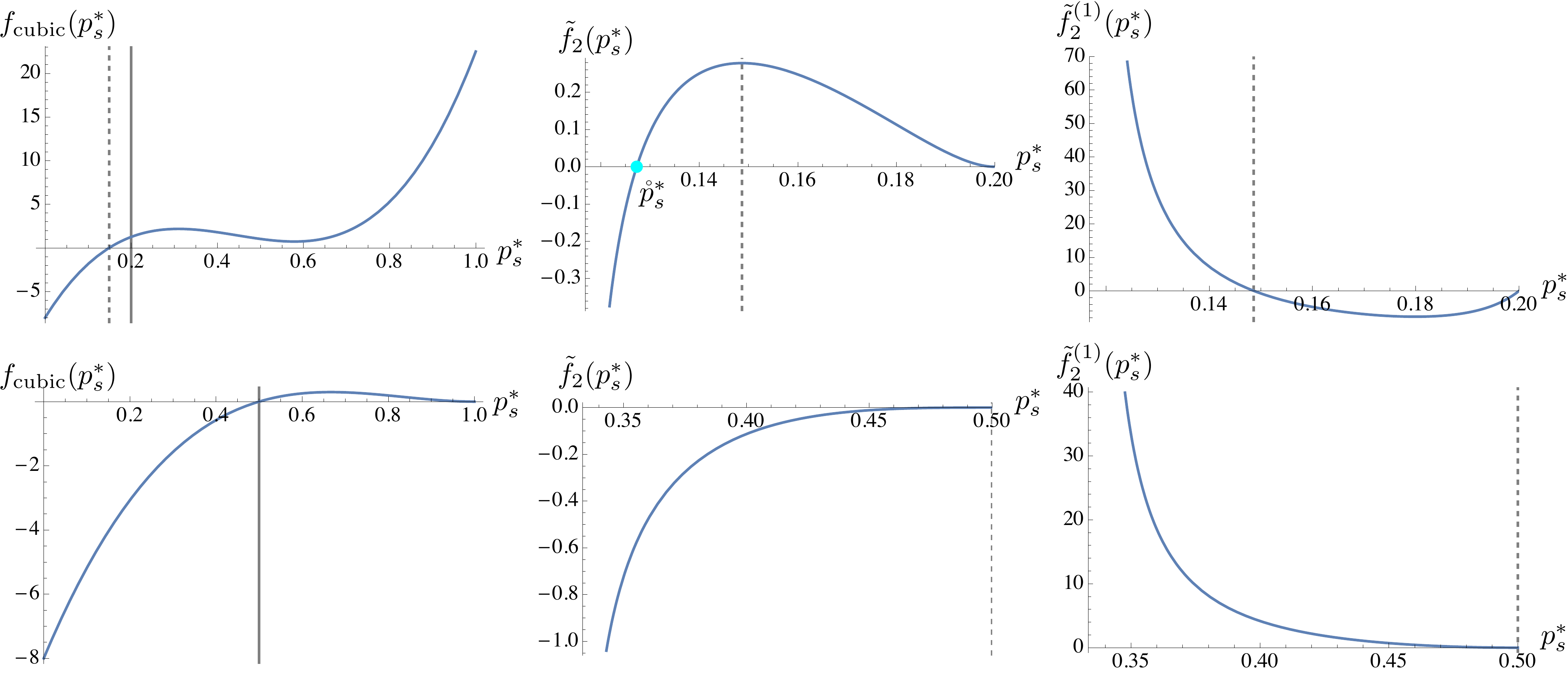}
\caption{Illustration of the proof of item $2)$ in \thmref{AlphaFairTFtradeoffProperties} regarding the thresholding $p_{s}^{*}$. Shown are the polynomial $f_{\rm cubic}(p_{s}^{*})$ \eqref{eq:cubicEquationForThresholdingps} ($1^{\rm st}$ column), the function $\tilde{f}_{2}(p_{s}^{*})$ ($2^{\rm nd}$ column), and its first derivative w.r.t.\ $p_{s}^{*}$ denoted $\tilde{f}_{2}^{(1)}(p_{s}^{*})$ ($3^{\rm rd}$ column), for $n = 5$  (top row) and $n = 2$ (bottom row) respectively. In both cases $\alpha = 1.5$. The solid gridlines indicate $1/n$; the dashed gridlines indicate the only stationary point of $\tilde{f}_{2}$ on $(0, 1/n)$ (also the unique real root of $f_{\rm cubic}$ on $(0, 1/n)$). Except $1^{\rm st}$ column, the plot ranges for the horizontal axis $p_{s}^{*}$ are $(p_{s-}^{*}, 1/n)$. For $n = 5$, $p_{s-}^{*} \approx 0.11683$ and the stationary point of $\tilde{f}_{2}$ is at $p_{s}^{*} \approx 0.1487$; for $n = 2$, $p_{s-}^{*} \approx 0.3333$ and the stationary point of $\tilde{f}_{2}$ is at $p_{s}^{*} = 0.5$. The thresholding $\mathring{p}_{s}^{*} \in (0, 1/n)$ exists only when $n > 2$. Here when $n = 5$ and $\alpha = 1.5$, we have $\mathring{p}_{s}^{*} \approx 0.1273$ solved from \eqnref{thresholdingps} and marked as the cyan dot in the top-middle figure where $\tilde{f}_{2}(p_{s}^{*})$ versus $p_{s}^{*}$ is shown.}
\label{fig:TildeTildefAndItsFirstAndSecondDerivativesWRTps}
\end{figure}

\underline{Subcase 1}: $n > 2$

Directly showing the desired monotonicity (change) of $f_{2}$ w.r.t.\ $p_{s}^{*}$ does not seem easy, as the derivative of $f_{2}$ w.r.t.\ $p_{s}^{*}$ has polynomials of $p_{s}^{*}$ further raised to the power of $\alpha$. Therefore we seek to show the following equivalent condition (recall $p_{s}^{*} \in (0, 1/n)$ and observe $Z \geq 0$). 
\begin{equation} 
\label{eq:fAlphaOriginal}
f_{2} \lessgtr 0 
 \iff 
 \frac{-Z + (1-p_{s}^{*}) }{(1-p_{s}^{*}) \left(\frac{Z}{1-(n-1)p_{s}^{*}} + 1\right)} 
\lessgtr  \left(\frac{x_{s}^{*}}{x_{l}^{*}}\right)^{\alpha}
\end{equation}
First consider the $f_{2} < 0$ case.  Define the following events
\begin{eqnarray}
E_{2} & = & \{ p_{s}^{*} \in (0, 1/n): ~ f_{2}(p_{s}^{*}) < 0 \}, \nonumber \\
\tilde{E}_{2} &  =  & \{ p_{s}^{*} \in (0, 1/n): ~ \tilde{f}_{2}(p_{s}^{*}) < 0 \}, \nonumber \\
E_{Z}  & =  & \{ p_{s}^{*} \in (0, 1/n): ~ -Z(p_{s}^{*}) + (1-p_{s}^{*}) \leq 0 \}, \nonumber \\
\overline{E}_{Z}  & =  & \{ p_{s}^{*} \in (0, 1/n): ~ -Z(p_{s}^{*}) + (1-p_{s}^{*}) > 0 \},
\end{eqnarray}
where
\begin{equation}
\tilde{f}_{2} (p_{s}^{*}) \equiv \frac{1}{\alpha} \log\left( \frac{-Z + (1-p_{s}^{*}) }{(1-p_{s}^{*}) \left(\frac{Z}{1-(n-1)p_{s}^{*}} + 1\right)}  \right) - \log \frac{x_{s}^{*}}{x_{l}^{*}}.
\end{equation}
Observe the following equivalence of events
\begin{equation}
\label{eq:separatingAlphaToOneSide}
E_{2} = E_{Z} \cup \left( \overline{E}_{Z}  \cap  \tilde{E}_{2} \right).
\end{equation}
By substituting the definition of $Z$ given in \eqref{eq:KeyTerm-AlphaFair}, the expression $-Z + (1-p_{s}^{*})$ can be expressed as a quadratic in $p_{s}^{*}$ (with a negative coefficient of the term $p_{s}^{*2}$) whose smaller ($p_{s-}^{*}$) and larger ($p_{s+}^{*}$) roots are 
\begin{equation}
p_{s \mp}^{*} = \frac{3 \alpha n - 1 \mp \sqrt{\alpha n (\alpha n + 4 n - 6) + 1}}{2 \alpha n^{2}}. 
\end{equation}
Therefore \eqnref{separatingAlphaToOneSide} is equivalent to 
\begin{equation}
\label{eq:fAlphaTransformed}
E_{2} = E_{Z}' \cup \left( \overline{E}_{Z}'  \cap  \tilde{E}_{2} \right),
\end{equation}
where
\begin{eqnarray}
E_{Z}'  =  \{ p_{s}^{*} \in (0, p_{s-}^{*})\}, ~ \overline{E}_{Z}'   =  \{ p_{s}^{*} \in (p_{s-}^{*}, 1/n) \},
\end{eqnarray}
because we can verify that $p_{s+}^{*} > \frac{3 \alpha n - 1}{2 \alpha n^{2}} > \frac{1}{n}$ for $\alpha \geq 1$, $n \geq 2$, and that $E_{Z}'  = E_{Z} $, $\overline{E}_{Z}'   = \overline{E}_{Z}$.

So we focus on the events $\left( \overline{E}_{Z}'  \cap  \tilde{E}_{2} \right)$ in \eqnref{fAlphaTransformed}.
Our goal now is to show there exists one and only one thresholding $\mathring{p}_{s}^{*} \in (p_{s-}^{*}, 1/n)$ upon which $\tilde{f}_{2}$ (and hence $f_{2}$, as implied by \eqnref{fAlphaTransformed}) changes its sign. We compute $\frac{\partial \tilde{f}_{2}}{\partial p_{s}^{*}}$ and find the stationary point(s) of $\tilde{f}_{2}$ is (are) the root(s) of the cubic equation
\begin{eqnarray}
\label{eq:cubicEquationForThresholdingps}
f_{\rm cubic} (p_{s}^{*}) & \equiv & p_s^{*3} \left(\alpha ^2 n^3-\alpha  n^3+2 \alpha  n^2-n^2+n\right) +   \nonumber \\
& &   p_s^{*2} \left(-5 \alpha ^2 n^2+2 \alpha  n^2+2 n^2-4 \alpha  n-3 n+3\right)  + \nonumber \\
& &  p_s^* \left(8 \alpha ^2 n-2 n\right)   + 1  - 4 \alpha ^2.
\end{eqnarray}
Lem.\ \ref{lem:cubicEquationHasOneValidRoot} shows this cubic equation has only one root on $(0, 1/n)$. It follows that $\tilde{f}_{2}$ has only one root on $(p_{s-}^{*}, 1/n)$. To see this, first note $\tilde{f}_{2}$ cannot have any root on $(0, p_{s-}^{*}]$ as otherwise $f_{2} < 0$ wouldn't hold for all $p_{s}^{*} \in (0, p_{s-}^{*}]$, contradicting \eqref{eq:fAlphaTransformed}.  Second, for the interval $(p_{s-}^{*}, 1/n)$, we prove by contradiction: assuming $\tilde{f}_{2}$ has two or more roots on $(p_{s-}^{*}, 1/n)$, since it can be verified that $\tilde{f}_{2}(1/n) = 0$, due to the continuity of $\tilde{f}_{2}$ and $\frac{\partial \tilde{f}_{2}}{\partial p_{s}^{*}}$, $\tilde{f}_{2}$ must necessarily have at least two stationary points on $(p_{s-}^{*}, 1/n)$ meaning $f_{\rm cubic}$ defined in \eqnref{cubicEquationForThresholdingps} has at least two roots on $(p_{s-}^{*}, 1/n) \subseteq (0, 1/n)$ contradicting Lem. \ref{lem:cubicEquationHasOneValidRoot}. In fact since it can be further verified that the second derivative of $\tilde{f}_{2}$ w.r.t.\ $p_{s}^{*}$ at $p_{s}^{*} = 1/n$ evaluates to a positive number $(1 -  2 p_{s}^{*}) (1 -  2 p_{s}^{*} + \alpha) / \left( (1 - p_{s}^{*})^{2} p_{s}^{*4} \right)$ meaning $\tilde{f}_{2}$ has a local minimum at $p_{s}^{*} = 1/n$: this implies if $\tilde{f}_{2}$ has more than one root on $(p_{s-}^{*}, 1/n)$, then it must necessarily have at least three roots on this interval as $\lim_{p_{s}^{*} \to p_{s-}^{*}} \tilde{f}_{2} = -\infty$. See Fig.\ \ref{fig:TildeTildefAndItsFirstAndSecondDerivativesWRTps} (top column) for an illustration.

The thresholding $\mathring{p}_{s}^{*}(\alpha>1,n>2)$ upon which $\tilde{f}_{2}$ changes its sign (namely the root of $\tilde{f}_{2}(p_{s}^{*})$) can be obtained by solving the following polynomial equation in $p_{s}^{*}\in \left(p_{s-}^{*}, \frac{1}{n}\right)$: 
\begin{equation} 
\label{eq:thresholdingps}
 -\alpha (n {p}_{s}^{*} - 2) (n {p}_{s}^{*} - 1) + (1-{p}_{s}^{*}) - \left(  \frac{(n-1) {p}_{s}^{*2}}{(1-{p}_{s}^{*}) (1-(n-1){p}_{s}^{*})}  \right)^{\alpha} (1-{p}_{s}^{*}) \left(\frac{\alpha (n {p}_{s}^{*} - 2) (n {p}_{s}^{*} - 1)}{1-(n-1){p}_{s}^{*}} + 1\right) = 0.
\end{equation}
This follows by substituting the definition of the terms given in the event $ \left( \overline{E}_{Z}  \cap  \tilde{E}_{2} \right)$ in \eqref{eq:separatingAlphaToOneSide} and recalling \eqref{eq:fAlphaOriginal} and \eqref{eq:d2FdT2ApplyingkStarEqNminusOne-AlphaFair}.

Finally, to obtain the thresholding $\mathring{\theta}_{\alpha}(n)$, we employ \eqnref{AlphaFair-TpsknprimeStar} which yields 
\begin{equation} 
\label{eq:thresholdingtheta}
\mathring{\theta}_{\alpha}(n) = T(\mathring{p}_{s}^{*}(\alpha,n), n-1, n),
\end{equation}
for $T(p_{s}, k, n')$ defined in \eqnref{signedtputgap}.

\underline{Subcase 2}: $n = 2$

In this case, we will show that the T-F curve is concave decreasing for all $p_{s}^{*} \in (0, 1/n)$. First, notice both $\tilde{f}_{2}$ and $\frac{\drm \tilde{f}_{2}}{\drm \alpha}$ simplify
\begin{eqnarray} 
\label{eq:TildeTildefAndItsDerivativeWRTAlphaWhennEqualsTwo}
\tilde{f}_{2} & = & \frac{1}{\alpha}\log \left(\frac{2}{1 + 2 \alpha -4 \alpha  p_{s}^{*}}-1\right)- 2 \log \left(\frac{p_{s}^{*}}{1-p_{s}^{*}}\right),  \nonumber \\
\frac{\drm \tilde{f}_{2}}{\drm \alpha} & = & -\frac{2 (2 \alpha -1) (2 \alpha +1) (1 -2 p_s^{*})^2}{(1 -p_s^{*}) p_s^{*} (-1-2 \alpha +4 \alpha  p_s^{*}) (1-2 \alpha +4 \alpha  p_s^{*})}.
\end{eqnarray}
Furthermore, since the smaller root $p_{s-}^{*}$ simplifies to $\frac{1}{4}\left( 2 - \frac{1}{\alpha} \right)$, it can be verified that $\frac{\drm \tilde{f}_{2}}{\drm \alpha}$ is continuous and positive on $(p_{s-}^{*}, 1/2)$ and $\frac{\drm }{\drm \alpha} \tilde{f}_{2}(1/2) = 0$, $\lim_{p_{s}^{*} \to p_{s-}^{*}} \frac{\drm \tilde{f}_{2}}{\drm \alpha} = \infty$.  This means $\tilde{f}_{2}$ is increasing on $p_{s}^{*} \in (p_{s-}^{*}, 1/2)$, from $-\infty$ (when $p_{s}^{*} \to p_{s-}^{*}$) to $0$ (when $p_{s}^{*}  = 1/2$). Also see Fig.\ \ref{fig:TildeTildefAndItsFirstAndSecondDerivativesWRTps} (bottom column) for an illustration. Therefore, according to \eqref{eq:fAlphaTransformed} and by recalling \eqnref{d2FdT2ApplyingkStarEqNminusOne-AlphaFair} through \eqnref{separatingAlphaToOneSide}, it means when $n = 2$, the T-F curve is concave for all $p_{s}^{*} \in (0, 1/n)$.

\textbf{Case 2: $\alpha = 1$}

In this case, according to \eqref{eq:d2FdT2ApplyingkStarEqNminusOne-AlphaFair}, the sign of $\frac{\drm^{2}}{\drm \theta^{2}} F_{\alpha}^{*}(\theta; n)$ is opposite to the sign of $f_{\rm quad}(p_{s}^{*}, \alpha, n)$. The symmetry axis of $f_{\rm quad}(p_{s}^{*}, \alpha, n)$ is given by $\frac{3}{2n}$ and it is decreasing on $p_{s}^{*} \in (0, 1/n)$ from $1$ to $2/n - 1$. Therefore, when $n = 2$, $f_{\rm quad}(p_{s}^{*}, \alpha, n)$ remains positive for all $p_{s}^{*} \in (0, 1/n)$ meaning the T-F curve is always concave.  When $n > 2$, the thresholding $\mathring{p}_{s}^{*}$ is the smaller root of this quadratic namely 
\begin{equation}
\mathring{p}_{s}^{*}(1,n > 2) = \frac{1}{2 n} \left( 3 - \sqrt{\frac{5 n - 9}{n - 1}}\right),
\end{equation}
which can be verified to be the same as obtained by solving \eqref{eq:thresholdingps} with $\alpha = 1$.

Item $3)$. 
We now investigate the dependence on $n$ while holding $\alpha \geq 1$ and target throughput $\theta \in (\theta_{n}, 1)$ both fixed. In this case it is clear from \thmref{mainresultAlphaFairWhenAlphaGreaterThanOrEqualToOne} that $F_{\alpha}^{*}$ in \eqnref{FAlphaStarThetaAlphaN} should be understood as 
\begin{equation}
F_{\alpha}(p_{s}^{*}(n), n) \equiv F_{\alpha} (\xbf(\pbf(p_{s}^{*}(n), n-1, n)); n). 
\end{equation}
In the following we compute the \textit{total derivative} of $F_{\alpha}(p_{s}^{*}(n), n)$ w.r.t.\ $n$ and show it is negative.
\begin{eqnarray}
\label{eq:TotalDerivativeDfDn}
\frac{\drm }{\drm n} F_{\alpha} (p_{s}^{*}, n)
&=& \frac{\partial}{\partial p_{s}^{*}}  F_{\alpha}(p_{s}^{*}, n) \frac{\drm p_{s}^{*}(n)}{\drm n} + \frac{\partial}{\partial n}  F_{\alpha}(p_{s}^{*}, n) \nonumber \\
& \stackrel{(a)}{=} & \frac{\partial}{\partial p_{s}^{*}} F_{\alpha}(p_{s}^{*}, n)  \left( - \frac{   \frac{\partial }{\partial n} T(p_{s}^{*}, n-1, n)   }{   \frac{\partial }{\partial p_{s}^{*}} T(p_{s}^{*}, n-1, n)  }\right) + \frac{\partial}{\partial n} F_{\alpha}(p_{s}^{*}, n),
\end{eqnarray}
where $(a)$ is by using the implicit function theorem, analogous to \eqnref{dpsdk}. We now address the cases $\alpha = 1$ and $\alpha > 1$ respectively. When $\alpha = 1$ \eqnref{TotalDerivativeDfDn} simplifies to
\begin{eqnarray}
\frac{\drm F_{1}}{\drm n} 
& = & \log x_{s}^{*} + \frac{(n-1) p_{s}^{* 2} + \log (1 - p_{s}^{*})}{p_{s}^{*} (1 - (n-1) p_{s}^{*})}  \nonumber \\
& < & \log x_{s}^{*} - 1 < 0,
\end{eqnarray}
where the first bounding is by applying \eqnref{InequalityOnLog} to $\log (1 - p_{s}^{*})$. When $\alpha > 1$ \eqnref{TotalDerivativeDfDn} can be shown to be 
\begin{eqnarray}
& & \frac{\drm F_{\alpha > 1}}{\drm n} 
= - \frac{x_{s}^{* (1-\alpha)} x_{l}^{* (- \alpha)} }{(\alpha -1) p_{s}^{*} (1 - n p_{s}^{*})} f_{1}(p_{s}^{*}, n) 
\end{eqnarray}
where
\begin{eqnarray}
f_{1}(p_{s}^{*}, n) &  \equiv & (\alpha -1) (1 - p_{s}^{*}) (p_{s}^{*} + \log (1 - p_{s}^{*})) \left( x_{s}^{* \alpha} - x_{l}^{* \alpha} \right)  + \alpha (1 - n p_{s}^{*}) p_{s}^{*} x_{l}^{* \alpha}.
\end{eqnarray}
As $p_{s}^{*} \in (0, 1/n)$, it follows that 
\begin{equation}
p_{s}^{*} + \log (1 - p_{s}^{*}) < p_{s}^{*} + (- p_{s}^{*}) = 0.
\end{equation}
As $x_{s}^{* \alpha} < x_{l}^{* \alpha}$, it follows that $f_{1}(p_{s}^{*}, n)$ is the summation of two positive numbers and hence $\frac{\drm F_{\alpha > 1}}{\drm n} < 0$.

\end{IEEEproof}